\documentclass[10pt]{article} 
\usepackage[preprint]{tmlr}

\usepackage{amsmath,amsfonts,bm}

\def\eqref#1{equation~\ref{#1}}

\def\1{\bm{1}}

\DeclareMathAlphabet{\mathsfit}{\encodingdefault}{\sfdefault}{m}{sl}
\SetMathAlphabet{\mathsfit}{bold}{\encodingdefault}{\sfdefault}{bx}{n}

\DeclareMathOperator*{\argmin}{arg\,min}

\usepackage{blindtext}
\usepackage{hyperref}
\usepackage{url}

\usepackage{amsmath}
\DeclareMathOperator*{\argminB}{argmin} 

\usepackage{subcaption}
\usepackage{bm}
\usepackage{dsfont}
\usepackage[export]{adjustbox}
\usepackage{algorithm}
\usepackage{algpseudocode}
\usepackage{placeins} 

\def\appropto{%
  \def\p{%
    \setbox0=\vbox{\hbox{$\propto$}}%
    \ht0=0.6ex \box0 }%
  \def\s{%
    \vbox{\hbox{$\sim$}}%
  }%
  \mathrel{\raisebox{0.7ex}{%
      \mbox{$\underset{\s}{\p}$}%
    }}%
}

\usepackage{lastpage}

\title{Multi-objective Bayesian optimization for Likelihood-Free inference in sequential sampling models of decision making}

\author{\name David Chen \email e1039688@u.nus.edu \\
       \addr Department of Statistics and Data Science \\
       National University of Singapore\\
       117546, Singapore
       \AND
       \name Xinwei Li \thanks{David Chen and Xinwei Li contributed equally to this work.} \email xinwei.li@u.nus.edu \\
       \addr Department of Civil and Environmental Engineering\\
       National University of Singapore\\
       117576, Singapore
       \AND
       \name Eui-Jin Kim \email euijin@ajou.ac.kr \\
       \addr Department of Transportation Systems Engineering\\
       Ajou University\\
       16499, Korea
       \AND
       \name Prateek Bansal \email prateekb@nus.edu.sg \\
       \addr Department of Civil and Environmental Engineering\\
       National University of Singapore\\
       117576, Singapore
       \AND
       \name David Nott \email standj@nus.edu.sg \\
       \addr Department of Statistics and Data Science \\
       National University of Singapore\\
       117546, Singapore
       }

\def\month{06}  
\def\year{2025} 
\def\openreview{\url{https://openreview.net/forum?id=hQjwDqfSzj}} 

\begin{document}

\maketitle

\begin{abstract}
~~~~Statistical models are often defined by a generative process for simulating synthetic data, but this can lead to intractable likelihoods. Likelihood free inference (LFI) methods enable Bayesian inference to be performed in this case. Extending a popular approach to simulation-efficient LFI for single-source data, we
propose Multi-objective Bayesian Optimization for Likelihood Free Inference (MOBOLFI) to perform LFI using multi-source data.  MOBOLFI models a multi-dimensional discrepancy between observed and simulated data, using a separate discrepancy for each data source. The use of a multivariate discrepancy allows for approximations to individual data source likelihoods in addition to the joint likelihood, enabling detection of conflicting information and deeper understanding of the importance of different data sources in estimating individual parameters.  
The adaptive choice of simulation parameters using multi-objective Bayesian optimization ensures simulation efficient approximation of likelihood components for all data sources.  We illustrate our approach in sequential sampling models (SSMs), which are widely used in psychology and consumer-behavior modeling.  SSMs are often fitted using multi-source data, such as choice and response time.  The advantages of our approach are illustrated in comparison with a single discrepancy for an SSM fitted to data assessing preferences of ride-hailing drivers in Singapore to rent electric vehicles. 

\end{abstract}
\section{Introduction}

~~~~In many scientific applications, the most natural way to incorporate subject matter knowledge into a statistical model is by specifying a generative algorithm (or simulator) for synthetic data. In this case, the likelihood function can sometimes be difficult to compute, and this makes traditional statistical inference methods hard to apply. In response to this challenge, researchers have developed likelihood-free inference (LFI) methods for 
conducting Bayesian inference using only simulation from the model, without any 
likelihood evaluations.  An alternative name for LFI is simulation-based inference (SBI), which stresses the role of simulating synthetic data.  The phrase ``likelihood-free'' should not be taken to mean that there is no likelihood.  Rather, it means that approximate likelihoods or approximate posterior sampling methods can be constructed from model simulations, without evaluating the likelihood. 

~~~~Our work considers settings where it is important to obtain accurate approximations
to the posterior density using as few simulations as possible, and where the computationally intractable likelihood function
factorizes into terms for multiple  data sources.    We extend a widely-used simulation
efficient method for single source data, Bayesian optimization for likelihood-free
inference (BOLFI) \citep{gutmann+c16}. BOLFI uses Bayesian 
optimization to sequentially choose the simulation parameters
to make best use of a limited computational budget. 
In the case of multi-source data, we consider a multi-objective extension 
which we call MOBOLFI, which measures similarity between synthetic and observed data by
a multi-dimensional discrepancy with a component for each data source.  The
multi-dimensional discrepancy is modelled, and this allows approximation of likelihood
terms for different data, which is important for understanding the importance
of different data sources for inference about individual parameters and for model
checking.  While BOLFI provides a closed form approximate likelihood for inference tasks, it does not provide approximations of likelihood terms for individual data sources for multi-source data.  In the multi-source setting,  
it is often natural to choose summary statistics and discrepancy metrics separately
for each data source, and with the classic BOLFI approach we then have to 
combine all discrepancies into a single one.  
If this is not done carefully, the information in the combined discrepancy may
not reflect correctly the relative importance of different data sources, resulting
in information loss and highly conservative uncertainty quantification.

~~~~There are many existing methods for LFI.  One of the most well-established methods is approximate Bayesian computation \citep[ABC,][]{sisson+fb18}, which approximates the posterior by parameter samples for which a simulated dataset has a discrepancy from the observed dataset below a certain tolerance. See Section~\ref{subsec:ABC} for a more detailed discussion. ABC implicitly uses a kernel density approximation to the likelihood for data summaries. A competitive approach, probability density approximation \citep[PDA,][]{turner2014generalized}, has also been widely used for over a decade. However, 
with high-dimensional summary statistics, both ABC and PDA can perform poorly, making
then unsuitable for many highly parameterized models. 

~~~~Structured density estimation approaches which deal better with high-dimensional summary statistics include synthetic likelihood, which is based on a working normal model for summary
statistics \citep{wood2010statistical,frazier2019bayesian}, flexible extensions of synthetic likelihood based on copulas 
or empirical saddle point approximation \citep{fasiolo+whb18,an2020robust} and neural density estimation
approaches \citep{papamakarios+m16,papamakarios+sm19,greenberg+nm19}.  
 However, 
these structured density estimation methods require a large number
of model simulations to obtain adequate approximations, which may be infeasible
if simulation from the model is computationally expensive.  A preferred
approach in such settings is to use active learning approaches such as
Bayesian optimization for Likelihood-Free
inference \citep[BOLFI,][]{gutmann+c16}, and extending these
methods is the focus of our work.  We review the BOLFI approach in Section~\ref{subsec:BOLFI}, and related existing literature 
in Section A.2 of the Appendix.  

~~~~Another challenge for LFI methods is model misspecification \citep{wilkinson13,frazier+rr20,frazier+d21,ward+cbfs22}. 
There is an extensive literature on this
issue, but here we focus on model checking with multi-source data, where 
the different data sources may bring conflicting information about the model
parameter \citep{presanis+osd13}.
Modern LFI methods have been used in the context of Sequential Sampling Models \citep{radev2023bayesflow} and for multi-source data \citep{schmitt2023fuse, bahg2020gaussian}, 
but questions of model adequacy have been previously addressed in a global way.  The MOBOLFI method proposed here 
has the capability to approximate both the joint likelihood as well
as likelihoods for individual data sources, enabling checks of the consistency
of information supplied by different parts of the data.  Examining this consistency
is essential for trustworthy analysis of complex multi-source data such
as those arising for Sequential Sampling Models (SSMs).  It is also valuable for understanding which data sources
are most informative for the estimation of individual SSM parameters.

~~~~This work makes three main contributions.  First, the multi-objective
aspect of our MOBOLFI method ensures efficient
exploration of the high-likelihood region 
for the separate likelihoods of all data sources.  Importantly, in the case of a computationally expensive simulator, the MOBOLFI approach can explore high-likelihood regions for all data source likelihoods in a single run.  This is particularly difficult to do with other approaches when the high-likelihood regions for individual data source likelihoods are disjoint.  Our second main contribution relates
to the difficulty of combining multiple discrepancies from individual data sources
into a single discrepancy, with consequent information 
loss and highly conservative uncertainty quantification if poor choices are made.  MOBOLFI avoids the need 
to do this, making it easier to use
in practice.  Finally, MOBOLFI is able to leverage likelihood approximations
for individual data sources to check their consistency 
and understand their importance for estimating individual SSM parameters.   

~~~~The paper is organized as follows. Section~\ref{sec:method} gives background
on Bayesian optimization and the BOLFI method of \citet{gutmann+c16}.  
Section~\ref{sec:MOBOLFI} then discusses multi-objective optimization, the
motivation for the MOBOLFI method, and its implementation.  
Section~\ref{sec:example} presents two instructive examples,
one involving a Brownian motion and the second involving a SSM. 
These examples illustrate the advantages of MOBOLFI compared to the 
BOLFI method with a single
discrepancy in settings involving
multi-source data. In the second example, we use an empirical choice-response time (choice-RT) dataset on consumer preferences for electric vehicles, and demonstrate MOBOLFI's application in establishing the importance of different data sources
for estimation of parameters of an SSM. Key takeaways and avenues for future research are discussed in section~\ref{sec:discussion}.

\section{Method}\label{sec:method}
~~~~Before explaining the MOBOLFI method, we give some necessary background on 
Bayesian optimization, ABC and the traditional BOLFI approach.  Describing our method requires
introducing some complex notation in Sections 2 and 3.
To help the reader, a
glossary of notation is given in Appendix A.  

\subsection{Approximate Bayesian computation}\label{subsec:ABC}

~~~~ABC methods approximate the likelihood by the probability of a synthetic
dataset being close to the observed data in terms of some discrepancy.  
The discrepancy is often defined in terms of
data summary statistics.  It is important to explain 
the ABC likelihood approximation, since this 
motivates the BOLFI approach discussed in subsection~\ref{subsec:BOLFI}. 
For a more extensive discussion of ABC than we give here see
\citet{sisson+fb18}.

~~~~Let $\theta$ be parameters in a statistical model for data $y\in\mathcal{Y}$
to be observed. Write $y_{\text{obs}}$ for the observed value and 
$p(y_{\text{obs}}|\theta)$ for the likelihood, where $p(y|\theta)$ 
is the density of $y|\theta$.  In ABC, inference is usually based on
a lower-dimensional summary of the original data of dimension $d$ say, 
defined by a mapping $S:{\mathcal Y}\rightarrow \mathbb{R}^d$.  The observed
summary statistic value $S(y_{\text{obs}})$ will be denoted by
$S_{\text{obs}}$.  

~~~~In ABC we first approximate the posterior density $p(\theta|y_{\text{obs}})\propto
p(\theta)p(y_{\text{obs}}|\theta)$ by the partial posterior density which conditions on $S_{\text{obs}}$ rather than $y_{\text{obs}}$, 
$$p(\theta|S_{\text{obs}})\propto p(\theta)p(S_{\text{obs}}|\theta).$$
~~~~If the summary statistics $S$ are sufficient, then there is no loss
in replacing $y_{\text{obs}}$ with $S_{\text{obs}}$, but non-trivial sufficient
statistics will not exist in most complex models of interest, and we
only require summary statistics to be informative about $\theta$.  Next, 
since $p(S_{\text{obs}}|\theta)$
is infeasible to compute if $p(y_{\text{obs}}|\theta)$ is, we replace
$p(S_{\text{obs}}|\theta)$ with the ``ABC likelihood'' 
denoted $p_t(S_{\text{obs}}|\theta)$ for some tolerance
$t>0$:  
\begin{align}
  p_t(S_{\text{obs}}|\theta) & \propto \int p(S|\theta)I(\Delta_\theta(S,S_{\text{obs}})<t)\,dS,  \label{abclikelihood}
\end{align}
where $\Delta_\theta(S,S_{\text{obs}})=\|S-S_{\text{obs}}\|$ and $\|\cdot \|$
is some distance measure.  $\Delta_\theta(S,S_{\text{obs}})$ is a measure
of the discrepancy between the simulated and observed summary statistics.  
As discussed further below, the ABC likelihood at $\theta$ can be thought of as 
proportional to the probability that a synthetic dataset $S\sim p(S|\theta)$ is within $t$ of the observed data in terms of the discrepancy.  
The ABC posterior is
\begin{align}
  p_t(\theta|S_{\text{obs}}) & \propto p(\theta)p_t(S_{\text{obs}}|\theta),   \label{abcposterior}
\end{align}
and there are a variety of methods for sampling from \eqref{abcposterior}, 
which are based on the fact that the right-hand side
of \eqref{abclikelihood} can be
estimated unbiasedly by $I(\Delta_\theta(S,S_{\text{obs}})<t)$
for $S\sim p(S|\theta)$.  The discussion above can be generalized by
replacing the indicator function with a more general kernel.

\subsection{Bayesian optimization} \label{subsec:BO}

~~~~Next we briefly describe Bayesian optimization (BO), which is used in the BOLFI method
discussed in subsection~\ref{subsec:BOLFI}.  A more detailed introduction can be found in \citet{garnett23}.
Bayesian optimization is used for finding a global optimum 
of a function 
$f(\theta)$, $\theta\in \Theta\subseteq \mathbb{R}^p$, where derivatives
of $f(\cdot)$ are not available and evaluations of $f(\cdot)$ may
be corrupted by noise.  Here we consider minimization problems, but changing
sign of the objective function turns minimization into maximization.
BO models noisy evaluations of $f(\theta)$ with a 
surrogate model 
describing our uncertainty about $f(\cdot)$ given the 
function evaluations made so far.  This surrogate model
is usually chosen to be a Gaussian
process \citep{rasmussen03}, and it guides the decision of where
further function observations should be made.

~~~~Since the surrogate model we use for BO is a Gaussian process (GP), we need
some background about GPs.  A random function $f(\cdot)$ defined on $\Theta$ is a GP
with mean function $\mu:\Theta\rightarrow \mathbb{R}$ and positive
definite covariance function $C:\Theta\times \Theta\rightarrow \mathbb{R}$ if,
for any $n$, and any $\theta_1,\dots, \theta_n\in\Theta$, 
the random vector $f(\theta_{1:n})=(f(\theta_1),\dots, f(\theta_n))^\top$ 
is multivariate normally
distributed, with mean vector $(\mu(\theta_1),\dots, \mu(\theta_n))$, and 
covariance matrix $C(\theta_{1:n},\theta_{1:n})=[C(\theta_i,\theta_j)]_{i,j=1}^n$. Suppose we observe the Gaussian process $f(\cdot)$ with noise at points 
$\theta_1,\dots, \theta_n\in \Theta$.  The noisy observations are
\begin{align}
  z_i & = f(\theta_i)+\epsilon_i,\;\;\;\;i=1,\dots, n, \label{noisyobs}
\end{align}
where $\epsilon_i\stackrel{iid}{\sim} N(0,\sigma^2)$, for
some variance $\sigma^2>0$.  Write $z_{\leq n}=(z_1,\dots, z_n)^\top$.  
We are interested in describing uncertainty about $f(\theta^*)$ for
some $\theta^*\in \Theta$, given the noisy observations 
$z_{\leq n}$.  The distribution of  
$f(\theta^*)|z_{\leq n}$ is Gaussian, $N(\mu_n(\theta^*),\sigma_n^2(\theta^*))$,
where the form of $\mu_n(\theta^*)$ and $\sigma_n^2(\theta^*)$
are given in Section A.1 in the Appendix. 

~~~~The uncertainty quantification provided by
the Gaussian process surrogate can be used to decide which $\theta^*$ 
should be used to obtain a further noisy observation 
$$z^*=f(\theta^*)+\epsilon^*,\;\;\;\; \epsilon^*\sim N(0,\sigma^2),$$ 
in our search for the minimizer of $f(\cdot)$.  
$\theta^*$ is usually chosen to minimize a so-called acquisition function. 
As an example,
in the BOLFI method described next, \citet{gutmann+c16}
suggested using the lower confidence bound acquisition function
\citep{cox+j97, srinivas2012information}, 
\begin{align}
 A_n(\theta) & = \mu_n(\theta)-\sqrt{\eta_n^2 \sigma_n^2(\theta)}, \hspace{0.2in}
\text{where}\hspace{0.2in} 
\eta_n^2=2\log \left(n^{\frac{p}{2}+2}\frac{\pi^2}{3\epsilon_\eta}\right), \label{lcb}
\end{align}
with the default value $\epsilon_\eta=0.1$.  
Intuitively, $\mu_n(\theta)$ is an estimate of $f(\theta)$ from
the noisy observations so far, and choosing $\theta^*$ to minimize
$A_n(\theta)$ encourages choosing $\theta$ where
this estimate is small, or where $\sigma_n^2(\theta)$ is large and
we are highly uncertain about $f(\theta)$.  Hence 
the Gaussian process model can help to manage an 
``exploration-exploitation trade-off''
in searching for the minimum.  
Other acquisition functions can also be used \citep{jarvenpaa+gpvm19}.
Forms for the mean and covariance function need to be specified, 
and any parameters, including the noise $\sigma^2$, estimated.  This is usually done using marginal maximum likelihood.   

\subsection{BOLFI} \label{subsec:BOLFI}
~~~~We now discuss the use of Bayesian optimization in the BOLFI method
of \citet{gutmann+c16}.  
A discussion of more recent work improving the BOLFI method is given
in Section A.2 of the Appendix.
We can rewrite the ABC likelihood 
\eqref{abclikelihood} as
\begin{align}
  p_t(S_{\text{obs}}|\theta) & \propto Pr(\Delta_\theta(S,S_{\text{obs}})<t), \label{bolfilikelihood}
\end{align}
for $S\sim p(S|\theta)$.  Hence the ABC likelihood can be approximated
if we know the distribution of $\Delta_{\theta}(S,S_{\text{obs}})$, 
$S\sim p(S|\theta)$.  This suggests we may be able to approximate the
distribution of $\Delta_{\theta}(S,S_{\text{obs}})$ as a function
of $\theta$ using regression, in order to obtain an approximation of the ABC
likelihood.  This is what the BOLFI method does, while using a sequential design 
approach based on BO to choose which $\theta$ to simulate from 
next for maximum benefit.  
The sequential design aspect 
allows simulation efficient exploration of the high likelihood region, making
BOLFI highly suited for the case of computationally demanding simulation models.

\citet{gutmann+c16} propose selecting parameter values for simulation 
using a BO algorithm to minimize the expected discrepancy function:
\begin{align}
  D(\theta) & = E(\Delta_\theta(S,S_{\text{obs}})),  \label{expectedD}
\end{align}
where the expectation is taken with respect to $S\sim p(S|\theta)$.  
In their approach, first some initial set of locations $\theta_i\in \Theta$, $i=1,\dots, n_0$, 
are chosen according to some space-filling design.  
Synthetic data are then simulated at these locations to obtain discrepancy values $\Delta_i=\Delta_{\theta_i}(S_i,S_{\text{obs}})$, $S_i\sim p(S|\theta_i)$, $i=1,\dots, n_0$.  We can write
\begin{align}
 \Delta_i & = D(\theta_i)+\epsilon_i,\;\;\;i=1,\dots, n_0,  \label{BOLFImodel}
\end{align}
where the $\epsilon_i$ are zero mean error terms.  In BOLFI it
is usually assumed, perhaps after a transformation of the $\Delta_i$, 
that $\epsilon_i\sim N(0,\sigma^2)$ for some variance parameter
$\sigma^2$.  

~~~~Next, model $D(\theta)$ as a Gaussian process, and
denote the training data used to fit the Gaussian process model required for BO as
$T_{n_0}=\{(\theta_i,\Delta_i):i=1,\dots, n_0\}.$
A new location $\theta_{n_0+1}$ is then chosen 
to simulate the next summary statistic value and obtain the discrepancy
$\Delta_{n_0+1}$.  This is done by optimization the BO acquisition
function, which uses the uncertainties of the Gaussian process model
to define the benefit of simulating a new discrepancy at any $\theta$.  
The Gaussian process is then refitted with training data $T_{n_0+1}=T_{n_0}\cup \{(\theta_{n_0+1},\Delta_{n_0+1})\}$.
The process of optimization of the acquisition function, simulation and
retraining is repeated until some computational budget $n_f>n_0$ of simulations
has been exhausted.  
The final Gaussian process model is then fitted to the training data 
$T_{n_f}$.  
Using the Gaussian process model and the noise assumption for \eqref{BOLFImodel}, we can approximate the distribution of a discrepancy value observed at any 
$\theta$.  
If we assume that 
$\Delta=D(\theta)+\epsilon$, and write $\mu_{n_f}(\theta), \sigma_{n_f}^2(\theta)$ 
for the mean and variance of the predictive distribution of $D(\theta)$ given
$T_{n_f}$, then $D(\theta)\sim N(\mu_{n_f}(\theta),\sigma_{n_f}^2(\theta))$ and $\epsilon\sim N(0,\sigma^2)$ independently.  
We can approximate the distribution of $\Delta$ as $\Delta\sim N(\mu_{n_f}(\theta),\sigma_{n_f}^2(\theta)+\sigma^2)$. Further details about the derivation
and form of $\mu_{n_f}(\theta)$ and $\sigma_{n_f}^2\theta)$ are given in Section A.1 in
the Appendix.  Employing this Gaussian
approximation to calculate the tail probability on the right-hand side of 
\eqref{bolfilikelihood}, we obtain 
$$p_t(S_{\text{obs}}|\theta) \appropto \Phi\left(\frac{t-\mu_{n_f}(\theta)}{\sqrt{\sigma_{n_f}^2(\theta)+\sigma^2}}\right),$$
where $\appropto$ denotes ``approximately proportional to''.  
This likelihood approximation can be used with MCMC sampling to draw
approximate posterior samples from $p_{t}(\theta|S_{\text{obs}})$.  
The choice of tolerance $t$ in our examples is discussed later.

\section{The MOBOLFI method} \label{sec:MOBOLFI}

~~~~Next we describe LFI with 
multi-source data with a discrepancy for each data source and
the method of approximating the likelihood we consider.  This is followed
by background on multi-objective optimization, and finally the description
of the new MOBOLFI method.  

\subsection{Likelihood approximation with multiple discrepancies} \label{subsec:LAMD}
~~~~To ease notation, we consider the
case of two data sources but the extension to three or more sources is
immediate. Suppose the data $y$ comprises $y=(x^\top,w^\top)^\top\in \mathcal{Y}=\mathcal{X}\times \mathcal{W}$ and decompose the joint
density for $y|\theta$ as $p(y|\theta)=p(x|\theta)p(w|x,\theta)$. The observed data is
$y_{\text{obs}}=(x_{\text{obs}}^\top,w_{\text{obs}}^\top)^\top$,
the likelihood is $p(y_{\text{obs}}|\theta)=p(x_{\text{obs}}|\theta)p(w_{\text{obs}}|x_{\text{obs}},\theta)$, 
and now there are summary statistics $S=(T^\top,U^\top)^\top:\mathcal{Y}\rightarrow \mathbb{R}^d$, where $S$ 
concatenates summary statistic mappings $T$ and $U$ for the
data sources $x$ and $w$ respectively, 
$T:\mathcal{X}\rightarrow \mathbb{R}^b$, 
$U:\mathcal{W}\rightarrow \mathbb{R}^c$, $d=b+c$.  Write $S_{\text{obs}}=S(y_{\text{obs}})$, 
$T_{\text{obs}}=T(x_{\text{obs}})$ and $U_{\text{obs}}=U(w_{\text{obs}})$.  

Similar to our previous discussion of ABC, we replace the likelihood 
$p(y_{\text{obs}}|\theta)$ with the summary statistic likelihood
\begin{align}
  p(S_{\text{obs}}|\theta) & =p(T_{\text{obs}},U_{\text{obs}}|\theta)=
  p(T_{\text{obs}}|\theta)p(U_{\text{obs}}|T_{\text{obs}},\theta).  \label{sslikelihood}
\end{align}
~~~~We then consider two discrepancies, $\Lambda_\theta(T,T_{\text{obs}})$ and
$\Psi_\theta(U,U_{\text{obs}})$, for simulated summary statistic values
$S=(T^\top,U^\top)^\top$, and formulate an ABC likelihood approximating
\eqref{sslikelihood} as 
\begin{align}
  p_t(S_{\text{obs}}|\theta) & \propto 
  \int p(S|\theta)I(\Lambda_\theta(T,T_{\text{obs}})<t_1)
  I(\Psi_\theta(U,U_{\text{obs}})<t_2)\,dS,  \label{mdsabclike}
\end{align}
where $t=(t_1,t_2)^\top$ is a discrepancy vector with $t_1,t_2>0$
Another way of writing \eqref{mdsabclike} is
\begin{align}
 p_t(S_{\text{obs}}|\theta) & \propto
 P(\Lambda_\theta(T,T_{\text{obs}})<t_1,
 \Psi_\theta(U,U_{\text{obs}})<t_2) \nonumber \\
 & = P(\Lambda_\theta(T,T_{\text{obs}})<t_1) \times 
 P(\Psi_\theta(U,U_{\text{obs}})<t_2|\Lambda_\theta(T,T_{\text{obs}})<t_1),  \label{mdsabclike2}
\end{align}
for $S=(T^\top,U^\top)^\top\sim p(S|\theta)$.

~~~~The first term on the right-hand side of \eqref{mdsabclike2} approximates
$p(T_{\text{obs}}|\theta)$, while the second approximates
$p(U_{\text{obs}}|T_{\text{obs}},\theta)$ (up to constants of proportionality).  
We could also approximate $P(U_{\text{obs}}|\theta)$ and $p(T_{\text{obs}}|U_{\text{obs}},\theta)$ by switching the two discrepancies
in \eqref{mdsabclike2}.  
Our extension of BOLFI will
model $\Delta_\theta(S,S_{\text{obs}})$
for $S\sim p(S|\theta)$ as a bivariate
Gaussian process.  We will use this bivariate process for sequential design
using multi-objective Bayesian optimization, and also for approximation
of data-source specific likelihood
contributions such as those shown in \eqref{mdsabclike2}.  

\subsection{Multi-objective optimization} \label{subsec:MOO}

~~~~Our MOBOLFI extension of BOLFI uses multi-objective optimization, and
we describe this now.  
Let $f(\theta)=(f_1(\theta),\dots, f_K(\theta))^\top$ be a multivariate
function, and suppose we wish to minimize the components of $f(\cdot)$.
There
need not be any common $\theta^*\in \Theta$ for all components where
a minimum is achieved, and multi-objective optimization
methods approximate the set of ``nondominated''
solutions which are not obviously inferior to any other solution.
We say that a value $\theta\in \Theta$ dominates $\theta'\in\Theta$
if $f_j(\theta)\leq f_j(\theta')$, $j=1,\dots, K$, with the inequality
being strict for at least one $j$.  The dominated solution
is inferior for minimizing $f(\cdot)$ along some dimensions and no better
for other dimensions. 
Multi-objective optimization
algorithms try to find the set of nondominated points in $\Theta$, 
the ``Pareto optimal set''. 
 
~~~~The Pareto optimal set is infinite in general, and 
numerical multi-objective optimization
methods obtain finite approximations to it.  The Pareto set is mapped by $f(\cdot)$ onto the Pareto frontier, the
set of optimal function values obtained by the points in the Pareto set.
Multi-objective Bayesian optimization \citep[Section 11.7]{garnett23} uses surrogate models
to implement multi-objective optimization for expensive to evaluate
functions, possibly observed with noise.  Similar to Bayesian optimization
with a scalar objective, the representation of uncertainty given by the surrogate
is used to efficiently decide where to perform the next function evaluation.

~~~~The surrogate model in our work is a multivariate Gaussian process, and we
need to explain what this means.  
Let $f(\theta)=(f_1(\theta),\dots, f_K(\theta))^\top$, $\theta\in \Theta$, be a multivariate
random function.  It is a multivariate Gaussian process with mean function
$\mu:\Theta\rightarrow \mathbb{R}^K$, 
$\mu(\theta)=(\mu_1(\theta),\dots, \mu_K(\theta))^\top$, and 
positive definite covariance function $C:\Theta\times \Theta\rightarrow \mathbb{R}^{K\times K}$, if for any $n$, and $\theta_1,\dots, \theta_n\in\Theta$, 
$f(\theta_{1:n})=(f(\theta_1)^\top,\dots, f(\theta_n)^\top)^\top$ is multivariate
Gaussian with mean $\mu(\theta_{1:n})=(\mu(\theta_1)^\top,\dots, \mu(\theta_n))^\top$, and covariance matrix $C(\theta_{1:n},\theta_{1:n})=[C(\theta_i,\theta_j)]_{i,j=1}^n$, with $K\times K$ block elements
$C(\theta_i,\theta_j)$.  Once again extending the discussion of subsection~\ref{subsec:BO}, suppose we observe
values of $f(\cdot)$ with noise at $\theta_1,\dots, \theta_n\in \Theta$, to obtain
\begin{align}
  z_i & = f(\theta_i)+\epsilon_i, \label{mvreg}
\end{align}
where now $z_i\in \mathbb{R}^K$ and $\epsilon_i\stackrel{iid}{\sim} N(0,\Sigma)$
where $\Sigma\in \mathbb{R}^{K\times K}$ is some positive definite
covariance matrix.  For any $\theta^*\in \Theta$, 
and writing $z_{\leq n}=(z_1^\top,\dots, z_n^\top)^\top$,
The distribution of $f(\theta^*)|z_{\leq n}$ is multivariate Gaussian, 
$N(\mu_n(\theta^*),\Sigma_n(\theta^*))$, where the form of $\mu_n(\theta^*)$
and $\Sigma_n(\theta^*)$ are given in Section A.3 in the Appendix.
Given the uncertainty quantification provided by the multivariate
Gaussian surrogate, an acquisition function can be defined.
If there is a finite set of points, say $\theta_1,\dots, \theta_n$, 
approximating the Pareto set, with corresponding approximation
$f_1,\dots, f_n$ of the Pareto frontier, one measure of performance 
that has been used is the volume of the space dominated by the current
approximation of the 
Pareto frontier and bounded below by a reference point, the so-called Pareto hypervolume.  Expected hypervolume improvement (EHVI) was first used as an acquisition function in multi-objective
Bayesian optimization by \citet{emmerich05}.  
For implementing the MOBOLFI method
of the next section, we use the noisy expected hypervolume improvement 
(NEHVI) method of \citet{daulton+bb21} which copes well with noisy function
evaluations and the trialing of batches of solutions in parallel with reduced
computational demands.  The method is implemented in the open source
python package {\tt BoTorch} \citep{balandat20}.  
To the best of our knowledge, no asymptotic guarantees of recovery of the Pareto front have been proven for the NEHVI approach.   \cite{daulton+bb21} derive a regret bound when trialing batches of samples chosen in a greedy fashion, and they also study theoretically the effects of approximating the acquisition function with a sample average function approximation to it.  Some random scalarization algorithms for multi-objective Bayesian optimization do come with theoretical guarantees (e.g. \citealt{zhang+g20}) and exploring such an approach in practice for LFI approximations with multi-source data is an interesting direction for future work.
We will not discuss
further the extensive literature on multi-objective Bayesian optimization, 
but refer the reader to \citet[Section 11.7]{garnett23} for an accessible
introduction.

\subsection{MOBOLFI} \label{subsec:MOBOLFI}

~~~~While BOLFI provides a closed form approximate likelihood for inference tasks, it cannot approximate likelihoods for individual data sources for multi-source data.  Multi-source data has become increasingly common in SSM design in recent years. 
Often it is natural to choose summary statistics and discrepancy metrics separately
for each data source, and with the classic BOLFI approach we then need to 
combine all discrepancies into a single one.  
If this is not done carefully, the information in the combined discrepancy may
not reflect correctly the relative importance of different data sources, resulting
in information loss and highly conservative uncertainty quantification.

~~~~Motivated by these issues for multi-source data, we 
develop our MOBOLFI extension of the original BOLFI method.  It
achieves simulation efficient likelihood
approximations in LFI for multi-source data 
by applying multi-objective BO methods to a vector
of data-source specific discrepancy functions.  Consider again the bivariate
setting and notation of subsection~\ref{subsec:LAMD} for simplicity, 
and define the vector of expected discrepancies
$$D(\theta)=E(\Delta_\theta(S,S_{\text{obs}}))=(D_1(\theta),D_2(\theta))^\top,$$ 
where $D_1(\theta)=E(\Lambda_\theta(T,T_{\text{obs}}))$, 
$D_2(\theta)=E(\Psi_\theta(U,U_{\text{obs}}))$, $S=(T,U)\sim p(S|\theta)$.  
A multi-objective Bayesian optimization algorithm applied to $D(\theta)$
efficiently explores the set of $\theta$ where both of the
data source discrepancy components is likely to be small, leading to
efficient approximations to the likelihood contributions from multiple
data sources.    

~~~~The optimization algorithm proceeds similarly to the case of a univariate
objective. Firstly, we choose some initial set of points $\theta_1,\dots, \theta_n\in \Theta$ according to some space filling design.  We then simulate discrepancy
values $\Delta_i=\Delta_{\theta_i}(S_i,S_{\text{obs}})$,  $S_i\sim p(S|\theta_i)$, $i=1,\dots, n_0$, and we can write
\begin{align}
  \Delta_i & = D(\theta_i)+\epsilon_i, \label{mboptreg}
\end{align}
where the $\epsilon_i$ are zero mean independent errors.  It will be assumed
that $\epsilon_i\sim N(0,\Sigma)$, for some covariance matrix $\Sigma$.  
Assuming a Gaussian process model for $D(\cdot)$, we fit
a Gaussian process surrogate model to the training data 
$T_{n_0}=\{(\theta_i,\Delta_i): i=1,\dots, n_0\}$, learning all
hyperparameters including $\Sigma$ from the data.  
We can then choose the next observation point $\theta_{n_0+1}$ to minimize
the NEHVI acquisition function (or perhaps choose a batch of points), retrain
the GP and acquire new points, continuing until we have $n_f$ training points
for the final fitted GP. In GP fitting, we use a constant mean function
and Mat\'{e}rn covariance kernel.

~~~~From the model \eqref{mboptreg} and the assumed $N(0,\Sigma)$ distribution of the errors, we can approximate the ABC likelihood
\eqref{mdsabclike2} up to a proportionality constant 
by a bivariate Gaussian probability given a well-chosen vector-valued tolerance $t=(t_1,t_2)$:
\begin{align}
  \widetilde{p}_t(S_{\text{obs}}|\theta) := & \Phi((t_1,t_2);\mu_{n_f}(\theta),\Sigma_{n_f}(\theta)+\Sigma),\label{disclikelihood}
\end{align}
where $\Phi(\cdot;\mu_{n_f}(\theta),\Sigma_{n_f}(\theta))$ denotes the
cdf of a normal $N(\mu_{n_f}(\theta),\Sigma_{n_f}(\theta)+\Sigma)$ distribution.  
Henceforth we will omit the ``up to a proportionality constant'' qualification
when we talk about likelihood approximations.  
It is also possible to decompose the bivariate probability \eqref{disclikelihood}
into marginal and conditional components, approximating
the terms $P(\Lambda_\theta(T,T_{\text{obs}})<t_1)$ and
$P(\Psi_\theta(U,U_{\text{obs}})<t_2|\Lambda_\theta(T,T_{\text{obs}})<t_1)$ which in turn approximate
$p(T_{\text{obs}}|\theta)$ and $p(U_{\text{obs}}|T_{\text{obs}},\theta)$.  
The discrepancies can also be swapped in the above expressions.
Write
$\mu_n(\theta)=(\mu_{n1}(\theta),\mu_{n2}(\theta))^\top$, and 
write $\Sigma_{nij}(\theta)$ and $\Sigma_{ij}$ for the $(i,j)$th entries of $\Sigma_n(\theta)$ and $\Sigma$ respectively, 
$j=1,2$.  We approximate $p(T_{\text{obs}}|\theta)$ by
\begin{align} 
  \widetilde{p}_t(T_{\text{obs}}|\theta) & := \Phi\left(\frac{t_1-\mu_{n1}(\theta)}{\sqrt{\Sigma_{n11}(\theta)+\Sigma_{11}}}\right),  \label{Tlikeapprox}
\end{align}
and $p(U_{\text{obs}}|T_{\text{obs}},\theta)$ by 
\begin{align}
  \widetilde{p}_t(U_{\text{obs}}|T_{\text{obs}},\theta) & := \Phi\left(\frac{t_2-\mu_{n2|1}(\theta)}{\sqrt{\Sigma_{n2|1}(\theta)}}\right), \label{UgTlikeapprox}
\end{align}
where 
$$\mu_{n2|1}(\theta)=\mu_{n2}(\theta)+\frac{\Sigma_{n12}(\theta)+\Sigma_{12}}{\Sigma_{n22}(\theta)+\Sigma_{22}}(T_{\text{obs}}-\mu_{n1}(\theta)),$$
and
$$\Sigma_{n2|1}(\theta)=\Sigma_{n22}(\theta)+\frac{(\Sigma_{n12}(\theta)+\Sigma_{12})^2}{\Sigma_{n22}(\theta)+\Sigma_{22}}.$$
The likelihood for $p(U_{\text{obs}}|\theta)$ can be approximated 
by 
\begin{align} 
  \widetilde{p}_t(U_{\text{obs}}|\theta) & := \Phi\left(\frac{t_2-\mu_{n2}(\theta)}{\sqrt{\Sigma_{n22}(\theta)+\Sigma_{22}}}\right).  \label{Ulikeapprox}
\end{align}

\citet{gutmann+c16} chose the tolerance $t$ in the univariate BOLFI method as the $q$-quantile of $\Delta_1,...,\Delta_{n_f}$, where $q \in (0,1)$. In the bivariate MOBOLFI method, we extend the choice of tolerance $t=(t_1,t_2)$ to the 2-dimensional vector $q$-quantile of $\Delta_1,...,\Delta_{n_f}$, where $q \in (0,1)^2$. The approximate likelihood in \eqref{disclikelihood} is sensitive to the value of $t$. For results in this paper, we set $q=0.05$. A comparison of different $q$-quantile tolerances is given for 3 different examples in the Appendix. The dependent noise covariance matrix $\Sigma$ is estimated by the covariance of a simulated sample $\{\Delta_{i_\Sigma,j}\}_{j=1}^{n_\Sigma}$, where $\Delta_{i_\Sigma,j} = D(\theta_{i_\Sigma})+\epsilon_{i_\Sigma}$ are simulated
noisy discrepancies for some $\theta_{i_\Sigma}$. For results in this paper, we set $n_\Sigma = 100$ and $\theta_{i_\Sigma} = \underset{(\theta_i,\Delta_i) \in T_{n_f}}{\argmin}   
(\Delta_i-\mu_{n_f}(\theta_i)^\top \Sigma_{n_f}(\theta_i)^{-1}(\Delta_i-\mu_{n_f}(\theta_i))$. 
For better performance and numerical stability, we apply scaling to control the magnitude of $\Delta$. Details of scaling are provided in Section A.3 of the Appendix. 
We have described the MOBOLFI method in detail for the case of two data sources.  While the extension to more than two data sources is conceptually simple, the multivariate Gaussian process surrogate calculations in multi-objective BO can become more expensive when there are many discrepancies which need to be modelled jointly.  With many data sources with corresponding discrepancies, and if model simulation is no longer the dominant computational expense for MOBOLFI, then BOLFI could be preferred if a good way of combining the discrepancies into a single one can be found, and 
if we are confident there is no conflict between different data sources.

 Estimating a covariance matrix between discrepancies in MOBOLFI in the noise model for the multivariate GP surrogate allows to capture dependence between discrepancies in a flexible way.  This could be especially important in situations where the data sources are dependent, as in our motivating SSM application with choice and response data.
It is difficult to give a theoretical analysis of the choice of distance in the BOLFI and MOBOLFI methods.  However, BOLFI attempts to approximate an ABC method, and in the case of ABC there is some relevant theory demonstrating the importance of scaling the ABC distance using the summary statistic covariance matrix.    See \citet[Section 3, Theorem 1]{li+f18a} where it is demonstrated that without appropriate scaling, the limiting distribution of the ABC posterior mean may not coincide with the limiting distribution of the true posterior mean given the summary statistics, in cases where
the number of summary statistics is larger than the number of parameters.  
   
Estimating summary statistic covariance matrices can be difficult to do reliably when the number of summary statistics is large, since we need to simulate a large number of summaries at a point estimate of the parameter, which is undesirable in the setting of our work with a computationally expensive simulator. Our approach is to scale the summary statistics so that they have roughly equal variance, and to capture the dependence between the summary statistics through the covariance matrix of the vector of discrepancies, rather than the vector of summary statistics.  The covariance matrix of discrepancies is low-dimensional, making estimation easier for a computationally
expensive simulator. If we combined information from vectors of summary statistics by a linear weighting of their corresponding discrepancies to obtain a scalar joint discrepancy, this would not be a flexible approach to allowing for the dependence between discrepancies.  For example, with $k$ discrepancies we have $k$ linear weights, but the covariance matrix of the discrepancies used in the MOBOLFI surrogate has $k(k+1)/2$ distinct parameters and provides greater flexibility and many transformations can be used in the noise model.  Furthermore, it is not obvious how to estimate good linear weights for combining discrepancies, whereas estimating the covariance matrix of discrepancies in MOBOLFI by simulation at a point estimate is relatively simple.

The above discussion is for the case where the model is correctly specified.  In the case where the model is misspecified and indivdual data source likelihoods are peaked in different parts of the parameter space, then BOLFI with a single discrepancy formed as a weighted sum of data source discrepancies will not explore the regions of high likelihood for all the data source likelihoods in a single run.  Hence a single joint discrepancy is undesirable in that setting, no matter how it might be chosen, which is a strong motivation for the multi-objective approach.  This is demonstrated in an example in Section 4.

\subsection{Checking consistency of different data sources}

~~~~There are a number of advantages in separately approximating likelihood
contributions in a problem with multi-source data. One 
of the most important
is the ability to detect conflicting information about the parameter from
different parts of the data.  Given a prior $p(\theta)$ we can use the
likelihood approximation \eqref{disclikelihood} to sample
from the approximate posterior
\begin{align}
 \widetilde{p}_t(\theta|S_{\text{obs}}) & \propto p(\theta)\widetilde{p}_t(S_{\text{obs}}|\theta).  \label{approxpost1}
\end{align}
~~~~Using MCMC to sample from this does not involve any further
(computationally expensive) simulation from the model.  This is true
for the other posterior approximations discussed below also.
If the two components of the likelihood are in conflict and induce modes in different regions of the parameter space, our method of multivariate approximation can capture some of that complexity by separately considering the likelihood contributions for the different data sources. 

~~~~If we want to know what information is contained in the first
data source only, we can compute the approximate posterior
\begin{align}
 \widetilde{p}_t(\theta|T_{\text{obs}}) & \propto p(\theta)\widetilde{p}_t(T_{\text{obs}}|\theta).   \label{approxpost2}
\end{align}
~~~~Comparing the posterior densities \eqref{approxpost1} 
and \eqref{approxpost2} tells us how the second data source changes the
inference.  We could also consider a weakly informative prior 
$p_W(\theta)$ and consider the information about the parameter
contained in each data source through the posterior approximations
\begin{align}
 \widetilde{p}_t(\theta|T_{\text{obs}}) & \propto p_W(\theta)\widetilde{p}_t(T_{\text{obs}}|\theta)  \label{approxpost3}
\end{align}
and
\begin{align}
 \widetilde{p}_t(\theta|U_{\text{obs}}) & \propto p_W(\theta)\widetilde{p}_t(U_{\text{obs}}|\theta).  \label{approxpost4}
\end{align}
~~~~The purpose of using a weakly informative prior here is to separate the information
in the data from that contained in the prior, in so far as that is possible.  

~~~~In this work we consider only informal comparisons of posteriors based
on different data and prior choices, but there are more formal methods to
check for conflict between priors and data, or more generally 
between information in different parts of a hierarchical model.  We
refer the reader to \citet{evans+m06}, \citet{marshall+s07} and 
\citet{presanis+osd13} for further discussion of these.

\section{Examples}\label{sec:example}
~~~~In this section, we implement MOBOLFI in several examples. First, we apply MOBOLFI to a simple synthetic data example where we can illustrate the advantages of MOBOLFI compared to BOLFI for multi-source data, including the situation where misspecification is present. Following this, we use MOBOLFI to estimate a sequential sampling model (SSM), the Multi-attribute Linear Ballistic Accumulator (MLBA), to highlight MOBOLFI's superior design for choice-RT joint data sources compared to BOLFI. MLBA is chosen from other SSMs because of its generality for diverse choice situations and its closed-form likelihood, which enables gold-standard comparisons to be made for the likelihood approximations used in MOBOLFI. MOBOLFI requires training a bivariate GP, which involves higher computational cost than the BOLFI training of a univariate GP.  However, if model simulation is computationally expensive, this will dominate the computation time for both methods.  Although the likelihood
is tractable for the MLBA model, summary statistic based Likelihood-Free inference
can be of interest when the assumed model is misspecified.  
In this case, conditioning
on insufficient statistics that discard information which cannot be
matched by the assumed model while matching important features can
be used to develop models that are ``fit for purpose'' - see for example \citet{lewis+ml21} for further discussion.  

~~The detailed setup of our experiments and further analyses are described in Sections B and C of the Appendix. Section D in the Appendix discusses an example on
bacterial transmission in day care centres \citep{gutmann+c16,numminen2013estimating}.
Code and detailed results are submitted to github: \url{https://github.com/DZCQs/Multi-objective-Bayesian-Optimization-Likelihood-Free-Inference-MOBOLFI.git}.

\subsection{Toy example}
\begin{sloppypar}~~~~Our first synthetic example uses the MOBOLFI method to infer a parameter $\theta=(\theta_1,\dots, \theta_{10})^\top\in \mathbb{R}^{10}$ for a model with two data sources.  \end{sloppypar} 
~~~~The example is modified from \citet{schmitt2023fuse}.  The first
data source consists of $N=20$ independent 10-dimensional normal observations with mean $\theta$, $X_n\sim N(\theta,I)$, $n=1,\dots, N$.  
The second data source consists of $M=50$ 10-dimensional observations
$W_m$, $m=1,\dots, 50$.  The $W_m$ are obtained by observing discretely the
following ten-dimensional Brownian motion (BM) with drift:  
$$dw(t)=\theta dt+\sigma dW(t),\;\;t\in [0,3],$$
where $w(0)$ is a ten-dimensional vector of zeros, $W(t)$ is a standard ten-dimensional BM, and $\sigma=0.5$, leading to
$W_m=w((m-1)\delta)$, $\delta=3/(M-1)$, $m=1,\dots, M$.  

~~~~For the dataset $X=\{X_n\}_{n=1}^N$, and corresponding observed data denoted
$\{X_n^o\}_{n=1}^N$, we write $\overline{X}=N^{-1}\sum_n X_n$, $\overline{X}^o=N^{-1}\sum_n X_n^o$ and the discrepancy for data source $X$ is
$\Delta_1(X,X^o)=\|\overline{X}-\overline{X}^o\|$, where $\|\cdot\|$ denotes the
Euclidean distance.  For data $W=\{W_m\}_{m=1}^M$, and observed
data $W^o=\{W^o_m\}_{m=1}^M$, we write $\Delta W_m=W_{m+1}-W_m$, $m=1,\dots, M-1$,
$\Delta W_m^o=W_{m+1}^o-W_m^o$, $m=1,\dots, M-1$, $\overline{\Delta W}=(M-1)^{-1}\sum_{m=1}^{M-1} \Delta W_m$ and $\overline{\Delta W^o}=(M-1)^{-1}\sum_{m=1}^{M-1} \Delta W_m^o$.  Then the discrepancy used for data
source $Y$ is $\Delta_2(W,W^o)=\|\overline{\Delta W}-\overline{\Delta W^o}\|$.

~~We generate the observed data $(X^o,W^o)$ for the analysis by simulation with 
$\theta^{\text{true}}=(-0.7,0.7,...,-0.7,0.7)^T$. For implementing BOLFI and MOBOLFI, 100 initial prior samples $\{\theta^{(i)},(X^{(i)},W^{(i)})\}_{i=1}^{100}$ are drawn, where $X^{(i)}=\{X^{(i)}_n\}_{n=1}^N$, $W^{(i)}=\{W^{(i)}_m\}_{m=1}^M$, 
$\theta^{(i)} \sim N(0,I)$, and $X^{(i)}$ and $W^{(i)}$ are the $i$th simulations
for $X$ and $W$ given $\theta^{(i)}$. The training data is $\{\theta^{(i)},(\Delta_1(X^{(i)},X^o),\Delta_2(W^{(i)},W^o))\}_{i=1}^{100}$.  For implementing the BOLFI approach, we get a single discrepancy by a weighted sum of the two
data source specific discrepancies, $0.4 \Delta_1(X,X^o)+\Delta_2(W,W^o)$.
The weight of 0.4 on the first discrepancy is chosen to put the two data source specific discrepancies on a similar scale. For implementing MOBOLFI, we use the vector discrepancy $(\Delta_1, \Delta_2)$. For both approaches, 150 acquisitions are made in the BO algorithm, and 250 model simulations are needed in total.  

\begin{figure}
\centering
\begin{subfigure}{0.3\linewidth}
         \includegraphics[width=1.0\linewidth,height=4cm]{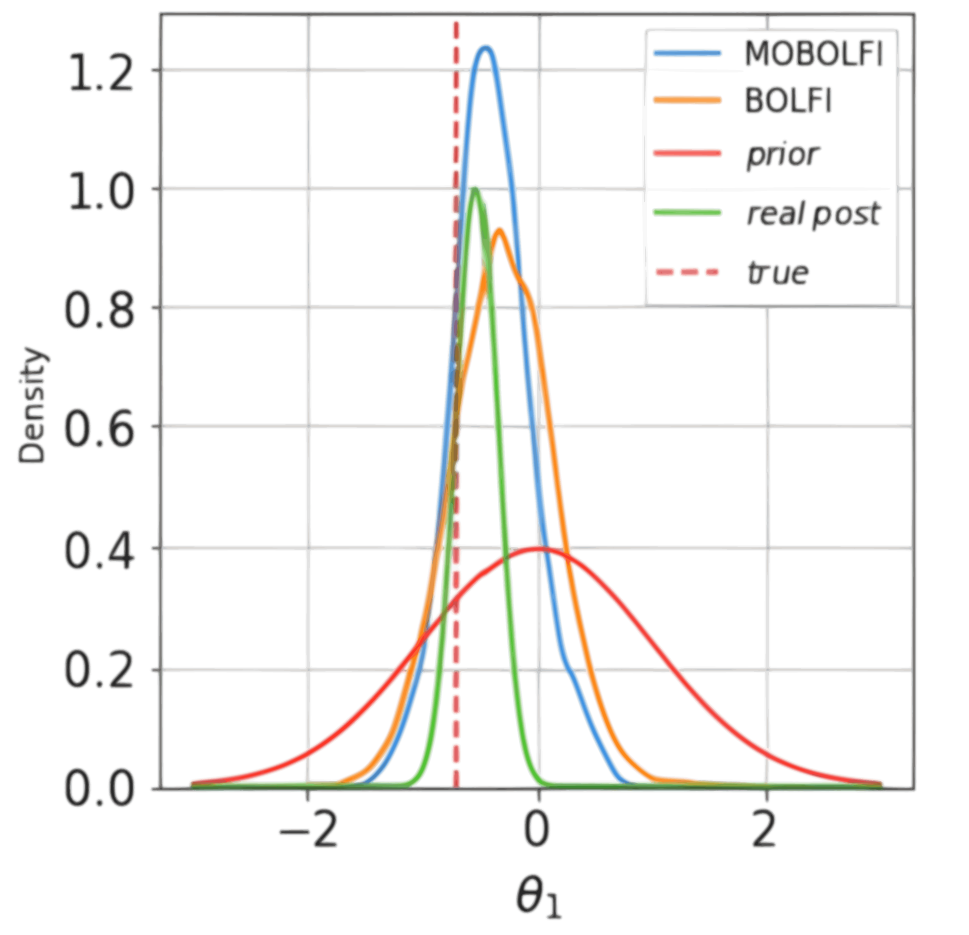}
         \caption{Joint Likelihood}
         \label{fig:BOLFI VS MOBOLFI}
\end{subfigure}
\hfill
\begin{subfigure}{0.3\linewidth}
         \includegraphics[width=1.0\linewidth,height=4cm]{figures/brownian_BOLFIvsMOBOLFIvsrealpost_X.pdf}
         \caption{Information in X}
         \label{fig:BOLFI VS MOBOLFI X}
\end{subfigure}
\hfill
\begin{subfigure}{0.3\linewidth}
         \includegraphics[width=1.0\linewidth,height=4cm]{figures/brownian_BOLFIvsMOBOLFIvsrealpost_W.pdf}
         \caption{Information in W}
         \label{fig:BOLFI VS MOBOLFI W}
\end{subfigure}
\caption{Approximate posteriors for the toy example. The left column shows approximate posteriors for the joint likelihood. The middle/right columns show
approximate posteriors conditioning only on $X/W$. The blue and orange curves are kernel density estimates obtained from HMC samples for MOBOLFI and BOLFI posteriors respectively. The green curves shows the true posteriors (joint/X/W on left/middle/right column respectively), and the red dash line is $\theta_1^{\text{true}}=-0.7$.}
\label{brownian_BOLFIvsMOBOLFIgroup}
\end{figure}

~~~~Given the symmetry of the model in the way that the components of $\theta$, $X$ and $W$ are generated, without loss of generality we present only results for inference about $\theta_1$. Figure \ref{brownian_BOLFIvsMOBOLFIgroup} compares MOBOLFI and BOLFI approximate posteriors of $\theta_1$.  We make three observations.  
First, the true posterior has smaller variance than the approximate posteriors. 
This is expected and mostly 
reflects the finite tolerance used and uncertainty in the likelihood approximation
from the surrogate model.  Secibd, the MOBOLFI approximations to the posterior conditional on $X$ or $W$ only show similar posterior location to the corresponding
true posteriors, but with somewhat larger variance. 
The MOBOLFI posteriors conditional on individual data sources are obtained 
without significant additional computation after the posterior approximation
for the joint posterior has been obtained. Third, in the left column of the figure, Hamiltonian Monte Carlo (HMC) samples from the MOBOLFI approximate posterior exhibit lower variance compared to those from the BOLFI posterior, and the MOBOLFI posterior is closer to the true posterior. This suggests that there is information loss from combining
the discrepancies from different data sources 
compared to MOBOLFI with multiple discrepancies unless the 
combined discrepancy is constructed very carefully.  
Section A.3 discusses the method used to scale discrepancies in the BOLFI 
and MOBOLFI algorithms.  Figure \ref{brownian_scaling} shows the effects of changing the scaling
and for BOLFI (left) there is greater sensitivity than for MOBOLFI (right).  In combining discrepancies for BOLFI with
multi-source data, a linear combination of the component discrepancies 
may be sub-optimal, and obtaining
a good combinination may not be as simple as choosing a linear weight.

Additional findings for this example are described in Section B.2 of the 
Appendix, where
we explore the effect of tolerance and the number of BO acquisitions on the inference.
Another experiment described there investigates the performance of MOBOLFI when parameters are present in only a subset of the data source-specific models.

\begin{figure}[h]
\centering
\begin{subfigure}{0.45\linewidth}
         \includegraphics[width=0.9\linewidth,height=4cm]{figures/brownian_scaling_BOLFI.pdf}
         \caption{BOLFI with different scaling}
         \label{fig:BOLFI scaling}
\end{subfigure}
\begin{subfigure}{0.45\linewidth}
         \includegraphics[width=0.9\linewidth,height=4cm]{figures/brownian_scaling_MOBOLFI.pdf}
         \caption{MOBOLFI with different scaling}
         \label{fig:MOBOLFI scaling}
\end{subfigure}
\caption{ Approximate posterior in the toy example given different scaling weights to $\Delta_1$. The left column shows the BOLFI approximate posteriors given $weight = 0.1/0.4/0.7/1.0$ scaling, while the training objectives for BO is defined to be $weight*\Delta_1 + \Delta_2$. The right column presents the MOBOLFI approximate posteriors given the same scaling.}
\label{brownian_scaling}
\end{figure}

\subsubsection{Misspecified simulator}
~~~~We continue the above example by modifying it to introduce misspecification.  
Our purpose is to demonstrate that MOBOLFI is useful for understanding misspecification when different data sources provide conflicting information.  
In the modified example the observations are one-dimensional, 
$X_n\sim N(\theta,1)$, $n=1,\dots, N$, and
$W_m$, $m=1,\dots, M$, with $W_m=w((m-1)\delta)$, $\delta=3/(M-1)$ with 
$w(t)$, $t\in [0,3]$ a univariate process defined by 
$$dw(t)=\theta dt+\sigma dW(t),\;\;\;w(0)=0,$$
with $W(t)$ a univariate standard BM, and $N$, $M$ and $\sigma$ as before.   

~~~~The true data generating process (DGP) is not the above assumed model.  Instead, 
in the true DGP there are different values of $\theta$ in the
models for $X$ and $W$, so that 
 $X_n\sim N(\theta_X,1)$ and $dw(t)=\theta_W dt+\sigma dW(t)$, 
where $\theta_X\neq \theta_W$.  We simulate the observed data for
the analysis using $\theta_X=0.3$, $\theta_W=-0.7$.  In the misspecified
assumed model the likelihood contributions for different data sources produce
conflicting information about $\theta$.  

~~~~Approximate posteriors for $\theta$ are presented in Figure \ref{brownian_BOLFIvsMOBOLFIgroupmis}. 
The MOBOLFI and BOLFI approximate posteriors are similar when conditioning on 
both $X$ and 
$W$, with the posterior mode lying between $\theta_X$ and $\theta_W$.  
The middle/right columns shows the MOBOLFI approximate posterior
conditioning on only $X$, or only $W$, compared to the corresponding
true posterior and the BOLFI posterior 
conditioning on both $X$ and $W$.  We make two observations.  
First, the MOBOFLI approximate posteriors conditioning on $X$ only and on $W$ only
produce good estimates of $\theta_X$ and $\theta_W$, and the conflicting
information in the two data sources is evident.  Second, the approximations
of the individual data source posteriors in MOBOLFI are obtained without 
substantial additional computation, since the joint likelihood and 
likelihoods for $X$ and $Y$ are obtained simultaneously in the MOBOLFI approach.

\begin{figure}
\centering
\begin{subfigure}{0.3\linewidth}
        \includegraphics[width=1.0\linewidth,height=3.6cm]{figures/brownian_mis.pdf}
         \caption{Joint Likelihood (mis)}
         \label{fig:BOLFI VS MOBOLFI mis}
\end{subfigure}
\hfill
\begin{subfigure}{0.3\linewidth}
         \includegraphics[width=1.0\linewidth,height=3.6cm]{figures/brownian_misX.pdf}
         \caption{Information in X (mis)}
         \label{fig:BOLFI VS MOBOLFI mis X}
\end{subfigure}
\hfill
\begin{subfigure}{0.3\linewidth}
         \includegraphics[width=1.0\linewidth,height=3.6cm]{figures/brownian_misw.pdf}
         \caption{Information in W (mis)}
         \label{fig:BOLFI VS MOBOLFI mis W}
\end{subfigure}
\caption{Approximate posteriors for the toy example under misspecification. The left column shows MOBOLFI and BOLFI approximate posterior densities conditional on both $X$ and $W$. The middle/right presents approximate posterior densities conditioning only on $X$/$W$. The blue and orange curves are kernel estimates of posterior densities from HMC samples for MOBOLFI and BOLFI respectively. On the left column, the green curve shows the prior while the red curve shows the true posterior; on the middle/right columns, the green curve shows the posterior conditional on $X/W$ while the red curve shows the prior. The two dash lines are values of parameter produced by data sources with conflict information, labeled as true and mis(specified). $\theta_X=0.3$ and $\theta_W=-0.7$. }
\label{brownian_BOLFIvsMOBOLFIgroupmis}
\end{figure}

\subsection{Multi-attribute Linear Ballistic Accumulator}
~~~~MLBA is a state-of-the-art SSM for understanding the human decision-making process. We will apply MOBOLFI to infer parameters of one variant of MLBA 
\citep{hancock2021accumulation} using a simulated dataset where $\theta$ is known, and 
for a real-world dataset on preferences of ridehailing drivers to rent electric vehicles in Singapore. 

\subsubsection{Simulator introduction}
\begin{figure}[h]
    \centering
    \includegraphics[width = 14cm]{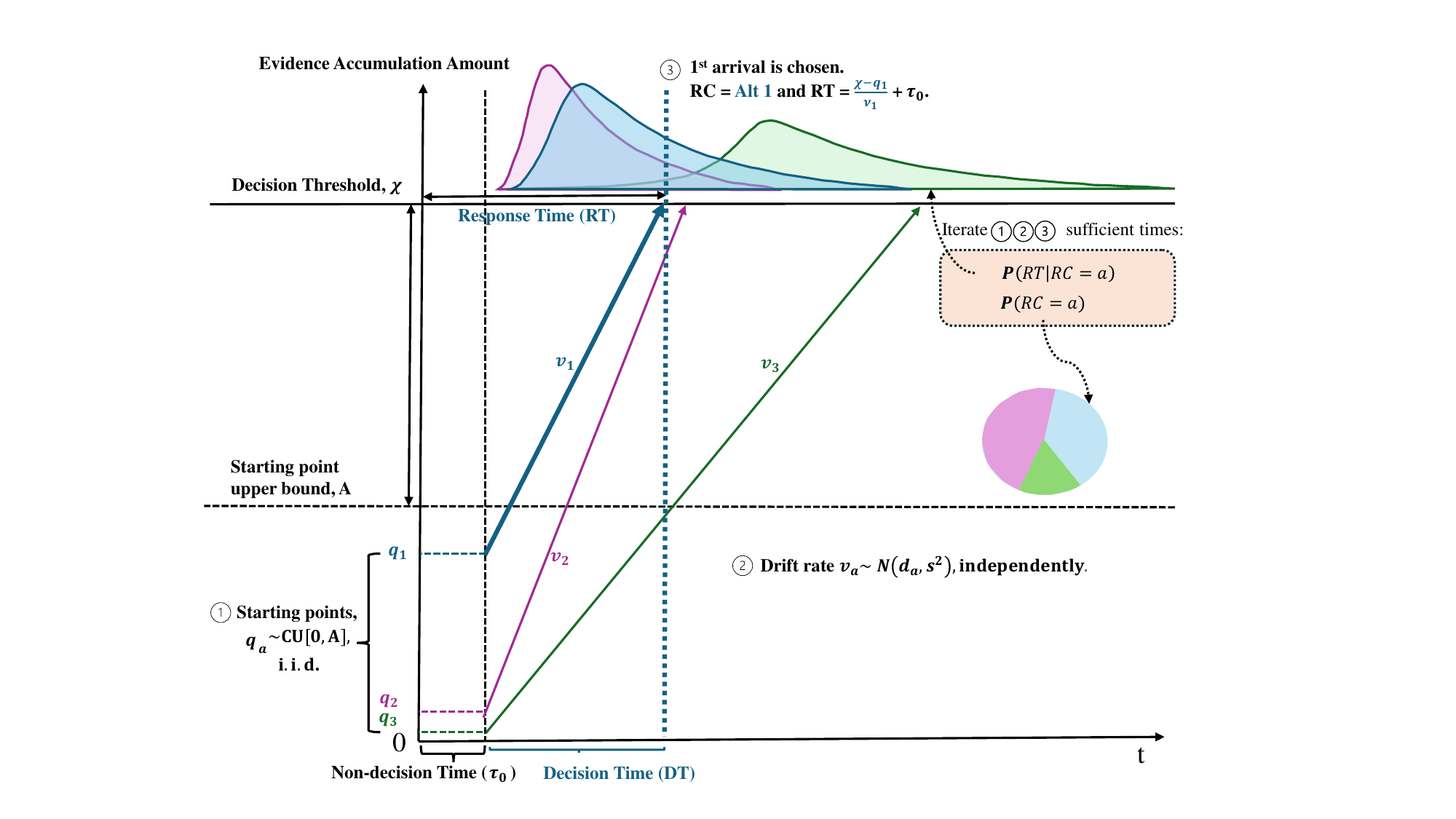}
    \caption{The simulation process of MLBA for three alternatives (M = 3). In this case, the final choice outcome is Alternative 1 (in blue and its accumulator is bold). The RT is the time for Alternative 1 to reach the threshold with non-decision time $\tau_0$. The directed solid lines demonstrate the realized evidence accumulation trajectory for each alternative. The MLBA simulator contains three steps annotated above. After iterating sufficient times, the choice proportion and its conditional RT distribution can be acquired.}
    \label{fig:mlba_illustration}
\end{figure}

~~~~MLBA assumes that decision-making can be seen as an evidence accumulation process, as illustrated in Figure~\ref{fig:mlba_illustration}. For a decision-maker facing $M$ alternatives, MLBA posits there exist $M$ independent evidence accumulators starting to collect supporting evidence for each corresponding alternative $a$ from a random starting point $q_a\overset{i.i.d.}{\sim }\mathcal{U}[0,A]$. Each alternative-specific accumulator evolves linearly at drift rate $v_a\overset{indep.}{\sim} N(d_a,s^2)$, where $a=1,\cdots, M$. $d_a$ depends on the pairwise attribute comparisons for all $K$ attributes of alternative $a$ with the remaining alternatives for each observation. $A$ and $s^2$ are constant over observations to measure the scale of initial preference and unobserved accumulation process noise, respectively. Finally, the alternative whose evidence accumulator reaches the common threshold $\chi$ first is considered the final choice, and the deliberation time is the same as this alternative's evidence accumulation time $\frac{\chi-q}{v}$. Strictly, the response time (RT) is defined as the sum of the deliberation time and the non-decision time, which stands for information encoding and decision execution time. However, for complex choice tasks with multi-attribute information, non-decision time can be omitted if it does not contribute significantly to RT. 

~~~~The MLBA model can be described in two parts.  The \textit{back-end part}, Linear Ballistic Accumulator (LBA), was originally proposed by \citet{brown_simplest_2008} for implicit choice situations (i.e., choice tasks without listed attribute values). To extend LBA to MLBA, \citet{trueblood_multiattribute_nodate} added the \textit{front-end part} to LBA by specifying alternative-specific drift rates. Several variants of the front-end part have been investigated to define the drift rate mean of MLBA. We choose \citet{hancock2021accumulation}'s specification because it can handle more than two attributes. For a choice set $\mathcal{C}$, the drift rate mean $d_a$ for the alternative $a$ is:
\begin{equation} \label{eq:d_mean}
    d_a = \max\{I_0+\delta_a+ \sum_{b\in\mathcal{C},b\neq a}\sum^K_{k=1}\omega_{abk}\beta_k(X_{ak}-X_{bk}),0\},
\end{equation}
where $X_{ak}$ is the value of attribute $k$ of alternative $a$ and \begin{equation}\label{eq:weight}
    \omega_{abk}
 =\begin{cases}
        \exp\{-\lambda_1|\beta_k( X_{ak}-X_{bk})|\}& \beta_k(X_{ak}-X_{bk})\ge 0\\
        \exp\{-\lambda_2|\beta_k(X_{ak}-X_{bk})|\} & \beta_k(X_{ak}-X_{bk})<0.
    \end{cases} 
\end{equation}

The final simulation output is a concatenation of RT and choice outcome:
\begin{equation} \label{eq:mlba_rtandchoice}
    (RT,CH) = (((\chi-g_{CH})/v_{CH})+\tau_0, \underset{a \in \mathcal{C}}{\argminB} (\chi-g_a)/v_a)
\end{equation}

$\Lambda:=(\lambda_1, \lambda_2)$ is interpreted as a vector of sensitivity parameters of positive/negative difference of attribute pairwise-comparisons respectively; $I_0$ is a common drift rate mean constant that prevents negative drift rate; $\bm{\beta} = (\beta_1,...\beta_K)$ are scaling parameters for the attributes; and $\bm{\delta}=(\delta_1,...,\delta_M)$ are alternative-specific constants for the drift rate mean. Thus, the estimable parameter is $\bm{\theta} = (\Lambda, I_0, \bm{\beta}, \bm{\delta}_{-1},\chi)$, while $(\mathcal{A},s,\tau_0,\delta_1)$ are fixed for parameter identification  and $\bm{\delta}_{-1}=(\delta_2,...,\delta_M)^T$. Following \citet{brown_simplest_2008} and \citet{terry_generalising_2015}, the joint likelihood of observing the response time $RT$ and the choice made $CH$ of multiple candidates has a closed form expression. It is presented in Section C.1 of the Appendix. We use the closed-form likelihood as a gold-standard to benchmark approximate posteriors obtained by LFI in subsequent experiments.

\subsubsection{Synthetic data} \label{subsec:MLBAsyndat}
~~~~We use the survey question design and alternative-specific attribute matrix for the real example of
section~\ref{subsec:MLBAreal}.  A choice question with $M=3$ alternatives is provided to candidates.  Each alternative offers values of $K=3$ attributes, and a respondent will make decisions by evaluating the alternative and its attributes. For data generation, the synthetic output data is simulated using $\theta = (\Lambda, I_0, \bm{\beta}, \bm{\delta}, \chi) = ((0.1,0.8),2,(-22,-5,-6),(0,3,1.5),100)$, 
resulting in $1280=320\cdot (3+1)$ observations, after concatenating $CH$ (in one-hot encoding) and $RT$ for 320 candidates. The size of the alternative-specific attribute matrix $X$ is $320\times 9$. For parameter estimation, the true values of parameters we target in inference are $\theta^{\text{true}} = (\lambda_1, \beta_1, \beta_2, \delta_2, \delta_3, \log(\chi-\mathcal{A}))^{\text{true}} = (0.1, -5, -6, 3, 1.5, \log(99))$. The parameters $\mathcal{A}, s$, $\tau_0$, $\delta_1$ are fixed at 1, 1, 0, and 0 for identifiability, and $\lambda_2 = 0.8, I_0=2, \beta_3 = -6$ are assumed to be known for reducing the computational cost in further training such as facilitating scaling of the estimable parameters. This implementation is common in empirical studies since MLBA is designed bottom-up so that some parameters are highly correlated with each other. Lastly, we have log-transformed $\chi-\mathcal{A}$ in order to increase sampling efficiency, since the magnitude of $\log(\chi-\mathcal{A})$ is much smaller than that of $\chi$.

\begin{figure}[h]
\centering
\begin{subfigure}{0.4\linewidth}
         \includegraphics[width=1.0\linewidth,height=4cm]{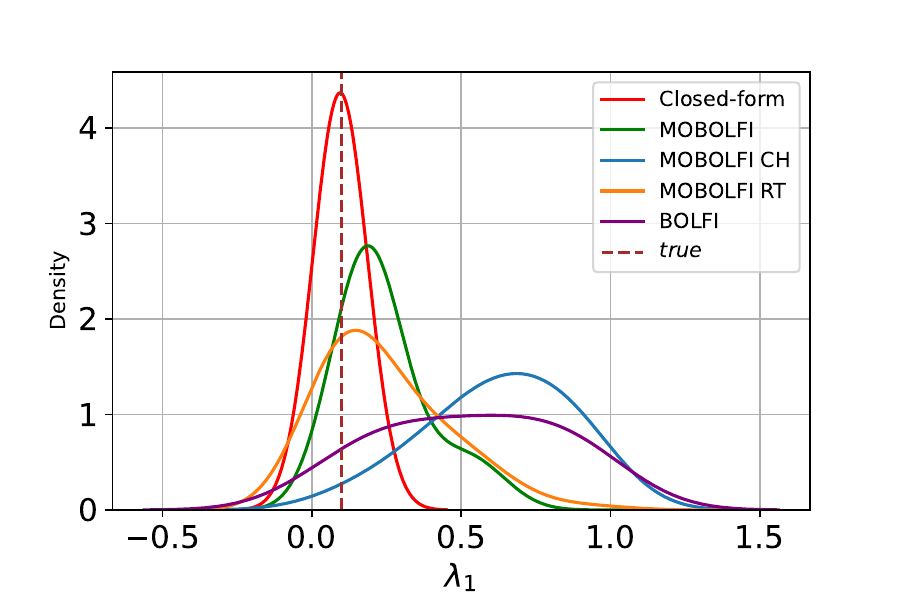}
         \label{fig: MLBA BOLFI VS MOBOLFI lambda1}
\end{subfigure}
\begin{subfigure}{0.4\linewidth}
         \includegraphics[width=1.0\linewidth,height=4cm]{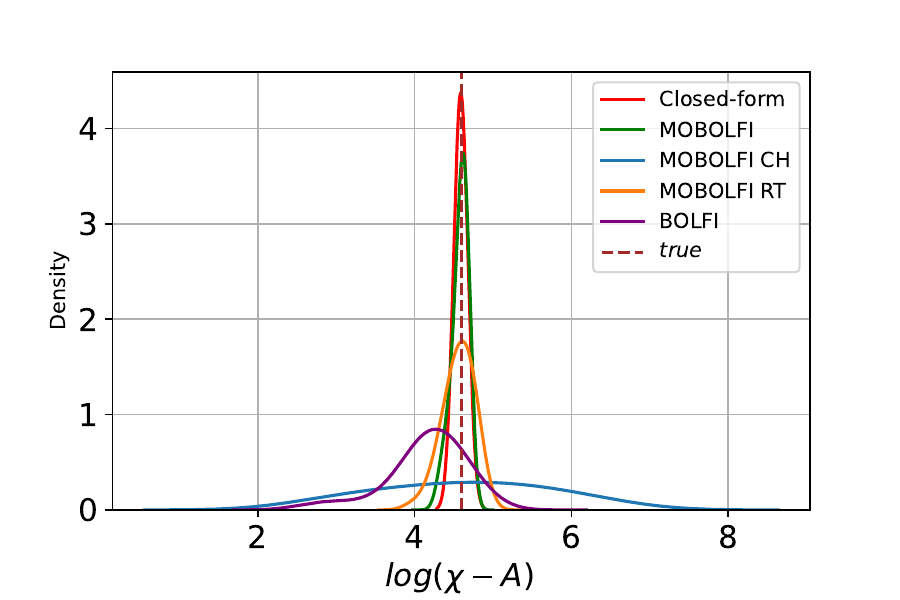}
         \label{fig: MLBA BOLFI VS MOBOLFI threshold}
\end{subfigure}

\begin{subfigure}{0.4\linewidth}
         \includegraphics[width=1.0\linewidth,height=4cm]{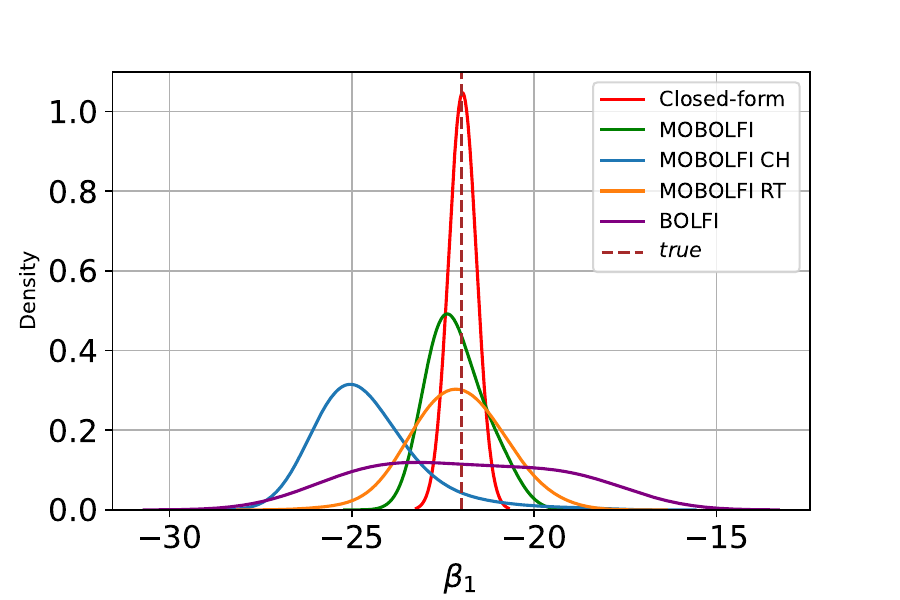}
         \label{fig: MLBA BOLFI VS MOBOLFI beta1}
\end{subfigure}
\begin{subfigure}{0.4\linewidth}
         \includegraphics[width=1.0\linewidth,height=4cm]{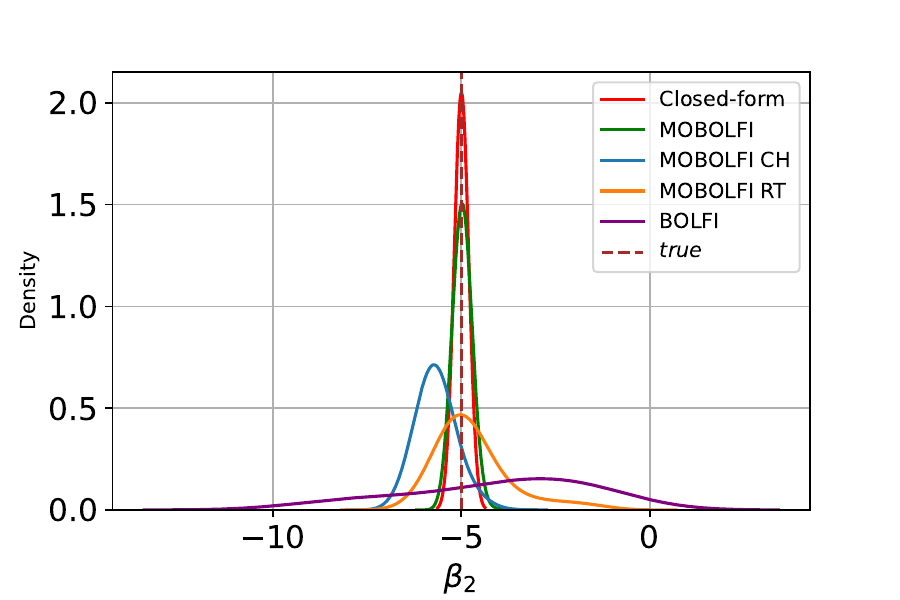}
         \label{fig: MLBA BOLFI VS MOBOLFI beta2}
\end{subfigure}

\begin{subfigure}{0.4\linewidth}
         \includegraphics[width=1.0\linewidth,height=4cm]{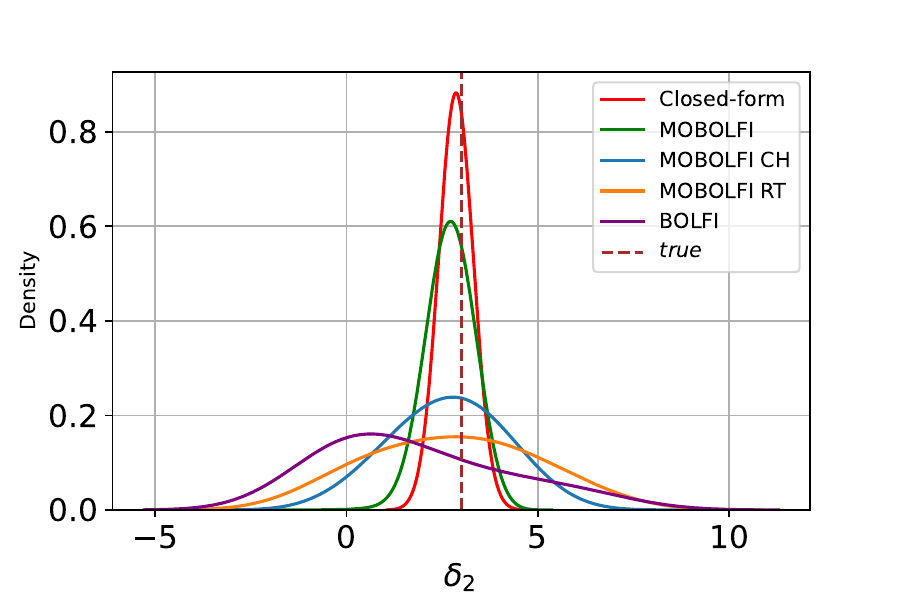}
         \label{fig: MLBA BOLFI VS MOBOLFI delta2}
\end{subfigure}
\begin{subfigure}{0.4\linewidth}
         \includegraphics[width=1.0\linewidth,height=4cm]{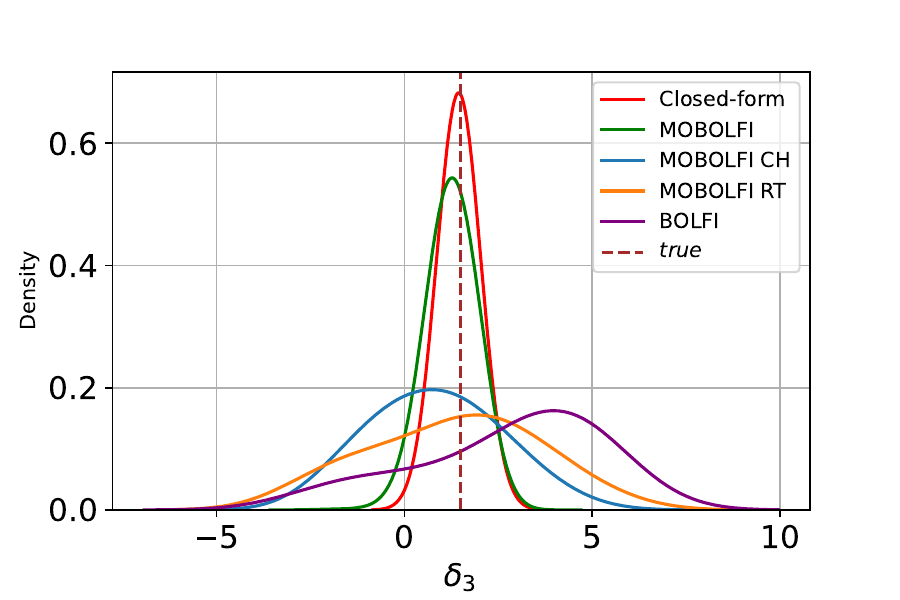}
         \label{fig: MLBA BOLFI VS MOBOLFI delta3}
\end{subfigure}
\caption{Approximate posteriors for MLBA example. The plots shows marginal posteriors for different methods and each parameter of interest. The green/purple curve represents the MOBOLFI approximate posterior/BOLFI approximate posterior. The red curve is the sample from the closed form posterior\protect\footnotemark , calculated using the closed form likelihood of MLBA. The orange/blue curves are the MOBOLFI approximate posteriors calculated by marginal approximate likelihood of $RT$/$CH$ data. The red dashed line is the value of $\theta^{\text{true}}$.}
\label{MLBA_BOLFIvsMOBOLFIvsmarginal}
\end{figure}

\begin{sloppypar}
~~~~The synthetic data $(RT^o,CH^o)$ is simulated by MLBA taking $\theta^{\text{true}}$ as input, where $RT$ and $CH$ are response time and choices. For implementing BOLFI and MOBOLFI, 100 samples from the prior are used
as initialization, $\{\theta^{(i)},(RT^{(i)},CH^{(i)})\}_{i=1}^{100}$, where $RT^{(i)}=\{RT^{(i)}_1,...,RT^{(i)}_{320}\}$, and $CH^{(i)}=\{CH^{(i)}_1,...,CH^{(i)}_{320}\}$ are the simulated data given $\theta^{(i)}$. 
For response times $RT$, write $\widetilde{RT}$ for the vector of order statistics of $RT$.  Similarly, $\widetilde{RT}^0$ is the vector of order statistics of $RT^0$.  
The training data is $\{\theta^{(i)},(\Delta_1(RT^{(i)},RT^o),\Delta_2(CH^{(i)},CH^o))\}_{i=1}^{100}$, where 
$\Delta_1(RT,RT^o)$ and $\Delta_2(CH,CH^o)$ are discrepancies for the 
response and choice data respectively.  The first discrepancy is defined as
\begin{equation}\label{RT_dis}
   \Delta_1(RT,RT^o) = \|\log(\widetilde{RT}^0)-\log(\widetilde{RT})\|_1, 
\end{equation} 
where $\|\cdot\|_1$ denotes the $L_1$ distance.  For the second
discrepancy, recall that the CH data is represented by 3 dimensional one-hot encoding of 3 alternatives, and we define 
\begin{equation}\label{CH_dis}
\Delta_2(CH^{(i)},CH^o)=\frac{1}{3}\|\frac{1}{320}\sum_{j=1}^{320} |CH^o_j-CH^{(i)}_j|\|_1.
\end{equation}
~~~~The prior assumes independence between components of $\theta$, with each marginal prior uniform on an interval of length 8 centred on the true value, with the exception of the parameter $\lambda$ which has a prior uniform
on $[0,1]$. Differential emission MCMC (De-MCMC, see \citet{turner2013method}) is used to sample from the approximate posterior. Further details of the experiment 
are given in Section C.2 of the Appendix.
\end{sloppypar}


~~~~Figure \ref{MLBA_BOLFIvsMOBOLFIvsmarginal} shows approximate
posterior distributions obtained using the BOLFI and
MOBOLFI methods. We make a number of observations.  First, 
the variance of the MOBOLFI approximate posterior is smaller
than that for BOLFI. In fact, MOBOLFI using only the
response time data produces more accurate approximations of the true posterior than BOLFI given both data sources.  
This suggests that when it is natural to define
discrepancies separately for different data sources, 
combining them linearly, even with weights, may result
in information loss.  Second, 
both MOBOLFI and BOLFI posterior variances are larger than
for the true posterior, which is expected, due to the irreducible information loss from approximation.  Third, when comparing the MOBOLFI posteriors calculated by likelihoods of different data sources, the posterior using both data sources is better than any posterior using only one data source. 
Fourth, for inference of $\beta$ and $\lambda_1$, the posterior using only $CH$ has its marginal maximum a posteriori probability (MAP) value far from the value of $\theta^{\text{true}}$; for inference of $\chi$ the posterior using only $CH$ has variance larger than that of the posterior using the joint likelihood, compared to using $RT$ only or the joint likelihood. In psychology and economics, researchers usually focus on the choice data $CH$ in parameter inference for SSMs like MLBA. 
The figure shows that using response time data in inference not only reduces the approximate posterior variance to make
it closer to the true posterior variance, but also helps locate the area of the global maximum. Such conclusions are likely to extend to other variants of MLBA and LBA. 
Section C.3 of the Appendix details additional
experiments on the factors affecting the performance
of MOBOLFI in this example.  

\subsubsection{Misspecified synthetic data}

~~~~We now extend the previous synthetic MLBA example
to a misspecified scenario, illustrating the advantages of MOBOLFI
for efficiently exploring the high likelihood region
for both data source likelihoods
when there is conflict.  
We define two sets of parameters $\theta^{\text{true}}_{RT} = (0.2,-25,-3.5,6,4,\log(99))$ and $\theta^{\text{true}}_{CH}= (0.05,-24,-6.5,3,1.5,\log(199))$ for simulating response time and choice data respectively. We simulate data $(RT^{rt},CH^{rt})$ by MLBA taking $\theta^{\text{true}}_{RT}$ as input, and data $(RT^{ch},CH^{ch})$ by MLBA taking $\theta^{\text{true}}_{CH}$ as input. The synthetic observed data $(RT^o,CH^o) = (RT^{rt},CH^{ch})$ concatenates the response time data generated using $\theta^{\text{true}}_{RT}$ and the choice data simulated using $\theta^{\text{true}}_{CH}$. Therefore, the two data sources provide conflicting
information about the parameter vector.

\begin{figure}[h]
   \centering
    \begin{subfigure}{0.4\linewidth}
         \includegraphics[height=3.98cm]{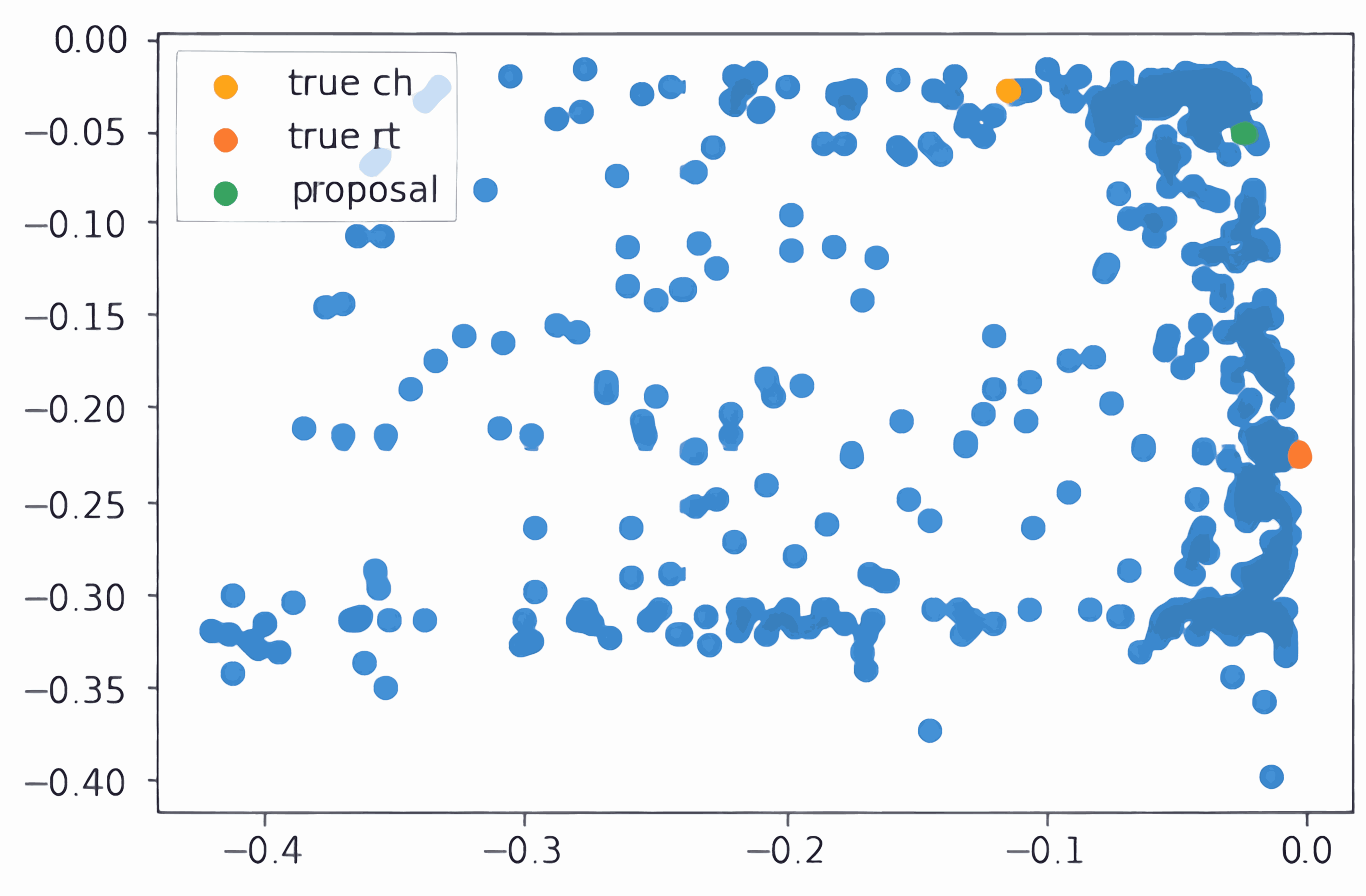}
         \label{fig: MLBA MOBOLFI mis BO}
    \end{subfigure}
    \begin{subfigure}{0.4\linewidth}
         \includegraphics[height=3.97cm]{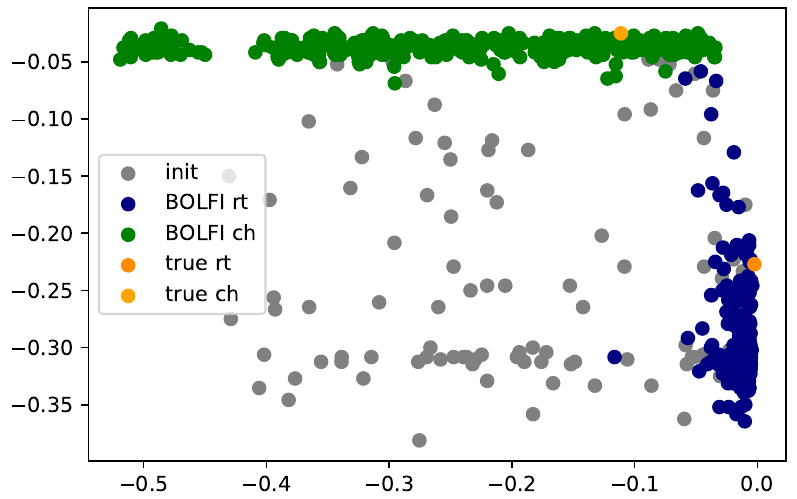}
         \label{fig: MLBA BOLFI mis BO}
    \end{subfigure}

\caption{Noisy Negative discrepancy values of Bayesian Optimization acquisition points. The horizontal axis denotes the negative RT discrepancy, the vertical axis denotes the negative CH discrepancy. The discrepancies have been scaled to a similar range.  The yellow/orange point represents simulated discrepancies obtained using $\theta^{\text{true}}_{RT}/\theta^{\text{true}}_{CH}$ as input. The green point is a simulated vector of negative discrepancies obtained at the mode of the MOBOLFI approximate posterior.}
\label{MLBA_misspeci_scatter}
\end{figure}

~~~~We follow the same design as in Sec 4.2.2 with some minor changes. Firstly, we remove the rejection criterion for sampling the initial training set for BO discussed
in Section C.2. With conflicting information, parameter samples are likely to obtain a low value of discrepancy on one data source and a high value on the other. Secondly, the number of iterations for the BO algorithm is set to 500.

~~~~Applying MOBOLFI to such $(RT^o,CH^o)$, we find that the multi-objective Bayesian Optimization explores a range of local optimum points near both $\theta^{\text{true}}_{RT}$ and $\theta^{\text{true}}_{CH}$. Although we cannot observe the expected discrepancy, Figure \ref{MLBA_misspeci_scatter} (left) shows the noisy negative discrepancy values of MOBOLFI acquisitions.  The plot 
shows that the algorithm explores
a range of parameter values leading to small
discrepancy for one discrepancy or the other, even
if both are not small simultaneously due to the conflict.
Dots on the plot show simulated discrepancies for 
$\theta_{RT}^{\text{true}}$ and $\theta_{CH}^{\text{true}}$.  The plot on the right in
the figure shows simulated discprepancies (for both
discrepancies) for BOLFI acquistions based on one data source.  Green dots are for BOLFI acquistions based
on choice data discrepancy, and dark blue dots are for
BOLFI acquistions based on response data discrepancy.  
Due to the conflict we see that the BOLFI acquisitions
cannot explore in a single run the full region where 
either one of the discrepancies is small.  Hence we cannot
obtain reliable estimates of the likelihood for both
data sources based on BOLFI acquistions for one data source only.  A similar problem would arise for any composite discrepancy linearly
combining the data-source specific discrepancies.  Plots of approximate marginal
posteriors are shown in Section C.3 of the appendices.

\subsubsection{Real-world data} \label{subsec:MLBAreal}  

~~~~Next we apply MOBOLFI and BOLFI to infer parameters of an MLBA model
for a real-world dataset from a consumer choice experiment. %
This experiment 
assesses the rental preferences of ride-hailing drivers in Singapore
for electric vehicles (EVs) via a street-intercept survey \citep{ding2024can}. Before the stated preference experiment, the driver's basic information such as working days per week, and information about currently rented internal combustion engine vehicles (ICEVs) is collected.  The drivers were asked to make a choice among three alternatives including ICEVs, Electric Vehicle Model A (EVA), and Electric Vehicle Model B (EVB) with three listed attributes, which are monthly rental cost (RC, in SGD), daily operating cost (OC, in SGD), and daily mileage (DR, in km). The time that elapses from the appearance of information to the confirmation of choice is recorded as RT. Note that EVA and EVB are assumed to be identical except for their values of three listed attributes. After data preprocessing, 149 participants with 584 valid observations are used in the parameter estimation. Since monthly rental cost (RC) and daily operating cost (OC) capture the monetary aspect, these two can be merged into one alternative specific attribute for the monthly total cost (TC, in SGD). 
\begin{equation}
    TC_{na} = RC_{na}+OC_{na}\times WF_{n}\times \frac{52}{12},
\end{equation}where $n$ and $a$ are the indices of observation and alternative respectively. $WF_{n}$ is the number of working days per week, hence $TC_{na}$ is the monthly total cost for alternative $a$ in observation $n$. Lastly, due to the heavy-tailed distribution of DR, it is transformed to log scale.  In summary, the size of the attribute matrix $X$ is $584\times 6$, and the real output data for the MLBA simulator $(RT,CH)$ is $584\times 4$.  

~~~~The unknown parameter is $\theta = (\lambda_1,\log{\beta_{TC}},\log{\beta_{\log{DR}}},\delta_{ICEV}, \log{(\chi-\mathcal{A})})$, and the rest of the parameters are fixed at known values. We set constants $\mathcal{A} = 45,s = 1,\delta_{EVB} = 0$, and $I_0 = 1$ to ensure parameter identification. The alternative-specific constant $\delta$ is often interpreted as the initial preference of the corresponding alternative before considering the attribute values in empirical studies. A mild and reasonable assumption made here is that $\delta_{EVA} = \delta_{EVB} =0$, since both are 
EVs. Furthermore, based on prospect theory that consumers pay more attention to loss than gain \citep{camerer1998prospect}, $\lambda_2$ is normally smaller than $\lambda_1$. Therefore, to simplify the estimation, we set $\lambda_2 = 0$ for the final empirical analysis.  Finally, similarly to synthetic data estimation, $\beta_{TC}$, $\beta_{DR}$ and $\chi-\mathcal{A}$ are log-transformed to increase sampling efficiency.

~~~~We use independent priors $\lambda_1\sim \mathcal{U}[0,5]$, $\delta_{ICEV}\sim\mathcal{U}[-3,5]$, $\log{\beta_{TC}}\sim\mathcal{U}[-3,5]$, $\log{\beta_{DR}}\sim \mathcal{U}[-3,5]$, and $\log{(\chi-\mathcal{A})}\sim\mathcal{U}[-2,6]$. 
For the implementation of MOBOLFI and BOLFI, 282 initial samples $\{\theta^{(i)}, (RT^{(i)}, CH^{(i)})\}_{i = 1}^{282}$ are obtained using a space-filling design \citep{Santner_2018}. 
The training data is $\{\theta^{(i)}, (\Delta^*_1(RT^{(i)}, RT^o), \Delta^*_2(CH^{(i)}, CH^o))\}_{i = 1}^{282}$, where $(RT^o, CH^o)$ are observed empirical data. Unlike the synthetic data experiment where a single simulated dataset is used for computing discrepancies at each parameter by \eqref{RT_dis} and \eqref{CH_dis}, $(RT^{(i)}, CH^{(i)}) = \{(RT^{(i)}_s, CH^{(i)}_s)\}_{s=1}^{S}$ is a collection of multiple simulations independently generated by the MLBA simulator with $\theta^{(i)}$ and attribute matrix $X$. The discrepancy for the RT data source $\Delta_1^*(RT^{(i)},RT^o)$ is the log-average of $1$-Wasserstein distance between the empirical distribution of log-transformed
$RT^o$ and the $S=50$ replicates of $RT^{(i)}_s$: 
\begin{equation}\label{eq:mlbarealdatart}
      \Delta_1^*(RT^{(i)},RT^o) = \log\Bigg(\frac{\sum_{s=1}^S \|\log(\widetilde{RT}^o)-\log(\widetilde{RT}^{(i)}_s))\|_1}{3S}\Bigg),
\end{equation}
where $\widetilde{RT}$ is the vector of order statistics of $RT$.

~~~~The discrepancy $\Delta_2^*(CH^{(i)},CH^o)$ for the choice data is the log-average of SSE of choice proportion over the $S$  replicates for the synthetic choice data and the observed choice outcomes (i.e. the average  squared difference between empirical choice proportions by simulations and observed outcomes): 
\begin{equation}\label{eq:mlbarealdatachBrier}
      \Delta^*_2(CH^{(i)},CH^o) = \log\Bigg(\frac{\sum_{s=1}^{S} \mathds{1}_N^T(CH_s^{(i)}-CH^o)(CH_s^{(i)}-CH^o)^T\mathds{1}_N}{9SN^2}\Bigg),
\end{equation}
where $N = 584$ is the number of observations and $\mathds{1}_N \in \mathbb{R}^{N\times 1}$ is the vector whose each element is 1. 
In addition, BOLFI's one-dimensional discrepancy for joint data sources is the summation of $\Delta_1^*(CH^{(i)},CH^o)$and $\Delta_2^*(RT^{(i)},RT^o)$ in \eqref{eq:mlbarealdatachBrier} and \eqref{eq:mlbarealdatart}. Therefore, the training data for BOLFI is  $\{\theta^{(i)}, (\Delta^*_1(RT^{(i)}, RT^o)+\Delta^*_2(CH^{(i)}, CH^o))\}_{i = 1}^{282}$. 

~~~~There are two reasons for adopting
$\Delta^*$ instead of $\Delta$ in \eqref{RT_dis} and \eqref{CH_dis} for the empirical experiment. 
First, log transformation on both discrepancies encourages the MOBOLFI and BOLFI to explore more in the high-density area (when $\Delta$ is close to 0), and therefore, letting the approximate log-likelihood be more sensitive (larger gradient) around the high-density area in parameter space. Second, the simulation data from $\mathcal{S}$ replicates provide a more robust discrepancy for each data source by reducing randomness brought by the MLBA simulator.  De-MCMC is used for the MOBOLFI/BOLFI approximate posterior and closed-form MLBA posterior estimation.   

\begin{figure}[hbt]
    \centering
    \begin{subfigure}{0.39\linewidth}
        
         \includegraphics[width = 1.0\linewidth,height = 4cm]{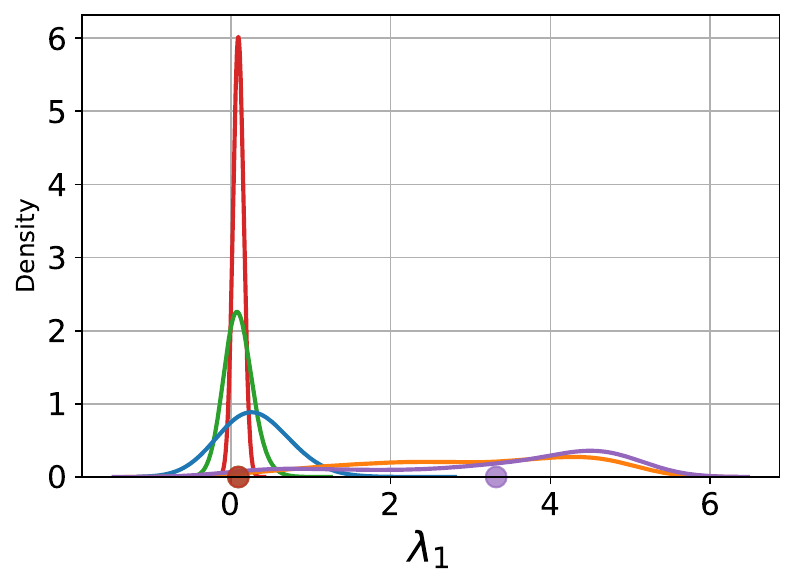}
         \label{fig: MLBA_em_lam}
    \end{subfigure}
    ~\begin{subfigure}{0.395\linewidth}
        
         \includegraphics[width = 1.0\linewidth,height = 4cm]{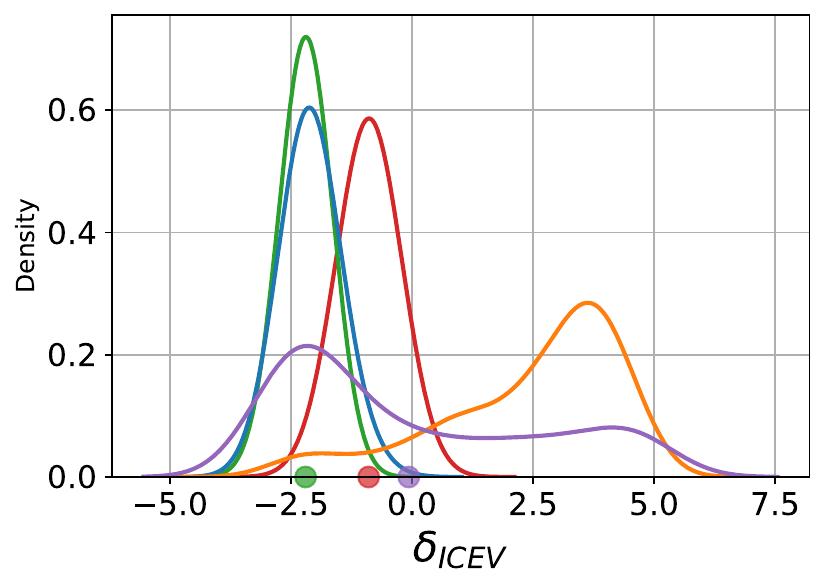}
         \label{fig: MLBA_em_delta}
    \end{subfigure}
    
    ~\begin{subfigure}{0.4\linewidth}
         \includegraphics[width=1.0\linewidth,height = 4cm]{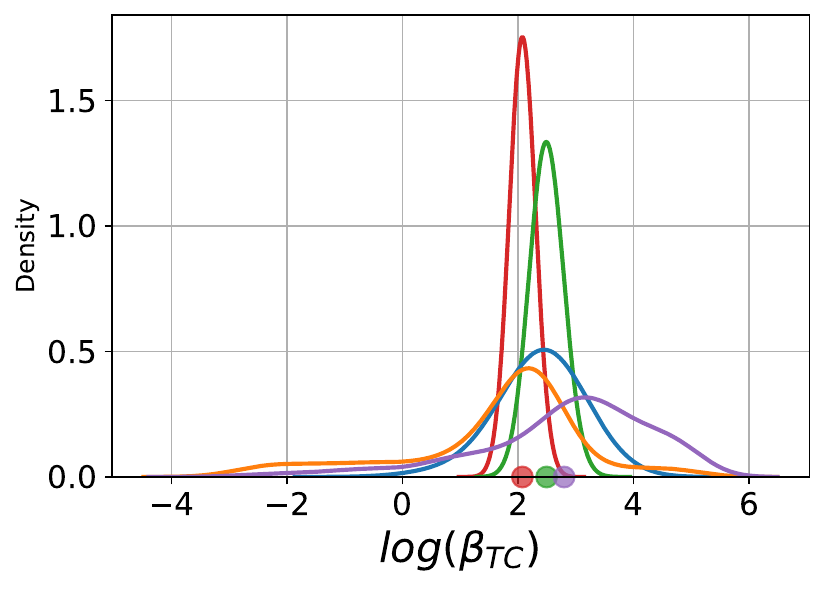}
         \label{fig: MLBA_em_betaTC}
    \end{subfigure}
     ~
    \begin{subfigure}{0.4\linewidth}
        \includegraphics[width=1.0\linewidth,height = 4cm]{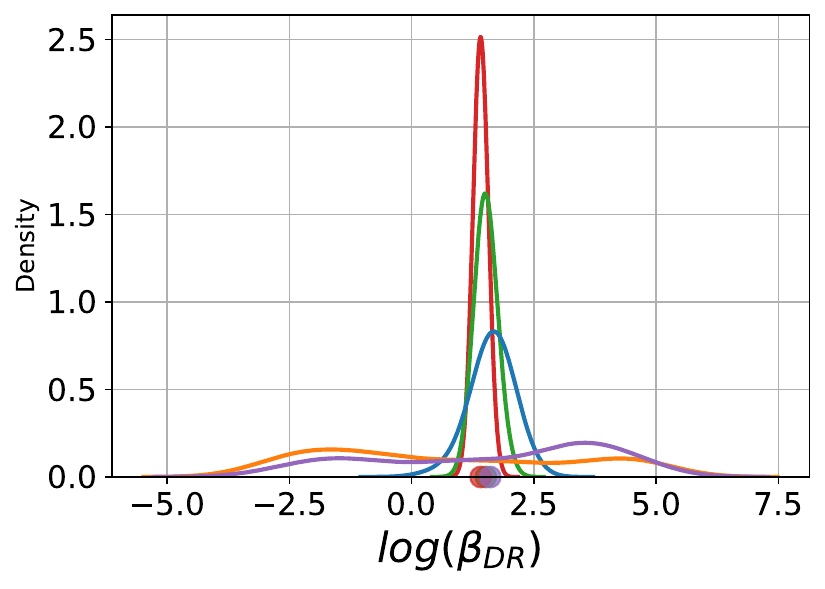}
         \label{fig: MLBA_em_betaDR}
    \end{subfigure}
   
    \begin{subfigure}{0.42\linewidth}
        
         \includegraphics[width=1.0\linewidth,height = 4cm]{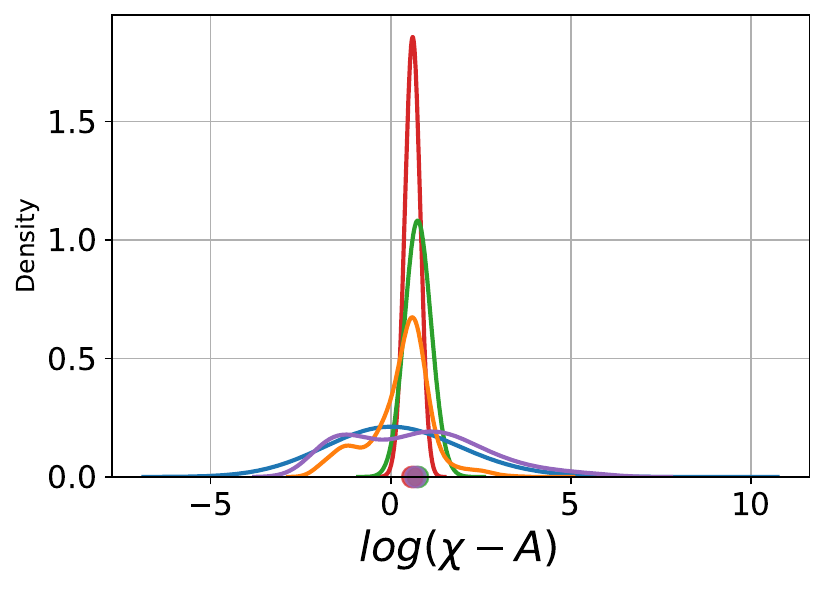}
         \label{fig: MLBA_em_chi}
    \end{subfigure}
~
    \begin{subfigure}{0.4\linewidth}
         \centering\includegraphics[trim={0.1cm 0.7cm 0.1cm 0.6 cm},clip,height = 3cm,raise=3.8em]{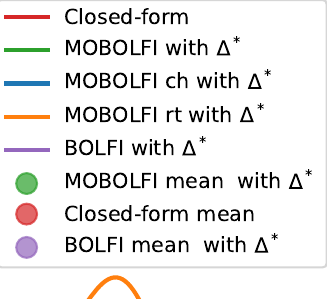}
    
    \end{subfigure}
  
    \caption{Approximate and closed-form posteriors for MLBA example. The plots show estimated marginal posteriors for different parameters. All posteriors are estimated using De-MCMC sampling and kernel density estimation.  The coloured lines and dots show the results for different methods.  The methods compared are estimates obtained with closed-form likelihood (red), MOBOLFI with both data sources (green), MOBOLFI with only choice data (blue), MOBOLFI with only response data (orange), and BOLFI with both data sources (purple). 
}
\label{fig: MLBA_em_post}
\end{figure}


~~~~Figure~\ref{fig: MLBA_em_post} shows approximate marginal
MOBOLFI and BOLFI posteriors for selected parameters and the
corresponding posteriors for the closed-form MLBA likelihood.  We make
two observations.  First, the MOBOLFI approximate posteriors (green curves) are much closer to the closed-form posteriors (red curves) than those of BOLFI (purple curves) for all parameters with the same training iterations, highlighting MOBOLFI's advantages for multi-source data. 
 Second, the figure demonstrates that the contributions of different data sources to the estimation of individual parameters can vary 
greatly according to the parameter, and MOBOLFI can 
reveal these relationships.  The $CH$ data contributed more to the estimation of the front-end parameters of MLBA ($\delta_{ICEV},\mathbf{\beta},\lambda_1$) related to the preference across alternatives and attributes, while the $RT$ data contributes more to estimation of the back-end parameters of MLBA ($\log{(\chi-\mathcal{A})}$). 
For example, the MOBOLFI marginal posterior of $\lambda_1$ and $\log{(\beta_{DR})}$ is very diffuse when conditioning on only response-time data, showing that the MOBOLFI posterior conditioning on both data sources is informative mostly because of the choice data.  In contrast, 
the $\log{(\chi-\mathcal{A})}$ MOBOLFI marginal posterior conditioning
on both choice and response time is more similar to its
marginal posterior conditional on response time only.  This is consistent with the interpretation that $\chi$ reflects the decision difficulty and respondent's cautiousness, which are associated with RT. Lastly, the approximate MOBOLFI posterior conditional on both data sources is always less dispersed than the MOBOLFI posterior conditional on a single data source only. This observation is aligned with the finding \citep{li2024importance} that joint choice-RT data provides a more efficient parameter estimate compared to an estimate applying marginal data sources (CH or RT only).

\begin{figure}[h]
    \centering
    \begin{subfigure}{0.38\linewidth}
\includegraphics[width = \linewidth]{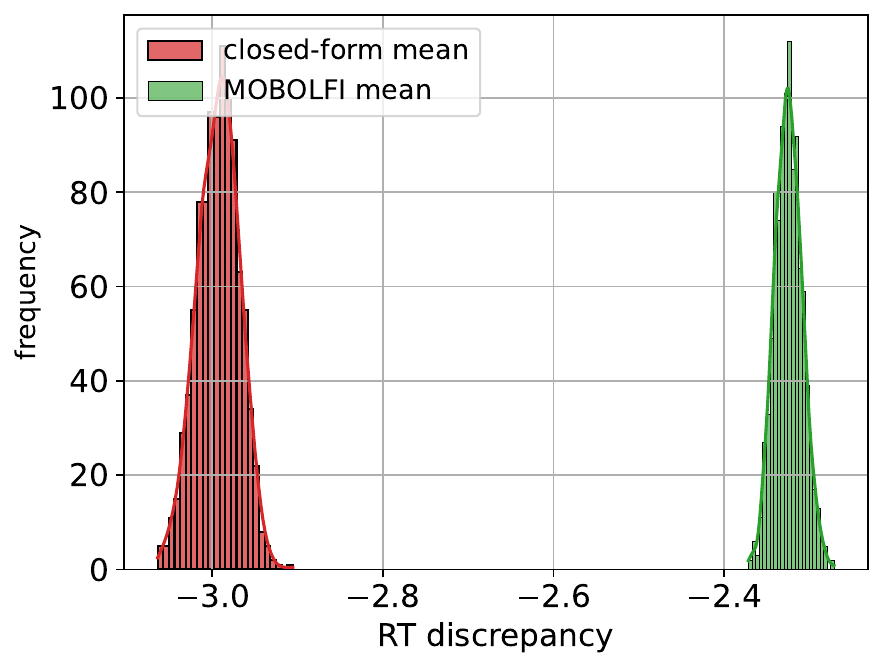}
         \label{fig: MLBA_em_ch2}
    \end{subfigure}
    ~\begin{subfigure}{0.38\linewidth}
\includegraphics[width = \linewidth]{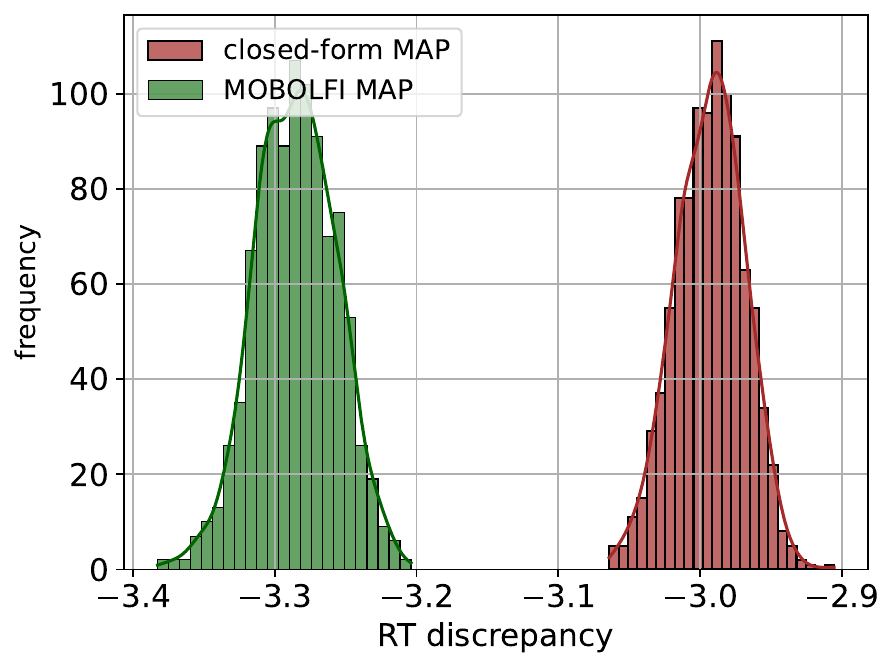}
         \label{fig: MLBA_em_ch1}
    \end{subfigure}

    \begin{subfigure}{0.38\linewidth}
\includegraphics[width = \linewidth]{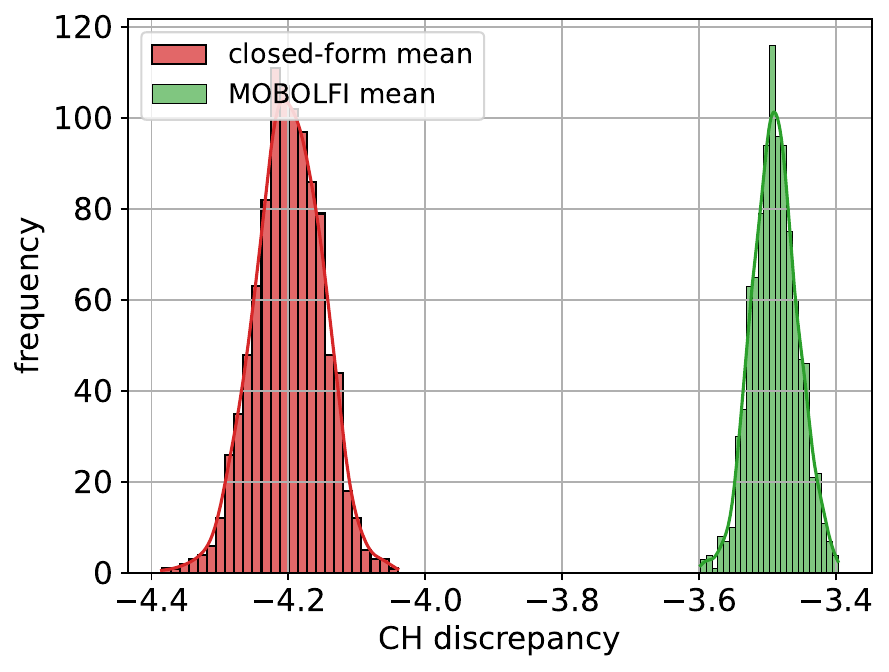}
         \label{fig: MLBA_em_rt2}
    \end{subfigure}
    ~
    \begin{subfigure}{0.38\linewidth}
\includegraphics[width = \linewidth]{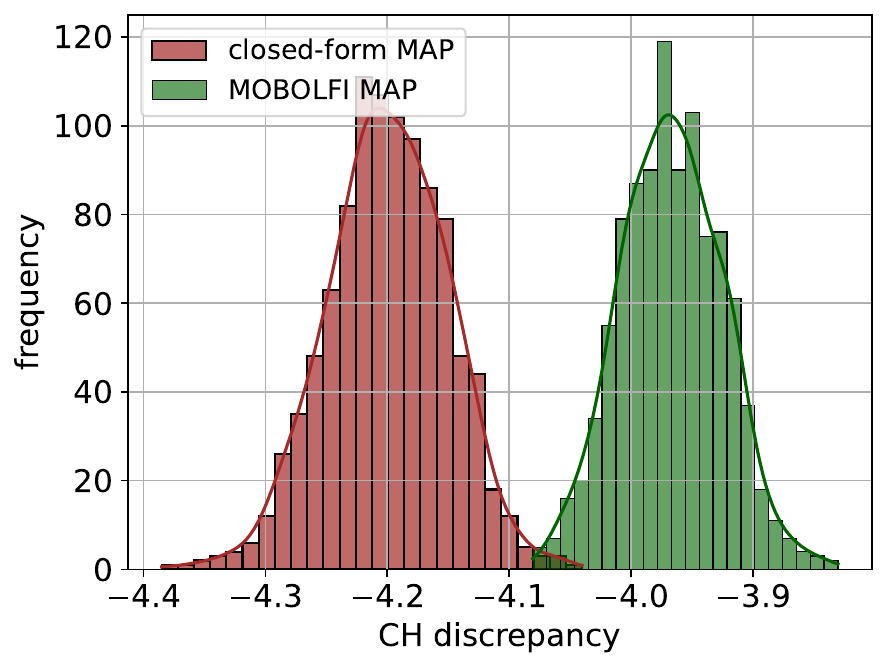}
         \label{fig: MLBA_em_rt1}
    \end{subfigure}
    
    \caption{The histogram of discrepancies with 1000 replicates based on different point estimates. The left column denotes the RT (top) and CH (bottom) discrepancy distributions for posterior means of MOBOLFI (green) and closed-form likelihood (red), respectively. The right column shows the analogous RT and CH discrepancy distributions of the posterior MAP based on MOBOLFI and closed-form.  }
    
    \label{fig: MLBA_em_dis}
\end{figure}

~~~~The posterior estimates of MOBOLFI are different from those from the closed-form likelihood in Figure~\ref{fig: MLBA_em_post}, mainly on $\log{\beta_{TC}}$ and 
$\delta_{ICEV}$.  There are several sources of error arising in the application of the MOBOLFI method.  One is the choice of discrepancy, which is 
part of the definition of the BOLFI or MOBOLFI approximate likelihoods, 
which differ from the true likelihood.  A second source of error is the choice of tolerance parameter in forming the approximation.   A third source of approximation is that the approximate likelihood is calculated from the BO Gaussian process surrogate.  When the number of model simulations is small, the estimate of the expected discrepancy provided by the surrogate and estimates of its noise parameters may be poor, and the approximate likelihood obtained from the surrogate may be inaccurate even if a good discrepancy has been used.  This source of error will be reduced as the number of model simulations increases.  A final major source of error can occur if the error model in the BO surrogate is inappropriate – this can be mitigated by more sophisticated error models, or transformations.  We have taken care in implementing
the BOLFI and MOBOLFI approaches to use sufficient simulations, to choose the
tolerance carefully and to ensure the adequacy of the error model.  In this example
we find that the most important factor in the performance of BOLFI and MOBOLFI
is the choice of discrepancy. 
Figure~\ref{fig: MLBA_em_dis} examines how well the observed
data are reproduced in the fitted model for point estimates from the posterior means and MAP estimates for the closed-from likelihood and MOBOLFI approximate posteriors. The left column in
the figure shows that log discrepancies for synthetic data simulated using the posterior mean of MOBOLFI are generally larger than those for the closed-form likelihood posterior mean value on both $CH$ and $RT$. The right column of the figure compares simulated
discrepancies for MOBOLFI and close-form MAP point estimates. The $RT$ log discrepancies simulated using MOBOLFI MAP are smaller, while $CH$ log discrepancies simulated using closed-form MAP are smaller. 

\section{Discussion}\label{sec:discussion}
~~~~We developed Multi-objective Bayesian Optimization for Likelihood-Free Inference (MOBOLFI) to address Bayesian inference with multi-source data in complex
models with intractable likelihood, such as SSMs. MOBOLFI extends the classic
BOLFI method, which acquires model simulations at the most beneficial parameter
values by using Bayesian optimization (BO) applied to minimization of an expected discrepancy between synthetic and observed data.  
The surrogate model used in BOLFI is also used for approximating the likelihood.  The MOBOLFI extension for multi-source data
considers a discrepancy for each data source, and considers multi-objective
BO for exploring the parameter values where any individual data source
likelihood is high in a simulation efficient manner.  Major advantages of the approach include avoiding information
loss from naive methods for combination of data-source specific discrepancies
into a single discrepancy, and the ability to approximate both the joint likelihood
and likelihood for individual data sources.  The latter is useful for
detecting conflict between data sources, and for understanding the importance
of the data sources for estimation of individual parameters.

~~~~Although the MOBOLFI method is motivated by the parameter inference of SSMs with multi-source data, it can be applied to Likelihood-Free inference of other
complex models. For example, in Section D of the Appendix we consider an example on bacterial infection in day care centres where parameters of a latent Markov process are inferred.  We partition 4 data summaries into two and apply MOBOLFI for inference. MOBOLFI is  competitive with BOLFI even in this example with only an individual data source.  For the case of SSMs, we use the MLBA simulator as the simulator of interest in our experiments, and MOBOLFI could be used in a wide range of SSM variants, especially for those whose likelihood functions are intractable like DDM, MDFT, etc.  An interesting direction for future work is 
to consider different choices of the acquisition function in the MOBOLFI approach.  
Multi-objective BO is an active area of research in Bayesian optimization, 
and new developments in BO methodology may translate into improved performance
in Likelihood-Free applications.  State-of-the-art computational methods for the MLBA
with closed-form likelihood, including variants of the model including random
effects, are discussed in \citet{gunawan+htkb20}.  \citet{evans19} addresses
the important issue of efficient simulation from various SSM models, which
is particularly important for likelihood-free inference.

~~~~In applying MOBOLFI there are a number of challenges.  One is that the initialization
of the algorithm can be very important, and an adequate initial
covering of the space is needed.  This seems to be especially crucial with
acquisition functions based on expected hypervolume improvement, where
the BO algorithm can keep proposing points in a small area of the space 
without a proper initialization. Another difficulty, common to many 
other LFI algorithms, is the choice of data summary statistics used
to define discrepancies.  
In Section C.3 of the Appendix we consider alternative choices of summary
statistics to the ones considered in section~\ref{subsec:MLBAsyndat} based
on auxiliary model approaches to summary statistic construction which are
widely used in the ABC literature \citep{drovandi+pl15}.  The choice of informative summary statistics remains challenging when working with complex models like SSMs. Further exploration of these issues is left to future work.




\bibliography{MOBOLFIbib}

\begin{thebibliography}{61}
\providecommand{\natexlab}[1]{#1}
\providecommand{\url}[1]{\texttt{#1}}
\expandafter\ifx\csname urlstyle\endcsname\relax
  \providecommand{\doi}[1]{doi: #1}\else
  \providecommand{\doi}{doi: \begingroup \urlstyle{rm}\Url}\fi

\bibitem[An et~al.(2020)An, Nott, and Drovandi]{an2020robust}
Ziwen An, David~J Nott, and Christopher Drovandi.
\newblock Robust {B}ayesian synthetic likelihood via a semi-parametric approach.
\newblock \emph{Statistics and Computing}, 30\penalty0 (3):\penalty0 543--557, 2020.

\bibitem[Aushev et~al.(2022)Aushev, Pesonen, Heinonen, Corander, and Kaski]{aushev+phck22}
Alexander Aushev, Henri Pesonen, Markus Heinonen, Jukka Corander, and Samuel Kaski.
\newblock Likelihood-free inference with deep {G}aussian processes.
\newblock \emph{Computational Statistics and Data Analysis}, 174:\penalty0 107529, 2022.

\bibitem[Bahg et~al.(2020)Bahg, Evans, Galdo, and Turner]{bahg2020gaussian}
Giwon Bahg, Daniel~G Evans, Matthew Galdo, and Brandon~M Turner.
\newblock Gaussian process linking functions for mind, brain, and behavior.
\newblock \emph{Proceedings of the National Academy of Sciences}, 117\penalty0 (47):\penalty0 29398--29406, 2020.

\bibitem[Balandat et~al.(2020)Balandat, Karrer, Jiang, Daulton, Letham, Wilson, and Bakshy]{balandat20}
Maximilian Balandat, Brian Karrer, Daniel Jiang, Samuel Daulton, Ben Letham, Andrew~G Wilson, and Eytan Bakshy.
\newblock Botorch: A framework for efficient {Monte-Carlo Bayesian} optimization.
\newblock In H.~Larochelle, M.~Ranzato, R.~Hadsell, M.F. Balcan, and H.~Lin (eds.), \emph{Advances in Neural Information Processing Systems}, volume~33, pp.\  21524--21538. Curran Associates, Inc., 2020.

\bibitem[Bliznyuk et~al.(2008)Bliznyuk, Ruppert, Shoemaker, Regis, Wild, and Mugunthan]{bliznyuk+rsrwm08}
Nikolay Bliznyuk, David Ruppert, Christine Shoemaker, Rommel Regis, Stefan Wild, and Pradeep Mugunthan.
\newblock Bayesian calibration and uncertainty analysis for computationally expensive models using optimization and radial basis function approximation.
\newblock \emph{Journal of Computational and Graphical Statistics}, 17\penalty0 (2):\penalty0 270--294, 2008.

\bibitem[Brown \& Heathcote(2008)Brown and Heathcote]{brown_simplest_2008}
Scott~D Brown and Andrew Heathcote.
\newblock The simplest complete model of choice response time: Linear ballistic accumulation.
\newblock \emph{Cognitive psychology}, 57\penalty0 (3):\penalty0 153--178, 2008.

\bibitem[Camerer(2000)]{camerer1998prospect}
Colin~F. Camerer.
\newblock Prospect theory in the wild: Evidence from the field.
\newblock In Daniel Kahneman and Amos Tversky (eds.), \emph{Choices, Values, and Frames}, pp.\  288–300. Cambridge University Press, 2000.

\bibitem[Cobb \& Jalaian(2021)Cobb and Jalaian]{cobb2020scaling}
Adam~D. Cobb and Brian Jalaian.
\newblock Scaling {Hamiltonian Monte Carlo} inference for {B}ayesian neural networks with symmetric splitting.
\newblock In Cassio de~Campos and Marloes~H. Maathuis (eds.), \emph{Proceedings of the Thirty-Seventh Conference on Uncertainty in Artificial Intelligence}, volume 161 of \emph{Proceedings of Machine Learning Research}, pp.\  675--685. PMLR, 2021.

\bibitem[Conrad et~al.(2016)Conrad, Marzouk, Pillai, and Smith]{conrad+mps16}
Patrick~R. Conrad, Youssef~M. Marzouk, Natesh~S. Pillai, and Aaron Smith.
\newblock Accelerating asymptotically exact {MCMC} for computationally intensive models via local approximations.
\newblock \emph{Journal of the American Statistical Association}, 111\penalty0 (516):\penalty0 1591--1607, 2016.

\bibitem[Cox \& John(1997)Cox and John]{cox+j97}
Dennis~D Cox and Susan John.
\newblock {SDO}: A statistical method for global optimization.
\newblock In \emph{Multidisciplinary Design Optimization: State of the Art}, pp.\  315--329. SIAM: Philadelphia, 1997.

\bibitem[Craig et~al.(1997)Craig, Goldstein, Seheult, and Smith]{craig+gss97}
Peter~S Craig, Michael Goldstein, Allan~H Seheult, and James~A Smith.
\newblock Pressure matching for hydrocarbon reservoirs: a case study in the use of {B}ayes linear strategies for large computer experiments.
\newblock In \emph{Case Studies in Bayesian Statistics: Volume III}, pp.\  37--93. Springer, 1997.

\bibitem[Daulton et~al.(2021)Daulton, Balandat, and Bakshy]{daulton+bb21}
Samuel Daulton, Maximilian Balandat, and Eytan Bakshy.
\newblock Parallel {B}ayesian optimization of multiple noisy objectives with expected hypervolume improvement.
\newblock In M.~Ranzato, A.~Beygelzimer, Y.~Dauphin, P.S. Liang, and J.~Wortman Vaughan (eds.), \emph{Advances in Neural Information Processing Systems}, volume~34, pp.\  2187--2200. Curran Associates, Inc., 2021.

\bibitem[Ding et~al.(2024)Ding, Kim, Maksimenko, and Bansal]{ding2024can}
Jiaxuan Ding, Eui-Jin Kim, Vladimir Maksimenko, and Prateek Bansal.
\newblock Can decoy effects nudge ride-hailing drivers’ preferences for electric vehicles?
\newblock \emph{Available at SSRN 4682413}, 2024.

\bibitem[Drovandi et~al.(2015)Drovandi, Pettitt, and Lee]{drovandi+pl15}
Christopher~C. Drovandi, Anthony~N. Pettitt, and Anthony Lee.
\newblock {Bayesian Indirect Inference Using a Parametric Auxiliary Model}.
\newblock \emph{Statistical Science}, 30\penalty0 (1):\penalty0 72 -- 95, 2015.

\bibitem[Emmerich(2005)]{emmerich05}
Michael Emmerich.
\newblock \emph{Single-and multi-objective evolutionary design optimization assisted by {G}aussian random field metamodels}.
\newblock PhD thesis, Fachbereich Informatik, University of Dortmund, 2005.

\bibitem[Evans \& Moshonov(2006)Evans and Moshonov]{evans+m06}
M.~Evans and H.~Moshonov.
\newblock Checking for prior-data conflict.
\newblock \emph{Bayesian Analysis}, 1:\penalty0 893--914, 2006.

\bibitem[Evans(2019)]{evans19}
Nathan~J. Evans.
\newblock A method, framework, and tutorial for efficiently simulating models of decision-making.
\newblock \emph{Behavior Research Methods}, 51\penalty0 (5):\penalty0 2390--2404, 2019.

\bibitem[Fasiolo et~al.(2018)Fasiolo, Wood, Hartig, and Bravington]{fasiolo+whb18}
Matteo Fasiolo, Simon~N. Wood, Florian Hartig, and Mark~V. Bravington.
\newblock {An extended empirical saddlepoint approximation for intractable likelihoods}.
\newblock \emph{Electronic Journal of Statistics}, 12:\penalty0 1544 -- 1578, 2018.

\bibitem[Fielding et~al.(2011)Fielding, Nott, and Liong]{fielding+nl11}
Mark Fielding, David~J. Nott, and Shie-Yui Liong.
\newblock Efficient {MCMC} schemes for computationally expensive posterior distributions.
\newblock \emph{Technometrics}, 53\penalty0 (1):\penalty0 16--28, 2011.

\bibitem[Frazier \& Drovandi(2021)Frazier and Drovandi]{frazier+d21}
David~T. Frazier and Christopher Drovandi.
\newblock Robust approximate {B}ayesian inference with synthetic likelihood.
\newblock \emph{Journal of Computational and Graphical Statistics}, 30\penalty0 (4):\penalty0 958--976, 2021.

\bibitem[Frazier et~al.(2020)Frazier, Robert, and Rousseau]{frazier+rr20}
David~T. Frazier, Christian~P. Robert, and Judith Rousseau.
\newblock Model misspecification in approximate {B}ayesian computation: consequences and diagnostics.
\newblock \emph{Journal of the Royal Statistical Society: Series B (Statistical Methodology)}, 82\penalty0 (2):\penalty0 421--444, 2020.

\bibitem[Frazier et~al.(2022)Frazier, Nott, Drovandi, and Kohn]{frazier2019bayesian}
David~T Frazier, David~J Nott, Christopher Drovandi, and Robert Kohn.
\newblock {B}ayesian inference using synthetic likelihood: asymptotics and adjustments.
\newblock \emph{Journal of the American Statistical Association}, 118\penalty0 (544):\penalty0 2821--2832, 2022.

\bibitem[Garnett(2023)]{garnett23}
Roman Garnett.
\newblock \emph{Bayesian Optimization}.
\newblock Cambridge University Press, 2023.

\bibitem[Greenberg et~al.(2019)Greenberg, Nonnenmacher, and Macke]{greenberg+nm19}
David~S. Greenberg, Marcel Nonnenmacher, and Jakob~H. Macke.
\newblock Automatic posterior transformation for likelihood-free inference.
\newblock In Kamalika Chaudhuri and Ruslan Salakhutdinov (eds.), \emph{Proceedings of the 36th International Conference on Machine Learning, {ICML} 2019, 9-15 June 2019, Long Beach, California, {USA}}, volume~97 of \emph{Proceedings of Machine Learning Research}, pp.\  2404--2414. {PMLR}, 2019.

\bibitem[Gunawan et~al.(2020)Gunawan, Hawkins, Tran, Kohn, and Brown]{gunawan+htkb20}
D.~Gunawan, G.E. Hawkins, M.-N. Tran, R.~Kohn, and S.D. Brown.
\newblock {New estimation approaches for the hierarchical Linear Ballistic Accumulator model}.
\newblock \emph{Journal of Mathematical Psychology}, 96:\penalty0 102368, 2020.

\bibitem[Gutmann \& Corander(2016)Gutmann and Corander]{gutmann+c16}
Michael~U. Gutmann and Jukka Corander.
\newblock Bayesian optimization for likelihood-free inference of simulator-based statistical models.
\newblock \emph{Journal of Machine Learning Research}, 17\penalty0 (125):\penalty0 1--47, 2016.

\bibitem[Hancock et~al.(2021{\natexlab{a}})Hancock, Hess, Marley, and Choudhury]{hancock_accumulation_2021}
Thomas~O. Hancock, Stephane Hess, A.A.J. Marley, and Charisma~F. Choudhury.
\newblock An accumulation of preference: {Two} alternative dynamic models for understanding transport choices.
\newblock \emph{Transportation Research Part B: Methodological}, 149:\penalty0 250--282, jul 2021{\natexlab{a}}.

\bibitem[Hancock et~al.(2021{\natexlab{b}})Hancock, Hess, Marley, and Choudhury]{hancock2021accumulation}
Thomas~O Hancock, Stephane Hess, Anthony~AJ Marley, and Charisma~F Choudhury.
\newblock An accumulation of preference: two alternative dynamic models for understanding transport choices.
\newblock \emph{Transportation Research Part B: Methodological}, 149:\penalty0 250--282, 2021{\natexlab{b}}.

\bibitem[J{\"a}rvenp{\"a}{\"a} \& Corander(2021)J{\"a}rvenp{\"a}{\"a} and Corander]{jarvenpaa+c21}
Marko J{\"a}rvenp{\"a}{\"a} and Jukka Corander.
\newblock Approximate {B}ayesian inference from noisy likelihoods with {G}aussian process emulated {MCMC}.
\newblock \emph{arXiv preprint arXiv:2104.03942}, 2021.

\bibitem[J{\"a}rvenp{\"a}{\"a} et~al.(2018)J{\"a}rvenp{\"a}{\"a}, Gutmann, Vehtari, and Marttinen]{jarvenpaa+gvm18}
Marko J{\"a}rvenp{\"a}{\"a}, Michael~U. Gutmann, Aki Vehtari, and Pekka Marttinen.
\newblock {Gaussian process modelling in approximate Bayesian computation to estimate horizontal gene transfer in bacteria}.
\newblock \emph{The Annals of Applied Statistics}, 12\penalty0 (4):\penalty0 2228 -- 2251, 2018.

\bibitem[J{\"a}rvenp{\"a}{\"a} et~al.(2019)J{\"a}rvenp{\"a}{\"a}, Gutmann, Pleska, Vehtari, and Marttinen]{jarvenpaa+gpvm19}
Marko J{\"a}rvenp{\"a}{\"a}, Michael~U. Gutmann, Arijus Pleska, Aki Vehtari, and Pekka Marttinen.
\newblock {Efficient Acquisition Rules for Model-Based Approximate Bayesian Computation}.
\newblock \emph{Bayesian Analysis}, 14\penalty0 (2):\penalty0 595 -- 622, 2019.

\bibitem[J{\"a}rvenp{\"a}{\"a} et~al.(2021)J{\"a}rvenp{\"a}{\"a}, Gutmann, Vehtari, and Marttinen]{jarvenpaa+gvm21}
Marko J{\"a}rvenp{\"a}{\"a}, Michael~U. Gutmann, Aki Vehtari, and Pekka Marttinen.
\newblock {Parallel Gaussian Process Surrogate Bayesian Inference with Noisy Likelihood Evaluations}.
\newblock \emph{Bayesian Analysis}, 16\penalty0 (1):\penalty0 147 -- 178, 2021.

\bibitem[Kandasamy et~al.(2015)Kandasamy, Schneider, and Poczos]{kandasamy+sp15}
K.~Kandasamy, J.~Schneider, and B.~Poczos.
\newblock Bayesian active learning for posterior estimation.
\newblock In \emph{Proceedings of 24th International Joint Conference on Artificial Intelligence (IJCAI '15)}, pp.\  3605 -- 3611, July 2015.

\bibitem[Kennedy \& O'Hagan(2001)Kennedy and O'Hagan]{kennedy+o01}
Marc~C. Kennedy and Anthony O'Hagan.
\newblock Bayesian calibration of computer models.
\newblock \emph{Journal of the Royal Statistical Society: Series B (Statistical Methodology)}, 63\penalty0 (3):\penalty0 425--464, 2001.

\bibitem[Lewis et~al.(2021)Lewis, MacEachern, and Lee]{lewis+ml21}
John~R. Lewis, Steven~N. MacEachern, and Yoonkyung Lee.
\newblock {Bayesian Restricted Likelihood Methods: Conditioning on Insufficient Statistics in Bayesian Regression (with Discussion)}.
\newblock \emph{Bayesian Analysis}, 16\penalty0 (4):\penalty0 1393 -- 2854, 2021.

\bibitem[Li \& Fearnhead(2018)Li and Fearnhead]{li+f18a}
Wentao Li and Paul Fearnhead.
\newblock {On the asymptotic efficiency of approximate Bayesian computation estimators}.
\newblock \emph{Biometrika}, 105\penalty0 (2):\penalty0 285--299, 2018.

\bibitem[Li \& Bansal(2024)Li and Bansal]{li2024importance}
Xinwei Li and Prateek Bansal.
\newblock The importance of response time in preference elicitation: Asymptotic results.
\newblock \emph{Available at SSRN 4782582}, 2024.

\bibitem[Marshall \& Spiegelhalter(2007)Marshall and Spiegelhalter]{marshall+s07}
E.~C. Marshall and D.~J. Spiegelhalter.
\newblock Identifying outliers in {B}ayesian hierarchical models: a simulation-based approach.
\newblock \emph{Bayesian Analysis}, 2:\penalty0 409--444, 2007.

\bibitem[Meeds \& Welling(2014)Meeds and Welling]{meeds+w14}
Edward Meeds and Max Welling.
\newblock {GPS-ABC:} gaussian process surrogate approximate {B}ayesian computation.
\newblock In Nevin~L. Zhang and Jin Tian (eds.), \emph{Proceedings of the Thirtieth Conference on Uncertainty in Artificial Intelligence, {UAI} 2014, Quebec City, Quebec, Canada, July 23-27, 2014}, pp.\  593--602. {AUAI} Press, 2014.

\bibitem[Numminen et~al.(2013)Numminen, Cheng, Gyllenberg, and Corander]{numminen2013estimating}
Elina Numminen, Lu~Cheng, Mats Gyllenberg, and Jukka Corander.
\newblock Estimating the transmission dynamics of streptococcus pneumoniae from strain prevalence data.
\newblock \emph{Biometrics}, 69\penalty0 (3):\penalty0 748--757, 2013.

\bibitem[Papamakarios \& Murray(2016)Papamakarios and Murray]{papamakarios+m16}
George Papamakarios and Iain Murray.
\newblock Fast $\epsilon$-free inference of simulation models with {B}ayesian conditional density estimation.
\newblock In D.~Lee, M.~Sugiyama, U.~Luxburg, I.~Guyon, and R.~Garnett (eds.), \emph{Advances in Neural Information Processing Systems}, volume~29. Curran Associates, Inc., 2016.

\bibitem[Papamakarios et~al.(2019)Papamakarios, Sterratt, and Murray]{papamakarios+sm19}
George Papamakarios, David Sterratt, and Iain Murray.
\newblock Sequential neural likelihood: fast likelihood-free inference with autoregressive flows.
\newblock In Kamalika Chaudhuri and Masashi Sugiyama (eds.), \emph{Proceedings of the Twenty-Second International Conference on Artificial Intelligence and Statistics}, volume~89 of \emph{Proceedings of Machine Learning Research}, pp.\  837--848. PMLR, 2019.

\bibitem[Presanis et~al.(2013)Presanis, Ohlssen, Spiegelhalter, and Angelis]{presanis+osd13}
A.~M. Presanis, D.~Ohlssen, D.~J. Spiegelhalter, and D.~De Angelis.
\newblock Conflict diagnostics in directed acyclic graphs, with applications in {B}ayesian evidence synthesis.
\newblock \emph{Statistical Science}, 28:\penalty0 376--397, 2013.

\bibitem[Price et~al.(2018)Price, Drovandi, Lee, and Nott]{price+dln16}
Leah~F. Price, Christopher~C. Drovandi, Anthony~C. Lee, and David~J. Nott.
\newblock Bayesian synthetic likelihood.
\newblock \emph{Journal of Computational and Graphical Statistics}, 27\penalty0 (1):\penalty0 1--11, 2018.

\bibitem[Radev et~al.(2023)Radev, Schmitt, Schumacher, Elsemüller, Pratz, Schälte, Köthe, and Bürkner]{radev2023bayesflow}
Stefan~T. Radev, Marvin Schmitt, Lukas Schumacher, Lasse Elsemüller, Valentin Pratz, Yannik Schälte, Ullrich Köthe, and Paul-Christian Bürkner.
\newblock Bayesflow: Amortized bayesian workflows with neural networks.
\newblock \emph{Journal of Open Source Software}, 8\penalty0 (89):\penalty0 5702, 2023.

\bibitem[Rasmussen(2003)]{rasmussen03}
Carl~Edward Rasmussen.
\newblock Gaussian processes to speed up hybrid {M}onte {C}arlo for expensive {B}ayesian integrals.
\newblock In \emph{Bayesian Statistics 7: Proceedings of the Seventh Valencia International Meeting}, pp.\  651--660. Oxford University Press, 2003.

\bibitem[Santner et~al.(2018)Santner, Williams, and Notz]{Santner_2018}
Thomas~J. Santner, Brian~J. Williams, and William~I. Notz.
\newblock \emph{The Design and Analysis of Computer Experiments}.
\newblock Springer New York, 2018.

\bibitem[Schmitt et~al.(2023)Schmitt, Radev, and B{\"u}rkner]{schmitt2023fuse}
Marvin Schmitt, Stefan~T Radev, and Paul-Christian B{\"u}rkner.
\newblock Fuse it or lose it: Deep fusion for multimodal simulation-based inference.
\newblock \emph{arXiv preprint arXiv:2311.10671}, 2023.

\bibitem[Sisson et~al.(2018)Sisson, Fan, and Beaumont]{sisson+fb18}
S.~A. Sisson, Y.~Fan, and M.~A. Beaumont (eds.).
\newblock \emph{Handbook of Approximate Bayesian Computation}.
\newblock Chapman {\&} Hall/CRC, 2018.

\bibitem[Srinivas et~al.(2009)Srinivas, Krause, Kakade, and Seeger]{srinivas2009gaussian}
Niranjan Srinivas, Andreas Krause, Sham~M Kakade, and Matthias Seeger.
\newblock Gaussian process optimization in the bandit setting: No regret and experimental design.
\newblock \emph{arXiv preprint arXiv:0912.3995}, 2009.

\bibitem[Srinivas et~al.(2012)Srinivas, Krause, Kakade, and Seeger]{srinivas2012information}
Niranjan Srinivas, Andreas Krause, Sham~M Kakade, and Matthias~W Seeger.
\newblock Information-theoretic regret bounds for {G}aussian process optimization in the bandit setting.
\newblock \emph{IEEE transactions on information theory}, 58\penalty0 (5):\penalty0 3250--3265, 2012.

\bibitem[Terry et~al.(2015)Terry, Marley, Barnwal, Wagenmakers, Heathcote, and Brown]{terry_generalising_2015}
Andrew Terry, A.A.J. Marley, Avinash Barnwal, E.-J. Wagenmakers, Andrew Heathcote, and Scott~D. Brown.
\newblock Generalising the drift rate distribution for linear ballistic accumulators.
\newblock \emph{Journal of Mathematical Psychology}, 68-69:\penalty0 49--58, Oct 2015.

\bibitem[Thomas et~al.(2020)Thomas, S{\'a}-Le{\~a}o, de~Lencastre, Kaski, Corander, and Pesonen]{thomas+sdkcp20}
Owen Thomas, Raquel S{\'a}-Le{\~a}o, Herm{\'\i}nia de~Lencastre, Samuel Kaski, Jukka Corander, and Henri Pesonen.
\newblock Misspecification-robust likelihood-free inference in high dimensions.
\newblock \emph{arXiv preprint arXiv:2002.09377}, 2020.

\bibitem[Trueblood et~al.(2014)Trueblood, Brown, and Heathcote]{trueblood_multiattribute_nodate}
Jennifer~S Trueblood, Scott~D Brown, and Andrew Heathcote.
\newblock The multiattribute linear ballistic accumulator model of context effects in multialternative choice.
\newblock \emph{Psychological review}, 121\penalty0 (2):\penalty0 179, 2014.

\bibitem[Turner \& Sederberg(2014)Turner and Sederberg]{turner2014generalized}
Brandon~M Turner and Per~B Sederberg.
\newblock A generalized, likelihood-free method for posterior estimation.
\newblock \emph{Psychonomic bulletin \& review}, 21:\penalty0 227--250, 2014.

\bibitem[Turner et~al.(2013)Turner, Sederberg, Brown, and Steyvers]{turner2013method}
Brandon~M Turner, Per~B Sederberg, Scott~D Brown, and Mark Steyvers.
\newblock A method for efficiently sampling from distributions with correlated dimensions.
\newblock \emph{Psychological methods}, 18\penalty0 (3):\penalty0 368, 2013.

\bibitem[Ward et~al.(2022)Ward, Cannon, Beaumont, Fasiolo, and Schmon]{ward+cbfs22}
Daniel Ward, Patrick Cannon, Mark Beaumont, Matteo Fasiolo, and Sebastian Schmon.
\newblock Robust neural posterior estimation and statistical model criticism.
\newblock In S.~Koyejo, S.~Mohamed, A.~Agarwal, D.~Belgrave, K.~Cho, and A.~Oh (eds.), \emph{Advances in Neural Information Processing Systems}, volume~35, pp.\  33845--33859. Curran Associates, Inc., 2022.

\bibitem[Wilkinson(2013)]{wilkinson13}
Richard Wilkinson.
\newblock Approximate {B}ayesian computation ({ABC}) gives exact results under the assumption of model error.
\newblock \emph{Statistical Applications in Genetics and Molecular Biology}, 12\penalty0 (2):\penalty0 129 -- 141, 2013.

\bibitem[Wilkinson(2014)]{wilkinson14}
Richard Wilkinson.
\newblock {Accelerating ABC methods using Gaussian processes}.
\newblock In Samuel Kaski and Jukka Corander (eds.), \emph{Proceedings of the Seventeenth International Conference on Artificial Intelligence and Statistics}, volume~33 of \emph{Proceedings of Machine Learning Research}, pp.\  1015--1023, Reykjavik, Iceland, 22--25 Apr 2014. PMLR.

\bibitem[Wood(2010)]{wood2010statistical}
Simon~N Wood.
\newblock Statistical inference for noisy nonlinear ecological dynamic systems.
\newblock \emph{Nature}, 466\penalty0 (7310):\penalty0 1102--1104, 2010.

\bibitem[Zhang \& Golovin(2020)Zhang and Golovin]{zhang+g20}
Richard Zhang and Daniel Golovin.
\newblock Random hypervolume scalarizations for provable multi-objective black box optimization.
\newblock In Hal~Daum{\'e} III and Aarti Singh (eds.), \emph{Proceedings of the 37th International Conference on Machine Learning}, volume 119 of \emph{Proceedings of Machine Learning Research}, pp.\  11096--11105. PMLR, 2020.

\end{thebibliography}
\bibliographystyle{tmlr}

\appendix
\section{Derivation of Methods \& details of implementation}

Table \ref{glossary} is a glossary of the main notation used in this paper.  
\begin{table}[H]
\caption{\label{glossary}Glossary of notation}
\begin{center}
\begin{tabular}{ll}
$\theta$ & The model parameter taking a value in parameter space $\Theta\subseteq \mathbb{R}^p$. \\
$y$ & Data to be observed taking a value in $\mathcal{Y}$\\
$p(y|\theta)$ & The sampling density of $y$ given $\theta$\\
$y_{\text{obs}}$ & The observed value of $y$ \\
$S$ & A $d$-dimesional summary statistic to be observed \\
$p(S|\theta)$ & The sampling density of the summary statistic given $\theta$\\
$S_{\text{obs}}$ & The observed value of $S$  \\
$p(\theta|y_{\text{obs}})$ & The posterior distribution given the observed $y$ \\
$p(\theta|S_{\text{obs}})$ & The partial posterior distribution given the observed $S$ \\
$p_t(S_{\text{obs}}|\theta)$ & ABC likelihood with tolerance $t>0$ \\
$\Delta_\theta(S,S_{\text{obs}})$ & A discrepancy between a simulated summary $S\sim p(S|\theta)$ and $S_{\text{obs}}$ \\
$D(\theta)$ & The expectation of $\Delta_\theta(S,S_{\text{obs}})$ for $S\sim p(S|\theta)$ \\
$y=(x^\top,w^\top)^\top$ & For two-source data, $y$ consists of component data sources $x$ and $w$ \\
$y_{obs}=(x_{\text{obs}}^\top,w_{\text{obs}})$ & $x_{\text{obs}}$ and $w_{\text{obs}}$ are the observed
values of $x$ and $w$ \\
$S=(T^\top,U^\top)^\top$ & For two-source data, the vector of summaries $S$ 
concatenates   \\
& \;\;\;\;summaries $T$ for data source $x$ and $U$ for data source $w$ \\
$T_{\text{obs}},U_{\text{obs}}$ & The observed values 
of the summary statistics $T$ and $U$ \\
$p(T_{obs}|\theta),p(U_{\text{obs}}|\theta)$, & Likelihoods for $T$, $U$, $U|T$ and $T|U$ for  \\
\;\;\;\;$p(U_{\text{obs}}|T_{obs},\theta),p(T_{\text{obs}}|U_{\text{obs}},\theta)$ &  \;\;\;\;two-source
summary statistic data\\
$\Lambda_\theta(T,T_{\text{obs}})$ & A discrepancy between a simulated summary $T\sim p(T|\theta)$ and $T_{\text{obs}}$ \\
$\psi_\theta(U,U_{\text{obs}})$ & A discrepancy between a simulated summary $U\sim p(U|\theta)$ and $U_{\text{obs}}$ \\
$D_1(\theta)$ & The expectation of $\Lambda_\theta(T,T_{\text{obs}})$ for $T\sim p(T|\theta)$ \\
$D_2(\theta)$ & The expectation of
$\psi_\theta(U,U_{\text{obs}})$ for $U\sim p(U|\theta)$ \\
$\widetilde{p}_t(S_{\text{obs}}|\theta),\widetilde{p}_t(T_{\text{obs}}|\theta),\widetilde{p}_t(U_{\text{obs}}|\theta)$, & MOBOLFI approximate
likelihoods for $S$, $T$, $U$, $U|T$ and $T|U$\\
\;\;\;\;$\widetilde{p}_t(U_{\text{obs}}|T_{\text{obs}},\theta),
\widetilde{p}_t(T_{\text{obs}}|U_{\text{obs}},\theta)$ & 
\;\;\;\; with
tolerance $t=(t_1,t_2)$.  
\end{tabular}
\end{center}
\end{table}
\subsection{Gaussian Processes and Bayesian optimization}
~~~~Next we provide further background on Gaussian processes and Bayesian optimization (BO).  To make the discussion self-contained, 
there is some repetition of definitions and concepts explained in the main text.
BO with a Gaussian process surrogate is used in the BOLFI method which
inspires the new MOBOLFI approach in our work.  

Bayesian optimization attempts to find a global optimum of a function.
We will consider Bayesian optimization for finding a global minimum.  
We want to minimize
$f(\theta)$, $\theta\in \Theta\subseteq \mathbb{R}^p$, where derivatives
of $f(\cdot)$ are not available, and evaluations of $f(\cdot)$ may
be corrupted by noise.  BO is most suitable for problems where
the dimension of $\theta$ is not too large, although high-dimensional BO
is an active area of current research.
We model the noisy evaluations of $f(\theta)$ with a 
``surrogate model''
describing uncertainty about $f(\cdot)$ given the 
function evaluations previously made.  A common surrogate model in BO is a Gaussian
process, and this is used in the BOLFI method.  

~~~~We establish some notation and definitions first.  
Suppose that $\widetilde{\Theta}$ is an $n\times p$ matrix, with $i$th row 
$\widetilde{\theta}_i\in \Theta$.  For any function 
$g:\Theta\rightarrow \mathbb{R}$, we write $g(\widetilde{\Theta})$ for
the vector obtained by applying $g(\cdot)$ to the rows of $\widetilde{\Theta}$, 
i.e. $g(\widetilde{\Theta})=(g(\widetilde{\theta}_1),\dots, g(\widetilde{\theta}_n))^\top$.  Suppose that $\overline{\Theta}$ is an $m\times p$ matrix, 
where the $i$th row is $\overline{\theta}_i\in \Theta$.  For any function
$h:\Theta\times \Theta\rightarrow \mathbb{R}$, we write
$h(\widetilde{\Theta},\overline{\Theta})$ for the $n\times m$ matrix with
$(i,j)$th entry $h(\widetilde{\theta}_i,\overline{\theta}_j)$.  
~~~~For $\theta_1,\dots, \theta_n\in \Theta$, denote by
$\theta_{1:n}$ the $n\times p$ matrix with $i$th row $\theta_i$.  
A random function $f(\cdot)$ defined on $\Theta$ is a Gaussian process
with mean function $\mu:\Theta\rightarrow \mathbb{R}$ and positive
definite covariance function $C:\Theta\times \Theta\rightarrow \mathbb{R}$ if,
for any $n$, and any $\theta_1,\dots, \theta_n\in\Theta$, 
the random vector $f(\theta_{1:n})$ is multivariate normally
distributed, $f(\theta_{1:n})\sim N(\mu(\theta_{1:n}),C(\theta_{1:n},\theta_{1:n}))$.  

~~~~Suppose we observe the Gaussian process $f(\cdot)$ with noise at points 
$\theta_1,\dots, \theta_n\in \Theta$.  The noisy observations are
\begin{align}
  z_i & = f(\theta_i)+\epsilon_i,\;\;\;\;i=1,\dots, n, \label{noisyobssup}
\end{align}
where $\epsilon_i\stackrel{iid}{\sim} N(0,\sigma^2)$, for
some variance $\sigma^2>0$.  In many applications, the Gaussian noise assumption
can be reasonable in the vicinity of the minimizer with an appropriate transformation.
Write $z_{\leq n}=(z_1,\dots, z_n)^\top$.  
We are interested in describing uncertainty about $f(\theta^*)$ for
some $\theta^*\in \Theta$, given the noisy observations 
$z_{\leq n}$.  Since $f(\cdot)$ is a 
Gaussian process with mean function $\mu(\cdot)$ and covariance
function $C(\cdot,\cdot)$, $(f(\theta_{1:n})^\top,f(\theta^*))^\top$ is
multivariate normal, and because of the independent normally distributed
noise in \eqref{noisyobssup}, $(z_{\leq n}^\top,f(\theta^*))^\top$ is easily
shown to be multivariate normal.  It follows that 
$f(\theta^*)|z_{\leq n}\sim N(\mu_n(\theta^*),\sigma_n^2(\theta^*))$, where
\begin{align}
  \mu_n(\theta^*) & = \mu(\theta^*)+C(\theta^*,\theta_{1:n})\left\{C(\theta_{1:n},\theta_{1:n})+\sigma^2 I\right\}^{-1}(z_{\leq n}-\mu(\theta_{1:n})), \label{prmean} \\
  \sigma_n^2(\theta^*) & = C(\theta^*,\theta^*)-C(\theta^*,\theta_{1:n})\left\{C(\theta_{1:n},\theta_{1:n})+\sigma^2 I\right\}^{-1}C(\theta_{1:n},\theta^*). \label{prvar}
\end{align}
~~~~The uncertainty quantification provided by
\eqref{prmean} and \eqref{prvar} can be used to decide which $\theta^*$ 
should be used to obtain a further noisy observation 
$$z^*=f(\theta^*)+\epsilon^*,\;\;\;\; \epsilon^*\sim N(0,\sigma^2),$$ 
having the most benefit in the search for the minimizer.  
$\theta^*$ is chosen through an optimization, of the ``acquisition function''. 
This acquisition function 
can be the expected loss for some formal decision problem where
the expectation is taken with respect to the the Gaussian process
uncertainty about $f(\cdot)$, or it might be chosen using more heuristic
reasoning.  


~~~~In the BOLFI method, \citet{gutmann+c16}
suggested using the lower confidence bound acquisition function
\citep{cox+j97}, 
\begin{align}
 A_n(\theta) & = \mu_n(\theta)-\sqrt{\eta_n^2 \sigma_n^2(\theta)}, \label{lcbsup}
\end{align}
where 
$$\eta_n^2=2\log \left(n^{\frac{p}{2}+2}\frac{\pi^2}{3\epsilon_\eta}\right),$$
with the default value $\epsilon_\eta=0.1$. With such design of the $\eta^2_n$, the cumulative regret of a GP using Upper Confidence bound as the acquisition is proved to be bounded with high probability \citep{srinivas2009gaussian}. In the day care center example discussed in Section 4, we implement BOLFI using $\eta^2_n$. However, in the two examples presented in the main text, the large number of parameters $p$ results in a large $\eta^2_n$. This increases the probability that the acquisition point is not close to the global optimum, necessitating more iterations and longer training times for the implementation of BOLFI. Additionally, this approach risks proposing points outside the range of the uniform prior, potentially leading to numerical problems. To ensure numerical stability and efficient training, for the two examples in main text, we replace $\eta^2_n$ with $$\eta^{2*}_n=2\log \left(n^{2}\frac{\pi^2}{3\epsilon_\eta}\right).$$ We posit that using $\eta^{2*}_n$ specifically for these examples strikes an effective balance between exploration and exploitation.

~~~~Further discussion of the intuitive basis for this acquisition function is 
given in the main text.  We have described the process of choosing a new
location for taking a new noisy function evaluation in BO algorithms.  
To completely specify a BO algorithm, we need to describe a method of initialization 
(where some initial function values are obtained, perhaps using a space-filling
design on $\Theta$), a stopping rule (most simply we might stop when a given 
budget of function evaluations is exhausted) and a method for estimating
Gaussian process hyperparameters, and updating estimates as the algorithm proceeds.  
For further details on these practical issues see \citep{garnett23}. For the two examples in the main text, we considered different initializations, with details given in Sections 2.1 and 3.2.

\subsection{Related work on BOLFI}

~~~~The BOLFI approach of \citet{gutmann+c16} has been very successful 
for simulation efficient 
estimation of posterior distributions with expensive simulators, and the
basic method has been further developed in a number of ways.  
\citet{jarvenpaa+gvm18} explores the importance of the Gaussian process
formulation used in the BOLFI framework, including transformations 
of the discrepancy, heteroskedastic or classifier Gaussian process formulations 
for likelihood approximations, and different utilities for Gaussian process 
model choice.  \citet{jarvenpaa+gpvm19} go beyond generic acquisition
functions from the Bayesian optimization literature and develop
alternatives tailored to simulation-based inference problems targeting the reduction of posterior uncertainty.
\citet{jarvenpaa+gvm21} consider Bayesian optimization 
for likelihood-free inference with noisy log-likelihood evaluations 
and batch sequential strategies amenable to parallel computation.
\citet{aushev+phck22} consider replacing the Gaussian process used
in BOLFI with a deep Gaussian process, and demonstrate that this can result
in more accurate posterior approximations, particularly when the target posterior
density is multi-modal.  \citet{kandasamy+sp15} uses Bayesian optimization with some novel acquisition functions to query an expensive to evaluate likelihood, before estimating the posterior using the cheap to evaluate surrogate.
A generalized Bayesian version of BOLFI, which is suitable for misspecified
models and when the summary statistic dimension is high, is described
by \citet{thomas+sdkcp20}.  

~~~~The use of a Gaussian process or other ``surrogate'' models to obtain a likelihood
approximation is not only used in the context of BO methods.  
\citet{wilkinson14} considers a Gaussian process surrogate for 
a synthetic likelihood \citep{wood2010statistical,price+dln16} in so-called history matching algorithms \citep{craig+gss97}.  
\citet{meeds+w14} considers Gaussian process surrogate models
for summary statistic means and variances 
for a synthetic likelihood, and adaptively acquire simulations in order to 
reduce uncertainty in a Metropolis-Hastings accept/reject decision for
posterior simulation.  A similar more sophisticated approach has
recently been considered by \citet{jarvenpaa+c21}, where the authors
use a Gaussian process surrogate for the log-likelihood itself, and 
are more explicit about acquisition rules for augmenting the training
set for the Gaussian process.
A surrogate model can be useful too in applications where 
exact likelihood calculations can be made, but are expensive.  See, for
example, \citep{kennedy+o01,rasmussen03,bliznyuk+rsrwm08,fielding+nl11,conrad+mps16} among many others.

\subsection{Implementation of MOBOLFI}
~~~~Here we give details of multi-objective Bayesian optimization which are not given in the main text. Multi-objective BO is used in implementing the MOBOLFI method developed in the main text.   
Once again, there is some repetition with definitions and concepts in the main text
to make the discussion self-contained.

~~~~Let $f(\theta)=(f_1(\theta),\dots, f_K(\theta))^\top$ be a multivariate
function, for which we are interested in minimizing the components of $f(\cdot)$.
In general there is no value $\theta^*\in \Theta$ where all components are minimized simultaneously.  Multi-objective optimization
methods approximate the set of ``non-dominated''
solutions which are not obviously inferior to other solutions.
A value $\theta\in \Theta$ dominates $\theta'\in\Theta$
if $f_j(\theta)\leq f_j(\theta')$, $j=1,\dots, K$, with the inequality
being strict for at least one $j$.  The dominated solution
is inferior in the sense that there is another point at which $f(\cdot)$ 
is strictly smaller along some dimensions and no larger for the other dimensions. 
Multi-objective optimization
algorithms try to find the Pareto optimal set of non-dominated points in $\Theta$. 
 
~~~~Numerical multi-objective optimization
methods obtain finite approximations to the Pareto set.  The Pareto frontier is the
set of optimal function values obtained by the points in the Pareto set.
Multi-objective Bayesian optimization \citet[Section 11.7]{garnett23} uses 
surrogate models to implement multi-objective optimization for expensive to evaluate
functions, possibly observed with noise.  Similar to Bayesian optimization
with a scalar objective, the representation of uncertainty given by the surrogate
is used to efficiently decide where to perform the next function evaluation.

~~~~Multivariate Gaussian processes are a common choice of surrogate for multi-objective
Bayesian optimization, and
we give some background and notation now, extending the discussion
of Section 2.3. Suppose that $g:\Theta\rightarrow \mathbb{R}^K$, and that $\widetilde{\Theta}$ is an $n\times p$ matrix with $i$th row
$\widetilde{\theta}_i\in \Theta$.  We write $g(\widetilde{\Theta})=(g(\widetilde{\theta}_1)^\top,\dots, g(\widetilde{\theta}_n)^\top)^\top\in \mathbb{R}^{Kn}$.  Let $h:\Theta\times \Theta\rightarrow
\mathbb{R}^{K\times K}$ be a $K\times K$ matrix-valued function.  
Let $\overline{\theta}$ be an $n\times m$ matrix with $i$th row $\overline{\theta}_i$.  We write $h(\widetilde{\theta},\overline{\theta})$ for the partitioned
matrix with $n$ block rows and $m$ block columns where the $(i,j)$th block
entry is $h(\widetilde{\theta}_i,\overline{\theta}_j)\in \mathbb{R}^{K\times K}$.  
A random function $f(\cdot)=(f_1(\cdot),\dots, f_K(\cdot))^\top$ is a 
multivariate Gaussian process with mean
function $\mu:\Theta\rightarrow \mathbb{R}^K$ and positive definite covariance
function $C:\Theta\times \Theta\rightarrow \mathbb{R}^{K\times K}$ if
for any $n$ and $\theta_1,\dots, \theta_n$, $f(\theta_{1:n})$ is multivariate
normal, $N(\mu(\theta_{1:n}),C(\theta_{1:n},\theta_{1:n}))$. 

Once again extending the discussion of Section 2.2, suppose we observe
values of $f(\cdot)$ with noise at $\theta_1,\dots, \theta_n\in \Theta$, to obtain
\begin{align}
  z_i & = f(\theta_i)+\epsilon_i, \label{mvregsup}
\end{align}
where now $z_i\in \mathbb{R}^K$ and $\epsilon_i\stackrel{iid}{\sim} N(0,\Sigma)$
where $\Sigma\in \mathbb{R}^{K\times K}$ is some positive definite
covariance matrix.  As in our discussion of the univariate case, 
for some $\theta^*\in \Theta$, 
and writing $z_{\leq n}=(z_1^\top,\dots, z_n^\top)^\top$, 
$(z_{\leq n}^\top,f(\theta^*)^\top)^\top$ is multivariate Gaussian and the
conditional density of $f(\theta^*)|z_{\leq n}$ is multivariate normal with mean
vector and covariance matrix
\begin{align}
  \mu_n(\theta^*) & = \mu(\theta^*)+C(\theta^*,\theta_{1:n})\left\{C(\theta_{1:n},\theta_{1:n})+D_n(\Sigma)\right\}^{-1}(z_{\leq n}-\mu(\theta_{1:n}) \label{mmean} \\
  \Sigma_n(\theta^*) & = C(\theta^*,\theta^*)-C(\theta^*,\theta_{1:n})
  \left\{C(\theta_{1:n},\theta_{1:n})+D_n(\Sigma)\right\}^{-1}C(\theta_{1:n},\theta^*), \label{mcov}
\end{align}
where $D_n(\Sigma)\in \mathbb{R}^{nK\times nK}$ is the block diagonal matrix
with the $K\times K$ diagonal block entries 
equal to $\Sigma$.  For reducing computation cost, in our later numerical
experiments we consider multivariate Gaussian processes where the
components of $f(\cdot)$ are independent, but we consider correlated
noise in \eqref{mvregsup} i.e. $\Sigma$ is not diagonal. 

~~~~The noise $\epsilon_i$ can be observed by the variation of repeated simulation $\{z_{i,j}\}_{j=1}$ give input $\theta_i$. Therefore, the dependent noise covariance matrix $\Sigma$ is estimated by the covariance of a repeated finite simulation sample $\{\Delta_{i_\Sigma,j}\}_{j=1}^{n_\Sigma}$, where $\Delta_{i_\Sigma,j} = D(\theta_{i_\Sigma})+\epsilon_{i_\Sigma}$ for some $\theta_{i_\Sigma}$ is a simulated bivariate noisy discrepancy. For results in this paper, we set $n_\Sigma = 100$ and 
$\theta_{i_\Sigma} = \underset{(\theta_i,\Delta_i) \in T_{n_f}}{\argmin}   
(\Delta_i-\mu_{n_f}(\theta_i)^\top \Sigma_{n_f}(\theta_i)^{-1}(\Delta_i-\mu_{n_f}(\theta_i))$.

~~~~Given the uncertainty quantification provided by \eqref{mmean} and 
\eqref{mcov}, if there is a finite set of points, say $\theta_1,\dots, \theta_n$, 
approximating the Pareto set, with corresponding approximation
$f_1,\dots, f_n$ of the Pareto frontier, expected hypervolume improvement (EHVI) measures the volume of the space dominated by the current
approximation of the Pareto frontier and bounded below by a reference point, the so-called Pareto hypervolume.  EHVI was firstly used as an acquisition function in multi-objective Bayesian optimization by \citet{emmerich05}, where an expectation
of the hypervolume improvement is taken with respect to the surrogate model
uncertainty to define the acquisition function. In the MOBOLFI method, 
multi-objective BO is applied to a vector of expected discrepancies, and
the simulated discrepancies are noisy.  For this reason, we use 
the noisy expected hypervolume improvement (NEHVI) \citep{daulton+bb21}
for the acquisition function.  NEHVI 
implements a Bayesian treatment when calculating EHVI, integrating uncertainty 
about the Pareto frontier.  This makes NEHVI more suitable than EHVI
as an acquisition function with noisy data.

~~~~\citet{gutmann+c16} chose the tolerance $t$ in the univariate BOLFI method as the $q$-quantile of $\Delta_1,...,\Delta_{n_f}$, where $q \in (0,1)$. In the bivariate MOBOLFI method, we extend the choice of tolerance $t=(t_1,t_2)$ to the 2-dimensional vector $q$-quantile of $\Delta_1,...,\Delta_{n_f}$, where $q \in (0,1)^2$. Given that the evaluation of ABC approximate likelihood involves $t$, we also did a comparison study over different $q$-quantile tolerance for 3 different examples, by letting $q=0.01/0.05/0.1/0.2$. Our visuals present the affect of tolerance on performance of inference, varying by examples.  

~~~~In multi-objective Bayesian Optimization, one practical difficulty is the scaling of objectives. For a fixed diagonal matrix $V$, instead of applying multi-objective
BO to a vector of noisy discrepancies $\Delta$, we could apply it to 
$V^{-1}\Delta$ instead, and in general the results are not invariant to the
choice of $V$.  Figure \ref{brownian_scaling} studies the effect of scaling on inference in a toy example with $V^{-1} = \text{diag}(w,1)$, where
$w$ is a scalar weight.  Since only two objectives involved, we follow the notation in Section 4.1, where $\Delta=(\Delta_1(X^{(i)},X^o),\Delta_2(W^{(i)},W^o))$ denotes the joint objective over two data sources $X$ and $W$. From the figure, we find that with different $w$, if we simply add the discrepancy to 
get a univariate discrepancy for the BOLFI posterior the approximate 
posterior differs markedly, while for MOBOLFI the approximate posterior is less 
sensitive to $w$. Not doing scaling is equivalent to setting the $V^{-1} = I_2$, i.e. the red curve in this figure, which is not the choice with the best performance. 

~~~~One classic scaling method in the machine learning literature is normalization by using Mean absolute deviation (MAD). Algorithm \ref{alg:scaling} presents the detailed steps of doing scaling of a joint noisy objective $\Delta$ for implementing BOLFI. They key idea of Algorithm \ref{alg:scaling} is to put each elements of $\Delta$ on a similar scale. The use of Algorithm \ref{alg:scaling} does not always result in the best performance. In the plot \ref{fig:BOLFI scaling}, the green curve with $w=0.7$ is chosen by Algorithm \ref{alg:scaling} (rounding to 1 decimal places). The scaling with best performance is example specific, depending on the information of each data source, the LFI method and the BO acquisition function. We leave further investigation of optimal scaling approaches to future work. For the results of this paper, by default we choose the scaling as the outcome (rounded to 1 decimal places) from algorithm \ref{alg:scaling} using $n=100$. The auxiliary model example (orange curve in Figure \ref{MLBA_MOBOLFIvsAUX}) defines a discrepancy as the score vector of an auxiliary model, and the scaling in this example is slightly different and will be discussed in Section C.3 of Appendix.

\begin{algorithm}
\caption{Scaling of discrepancies in (MO)BOLFI}\label{alg:scaling}

\begin{algorithmic}
\Require Prior $\pi(\cdot)$ of $\theta$, target function $\Delta=(\Delta_1,\cdots,\Delta_K)^T$ to scale, number of sample size $n$
\State Sample $\theta^{(i)} \sim \pi(\theta) , i=1,\cdots,n$
\State For each $\theta_i$, evaluate the corresponding target function $\Delta(\theta^{(i)}) = (\Delta_1(\theta^{(i)}),\cdots,\Delta_K(\theta^{(i)}))^T$
\State For $j=1,\cdots,K$, write $v^{(j)} = \text{MAD}\{\Delta_j(\theta^{(1)}), \cdots, \Delta_j(\theta^{(n)})\}$ 
\State Write $V=\text{diag}(v^{(1)}, \cdots, v^{(K)})$
\State Scale the target function $\Delta_{\text{scale}} = V^{-1}\Delta$
\State (In BOLFI, $\Delta=\Delta_1+\cdots+\Delta_K$, and the scaled
discrepancy is $\Delta_{\text{scale}} = V^{-1}_{11}\Delta_1+\cdots+V^{-1}_{KK}\Delta_K$)
\end{algorithmic}
\end{algorithm}

\section{Experiment setup and extra findings - Toy example}

~~~~We give some further details of the implementation of MOBOLFI and some additional
experiments that were not included in the main text due to space limitations. Code 
to implement all experiments can be obtained at \begin{sloppypar} \url{https://github.com/DZCQs/Multi-objective-Bayesian-Optimization-Likelihood-free-Inference-MOBOLFI.git}.
\end{sloppypar}

\subsection{Experiment setup - toy example}

\begin{sloppypar}~~In the toy example in the main text initial training data of 100 observations was used, and BOLFI/MOBOLFI was used to train 1-dimensional/2-dimensional surrogate GP models respectively with 200 BO acquisitions.  Parameter samples are drawn from the approximate posterior distribution (Section 2.3) using Hamiltonian Monte Carlo (HMC) with tolerance $t$ set to be the 1\% quantile of the training data discrepancies.  This is done element-wise for each discrepancy to get the vector of tolerances for MOBOLFI.  We implement HMC using the {\texttt hamiltorch} package \citep{cobb2020scaling}, running four chains for 8,000 iterations each, using step size 0.1 and one step within each proposal (which corresponds to a Metropolis-Hastings adjusted Langevin algorithm). 

~~~~For MOBOLFI, the objective function optimized is 
$$D(\theta)=(E(\Delta_1(X,X^o)),E(\Delta_2(W,W^o))^\top,$$
where $X$ is the synthetic data and $X^o$ is the observed
data for the first data source, $\Delta_1(X,X^o)$ is
the discrepancy for the first data source, $W$ is synthetic
data and $W^o$ is the observed data for the second
data source, and $\Delta_2(W,W^o)$ is the discrepancy
for the second data source.  The definition of the discrepancies is given in the main text.  For implementing BOLFI, the training data is $\{\theta_i,w\cdot\Delta_1(X^{(i)},X_o)+\Delta_2(W^{(i)},W_o)\}_{i=1}^{100}$, where $w=0.4$ is the scaling with best performance from a number of alternatives (see figure \ref{fig:BOLFI scaling} in Section A.3 of this Appendix). We use the lower confidence bound acquisition function for BOLFI with $\eta=0.1$ in \eqref{lcbsup}.  For implementing MOBOLFI, the training data is 
$\{\theta_i,(w\cdot \Delta_1(X^{(i)},X_o),\Delta_2(W^{(i)},W_o))\}_{i=1}^{100}$. We use the NEHVI acquisition
function \citep{daulton+bb21} with a reference point $\min_i((\Delta_1(X^{(i)},X^o),\Delta_2(W^{(i)}, W^o)))-0.1$. Each optimization in the multi-objective
BO iterations approximates the optimum of the acquisition function from 100 candidate samples in 10 restarts.
To avoid observations with discrepancy values which are too large (which affects GP training), we set a mild rejection criterion to filter out observations higher
than a threshold, i.e. the $99\%$-quantile (rounded to 1 decimal places) obtained from simulated discrepancies that are drawn without any filtering criterion. \end{sloppypar}

~~~~The botorch package maximizes the supplied objective function by default, and so in implementation of BOLFI we maximize the negative expected discrepancy.  \citet{gutmann+c16} minimizes the lower confidence bound as (\ref{lcbsup}) in their work, and 
when maximizing the negative expected discrepancy it is equivalent to use the upper confidence
bound as acquisition function, 
\begin{align}
 A_n(\theta) & = \mu_n(\theta)+\sqrt{\eta_n^2 \sigma_n^2(\theta)}. \label{ucb}
\end{align} 

\subsection{Extra findings - toy example}
~~~~In addition to the experiments in the main text, 
Figure \ref{brownian_iterandtol} shows MOBOLFI approximate posteriors of $\theta_1$ for different choices
of the number of iterations in the BO algorithm and for
different choices of tolerance.  The approximate 
posteriors are kernel density estimates from HMC samples.  
The left-hand column of the figure demonstrates
that the MOBOLFI approximate posterior variance decreases with more BO iterations.  This is expected, 
since the surrogate model uncertainty contributes to
the uncertainty in the MOBOLFI approximate posterior.  
The right-hand column of the figure demonstrates that
the MOBOLFI approximate posterior is closer to the true 
posterior when a smaller quantile of the training data
is used for the tolerance. In Figure \ref{brownian_iterandtol} (b), 200 
BO acquisitions were used. 
\citet{gutmann+c16} suggested using the 5\% quantile of the training data, but a 1\% quantile attains comparable performance in this example. 
Both the selection of tolerance levels in the likelihood approximation and the number of BO iterations used are crucial for achieving accurate approximate posterior distributions using MOBOLFI.

\begin{figure}
\centering
\begin{subfigure}{0.4\linewidth}
         \includegraphics[width=0.8\linewidth,height=4.5cm]{figures/brownian_vsiter.pdf}
         \caption{Different iterations for GP training}
         \label{fig:MOBOLFI vs iter}
\end{subfigure}
\begin{subfigure}{0.4\linewidth}
         \includegraphics[width=0.8\linewidth,height=4.5cm]{figures/brownian_vstol.pdf}
         \caption{Different choices of tolerance}
         \label{fig:MOBOLFI vs tol}
\end{subfigure}
\caption{Approximate posterior for the toy example given different numbers of training iterations and tolerances. The left column shows the MOBOLFI approximate posteriors given 50/100/150/200 BO iterations. The right column presents the approximate posteriors for threshold $t$ set to 20\%/5\%/1\% quantiles of the training data discrepancies.  All approximate
posterior densities are kernel density estimates from HMC samples.  The red dashed line is at $\theta_1^{\text{true}}=-0.7$. }
\label{brownian_iterandtol}
\end{figure}

~~~~MOBOLFI differs from BOLFI by using multiple discrepancies in the BO algorithm. To explore the sensitivity of inference to including multiple data sources when the model for some data sources does not depend on some parameters, we modify the simulator as follows. The first 8 elements of $\theta_{\text{true}}$ are shared by both X and Y. However, the 9th element of $\theta_{\text{true}}$ contributes to X only, and the 10th element of $\theta_{\text{true}}$ contributes to Y only. Specifically,
\begin{equation}
\begin{aligned}
 & \theta \sim \mathbb{N}(\theta|0,I) \\
 & X_n \sim \mathbb{N}(x|\theta,I), n = 1,...,N \\ 
 & w(t)=\theta dt + \sigma dW(t) \\
 & \theta_X = (\theta_{\text{true},1},...,\theta_{\text{true},8}, \theta_{\text{true},9})^T \\ 
 & \theta_W = (\theta_{\text{true},1},...,\theta_{\text{true},8}, \theta_{\text{true},10})^T 
\end{aligned}
\label{brownian_noshare}
\end{equation}

~~~~We apply MOBOLFI with the same experiment setup as in Section B.1 of this Appendix. The MOBOLFI and BOLFI approximate posteriors are presented in Figure \ref{brownian_nosharegroup}. We focus on the parameters $\theta_9, \theta_{10}$, which are parameters influencing the distribution of only one of the data soruces. In plot \ref{fig:MOBOLFI vs BOLFI noshare}, the MOBOLFI approximate posterior, leveraging correlated noise between data sources, still outperforms BOLFI in inference of $\theta_9, \theta_{10}$. In plot \ref{fig:MOBOLFI vs BOLFI X noshare}, for $\theta_{10}$ not depending on $X$, the MOBOLFI approximate posterior conditional on of $X$ does not obtain smaller variance than the BOLFI approximate posterior. That is not surprising to see, since information from $X$ is viewed as useless and redundant for inference of $\theta_{10}$. On the other hand, MOBOLFI still performs better than BOLFI in the inference of $\theta_9$. Plot \ref{fig:MOBOLFI vs BOLFI W noshare} shows that approximate MOBOLFI posterior conditional on $W$ only does not obtain smaller variance than the BOLFI approximate posterior, for $\theta_9$ not depending on $W$.

\begin{figure}[h]
\centering
\begin{subfigure}{0.6\linewidth}
        \includegraphics[width=1.0\linewidth,height=4cm]{figures/brownian_noshare.pdf}
        \caption{MOBOLFI vs BOLFI}
        \label{fig:MOBOLFI vs BOLFI noshare}
\end{subfigure}
\begin{subfigure}{0.6\linewidth}
        \includegraphics[width=1.0\linewidth,height=4cm]{figures/brownian_noshareX.pdf}
        \caption{Information in X}
        \label{fig:MOBOLFI vs BOLFI X noshare}
\end{subfigure}
\begin{subfigure}{0.6\linewidth}
        \includegraphics[width=1.0\linewidth,height=4cm]{figures/brownian_noshareW.pdf}
        \caption{Information in W}
        \label{fig:MOBOLFI vs BOLFI W noshare}
\end{subfigure}
\caption{Approximate posterior for the updated toy example with two parameters $\theta_9, \theta_{10}$ solely depending on two data sources $X,W$, respectively. The row \ref{fig:MOBOLFI vs BOLFI noshare} shows the MOBOLFI and BOLFI approximate posteriors under the updated simulator, colored by blue/orange. The green curve is the real posterior, the red curve is the prior and the dash red line is the true value of $\theta_9, \theta_{10}$. In row \ref{fig:MOBOLFI vs BOLFI X noshare}/\ref{fig:MOBOLFI vs BOLFI W noshare} the MOBOLFI and BOLFI approximate posteriors calculated by marginal likelihood of $X/W$ are given by blue/orange lines. The green curve is the real posterior of $X/W$ respectively. }
\label{brownian_nosharegroup}
\end{figure}

\section{Experimental setup and extra findings: MLBA example}
\subsection{MLBA closed form likelihood function}
~~~~Consider decision-making for a single individual first. For a drift rate $v_a \sim N(d_a,s^2)$, $a=1,...,M$, the probability density function (pdf) of the time $t$ taken for the accumulator $a$ to reach the threshold $\chi$ is (see \citet{brown_simplest_2008}, Appendix A for a derivation):
{\small
\begin{equation}\label{eq:f_a}
    f_a (t)  = \frac{1}{\mathcal{A}}\left[-d_a\Phi\left(\frac{\chi- \mathcal{A}-td_a}{ts}\right)+s\phi\left(\frac{\chi- \mathcal{A}-td_a}{ts}\right)+d_a\Phi\left(\frac{\chi-td_a}{ts}\right)-s\phi\left(\frac{\chi-td_a}{ts}\right)\right],
\end{equation}} where $\phi(.)$ and $\Phi(.)$ are the pdf and cdf of the standard normal distribution.

~~~~Its associated cumulative density function (cdf) is:
\begin{equation}
\begin{split}\label{eq:F_a}
    F_a (t) &= 1+ \frac{\chi- \mathcal{A}-td_a}{\mathcal{A}}\Phi\left(\frac{\chi-\mathcal{A}-td_a}{ts}\right)-\frac{\chi-td_a}{\mathcal{A}}\Phi\left(\frac{\chi-td_a}{ts}\right)\\
     &+\frac{ts}{\mathcal{A}}\phi\left(\frac{\chi-\mathcal{A}-td_a}{ts}\right)-\frac{ts}{\mathcal{A}}\phi\left(\frac{\chi-td_a}{ts}\right).
\end{split}
\end{equation}

~~~~The joint pdf of $CH = a$ and $RT = t+\tau_0$ is
\begin{equation}
    \text{MLBA}_{joint}(CH = a, RT = t+\tau_0) = f_a(t)\Pi_{b\neq a} (1-F_b(t)),
    \label{eq:mlbajointllk}
\end{equation}
and
the marginal pdf of $CH = a$ is
\begin{equation}
    \text{MLBA}_{choice}(CH = i) =\int_0^\infty f_a(t)\underset{b\in \mathcal{C},b\neq a}{\Pi}(1-F_b(t))dt,
\end{equation} where $\mathcal{C}$ is the choice set.

~~~~We follow the adjustment suggested by \citet{terry_generalising_2015}, where the drift rate follows a truncated normal distribution: $v_a\sim TN(d_a,s^2,0,\infty)$, with 0 and $\infty$ as lower and upper bounds of the support. This distribution helps correct the original pdf and cdf in \eqref{eq:f_a} and \eqref{eq:F_a}. Moreover, the truncated normal distribution resulted in superior performance in empirical experiments conducted by \citet{hancock_accumulation_2021}. Therefore, both $f_a(t)$ and $F_a(t)$ are additionally divided by a factor $\Phi(\frac{d_a}{s})$ in \eqref{eq:f_a} and \eqref{eq:F_a}.

\subsection{Experimental setup - MLBA}
\begin{sloppypar}
~~~~When working on synthetic data, we firstly construct a training dataset $\{\theta^{(i)},(RT^{(i)},CH^{(i)})\}_{i=1}^{100}$ of 100 observations. \end{sloppypar}
~~~~In BOLFI/MOBOLFI, a univariate/bivariate GP is trained with 1000 iterations. Likelihood approximations for BOLFI/MOBOLFI use tolerance $t$ given by the 1\% quantile of the training discrepancies, and samples were generated from the approximate
posterior distributions using De-MCMC. These samples are compared to samples from the posterior with
the closed form likelihood \ref{eq:mlbajointllk} above. 

~~~~For MOBOLFI, the objective function optimized is 
$$D(\theta)=(E(\Delta_1(RT,RT^o)),E(\Delta_2(CH,CH^o))^\top,$$
where $RT,CH$ are synthetic response time and choice data respectively, $RT^o$ and $CH^o$ are the corresponding
observed data, $\Delta_1(RT,RT^o)$ is
the discrepancy for the response time data and $\Delta_2(CH,CH^o)$ is the discrepancy
for the choice data.  The definition of the discrepancies is given in the main text. In BOLFI, the training objective is $w\cdot\Delta_1(RT,RT^o)+\Delta_2(CH,CH^o)$, where $w=0.7$. In MOBOLFI, the training objective is the 2-dimensional vector $(w\cdot\Delta_1(RT,RT^o), \Delta_2(CH,CH^o))$. In BOLFI, the acquisition function is defined to be the Lower Confidence bound with variance weight $\eta=0.1$ defined in \eqref{lcbsup}. \begin{sloppypar} ~~In MOBOLFI the Bayesian Optimization acquisition function is chosen to be the qNEHVI (see \url{https://botorch.org/api/acquisition.html#botorch.acquisition.multi_objective.monte_carlo.qNoisyExpectedHypervolumeImprovement} for details). Both MOBOLFI and BOLFI surrogate models are trained with 1000 BO acquisitions. \end{sloppypar}

~~~~For faster convergence, we set different hyperparameters for running De-MCMC to sample from MOBOLFI/BOLFI and the posterior with closed form likelihood. When sampling from the posterior for closed form likelihood, we set 9 chains, sample size 20000, burn-in size 18000 and migration rate 0.5. When sampling from BOLFI/MOBOLFI approximate posteriors, we set 9 chains, sample size 16000 and burn-in 13000. The coefficient $\gamma$ in the De-MCMC algorithm is set to be the fixed constant $2.38/\sqrt{2\cdot n_{\theta}}$, where $n_{\theta}=6$ is the number of parameters we infer.

~~~~Similar to the toy example, we set some rejection criteria in sampling the initial training set for BO. Points with values of any one of the discrepancies greater than a threshold, i.e. the $99\%$-quantile obtained from simulated discrepancies that are drawn without any filtering criterion, are discarded.  In MLBA simulation we observe that there are some prior samples $\theta^{(i)}$ that simulate $RT^{(i)}=\{RT^{(i)}_1,\cdots,RT^{(i)}_{320}\}$ where $RT^{(i)}_j \approx \overline{RT^o}, \forall j \in \{1,\cdots,320\}$, where $\overline{RT^o}$ denotes the sample mean of $RT^o$. Ideally the prior should be constrained to avoid such degenerate
regions of the parameter space, but these regions are not easily characterized
analytically.  Hence we also filter out $\theta^{(i)}$ such that 
$\text{Var}(RT^{(i)})<\text{Var}(RT^{o})\cdot 0.7$.

\begin{figure}[h]
\centering
\begin{subfigure}{0.38\linewidth}
         \includegraphics[width=1.0\linewidth,height=4cm]{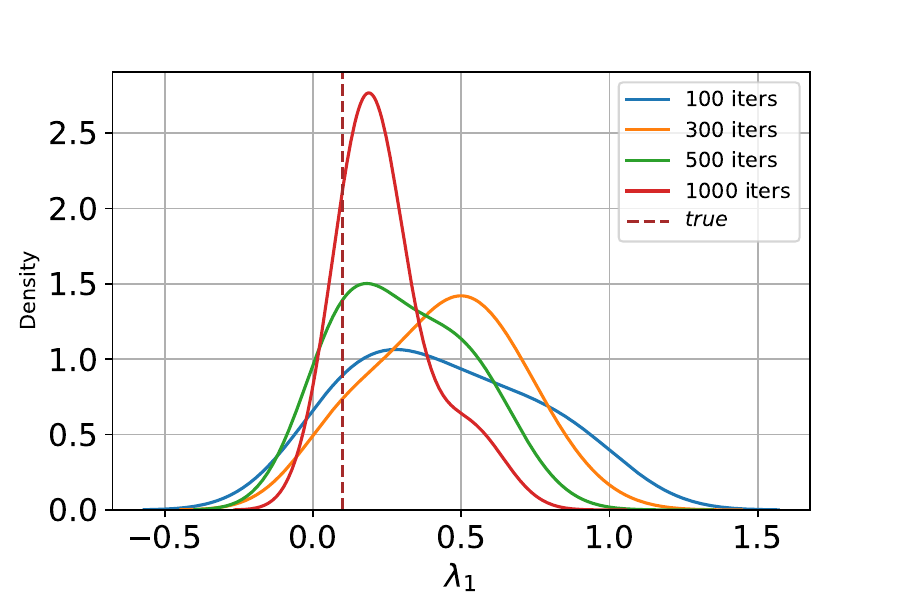}
         \label{fig: MLBA MOBOLFI vs iter lambda1}
\end{subfigure}
~
\begin{subfigure}{0.38\linewidth}
         \includegraphics[width=1.0\linewidth,height=4cm]{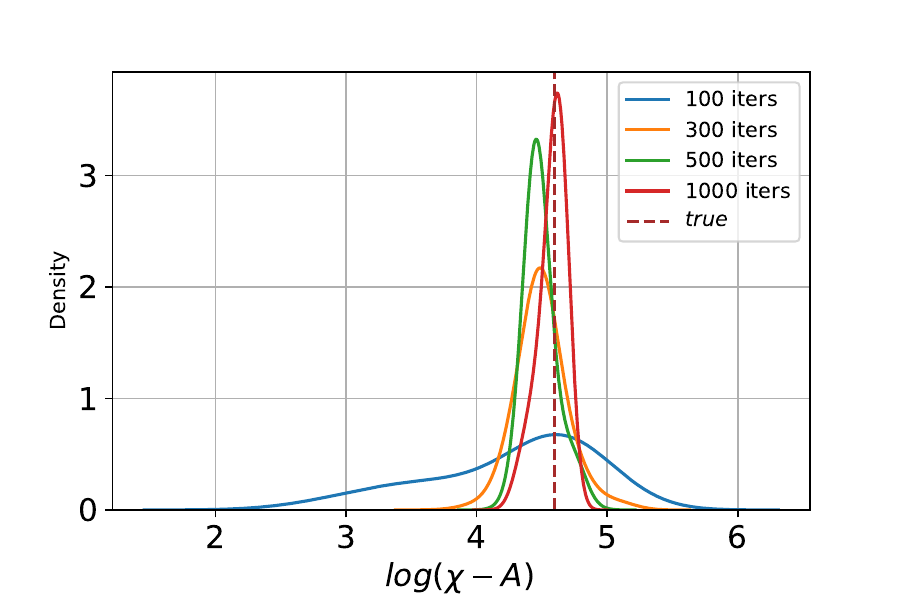}
         \label{fig: MLBA MOBOLFI vs iter threshold}
\end{subfigure}

\begin{subfigure}{0.38\linewidth}
         \includegraphics[width=1.0\linewidth,height=4cm]{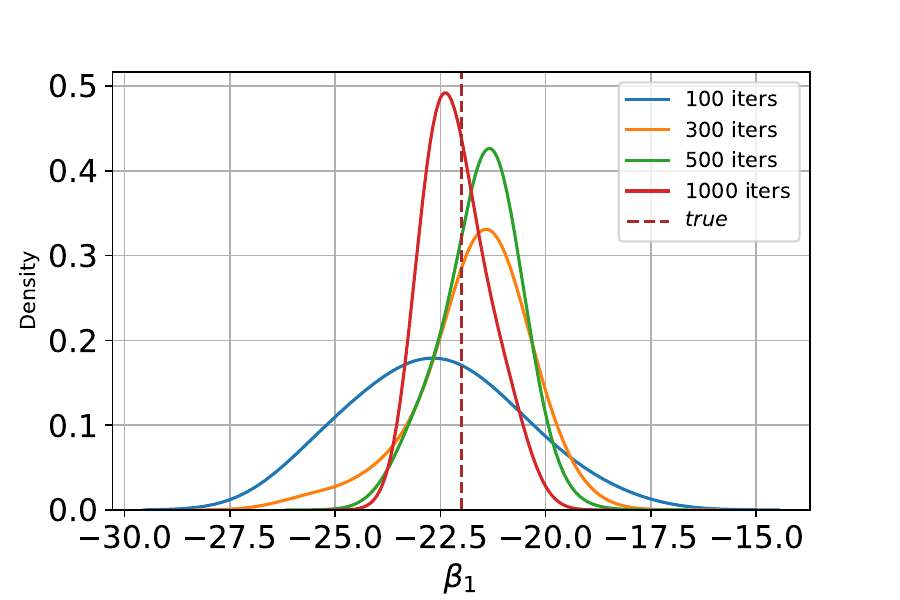}
         \label{fig: MLBA MOBOLFI vs iter beta1}
\end{subfigure}
~
\begin{subfigure}{0.38\linewidth}
         \includegraphics[width=1.0\linewidth,height=4cm]{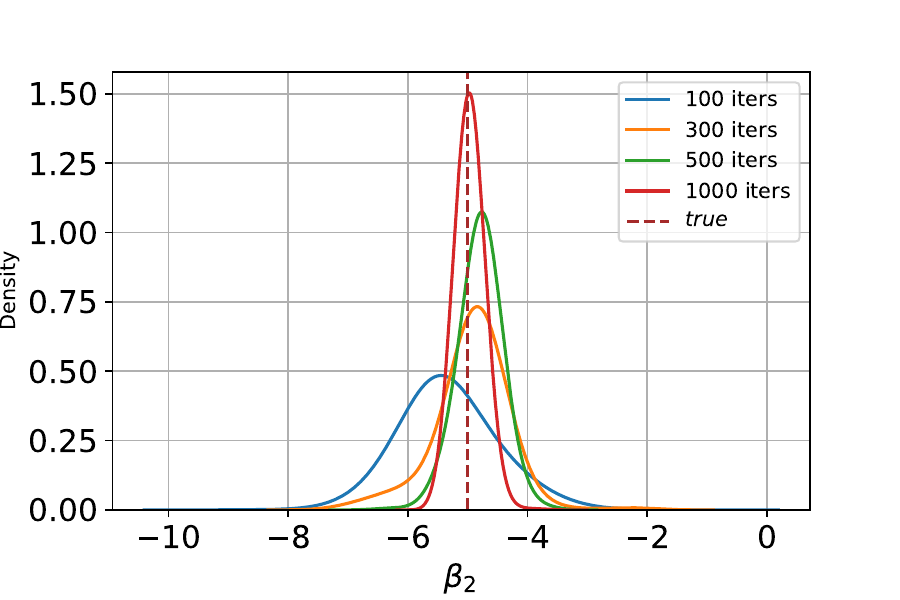}
         \label{fig: MLBA MOBOLFI vs iter beta2}
\end{subfigure}

\begin{subfigure}{0.38\linewidth}
         \includegraphics[width=1.0\linewidth,height=4cm]{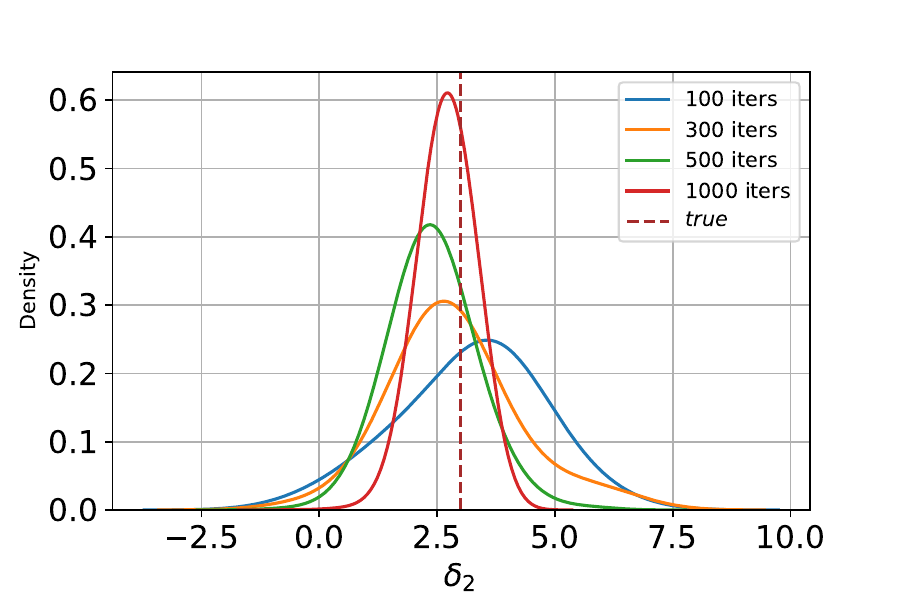}
         \label{fig: MLBA MOBOLFI vs iter delta2}
\end{subfigure}
~
\begin{subfigure}{0.38\linewidth}
         \includegraphics[width=1.0\linewidth,height=4cm]{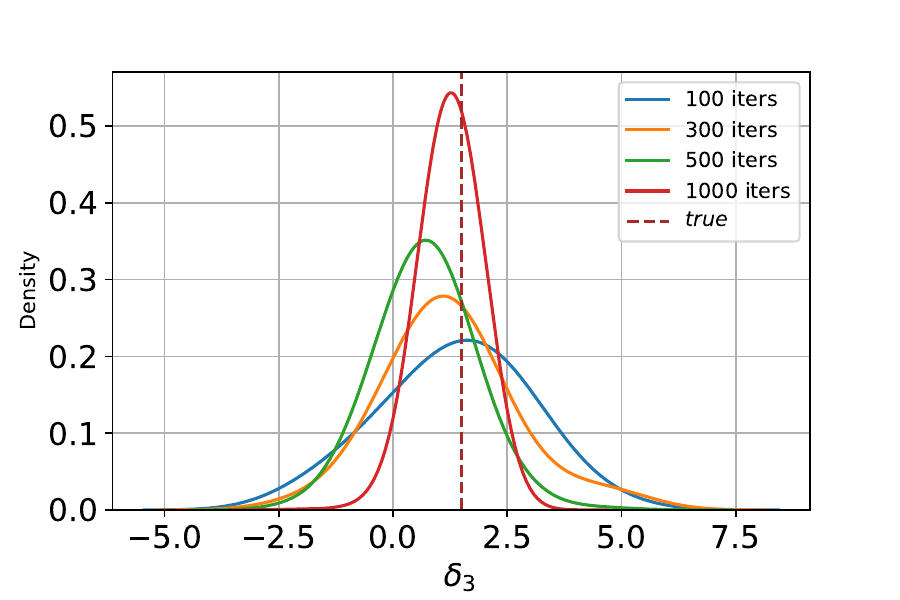}
         \label{fig: MLBA MOBOLFI vs iter delta3}
\end{subfigure}
\caption{Approximate posteriors for MLBA example given different numbers of iterations for the BO algorithm. Each plot shows approximate marginal posteriors for one parameter of interest. The densities shown are kernel density estimates obtained from MCMC samples. The dashed red line shows the location of the corresponding true parameter.}
\label{MLBA_MOBOLFIvsiter}
\end{figure}

\subsection{Extra findings - MLBA example (Synthetic data)}
~~~~Similar to the toy example, we build MOBOLFI approximate posteriors given different number of iterations for the BO algorithm. Figure \ref{MLBA_MOBOLFIvsiter} compares kernel 
density estimates obtained from samples from the approximate posteriors. Due to the complexity of the MLBA simulator, we need many more iterations in the BO algorithm than for the toy example 
to obtain good approximations.
The Figure shows that 100 iterations in the BO algorithm
is too little, but the approximation to the marginal posteriors has mostly stabilized after 300 iterations for most parameters.    

\begin{figure}[h]
\centering
\begin{subfigure}{0.4\linewidth}
         \includegraphics[width=1.0\linewidth,height=4cm]{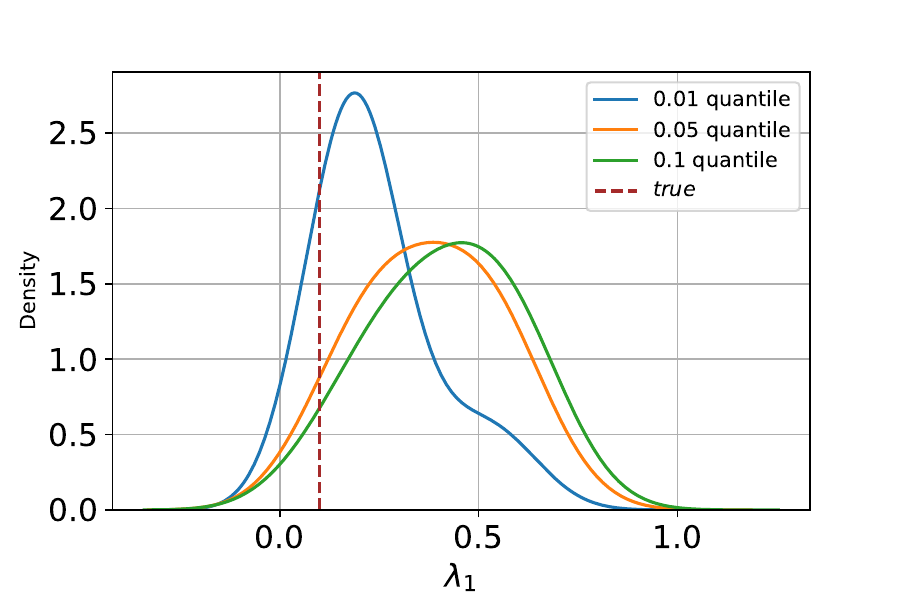}
         \label{fig: MLBA MOBOLFI vs tol lambda1}
\end{subfigure}
\begin{subfigure}{0.4\linewidth}
         \includegraphics[width=1.0\linewidth,height=4cm]{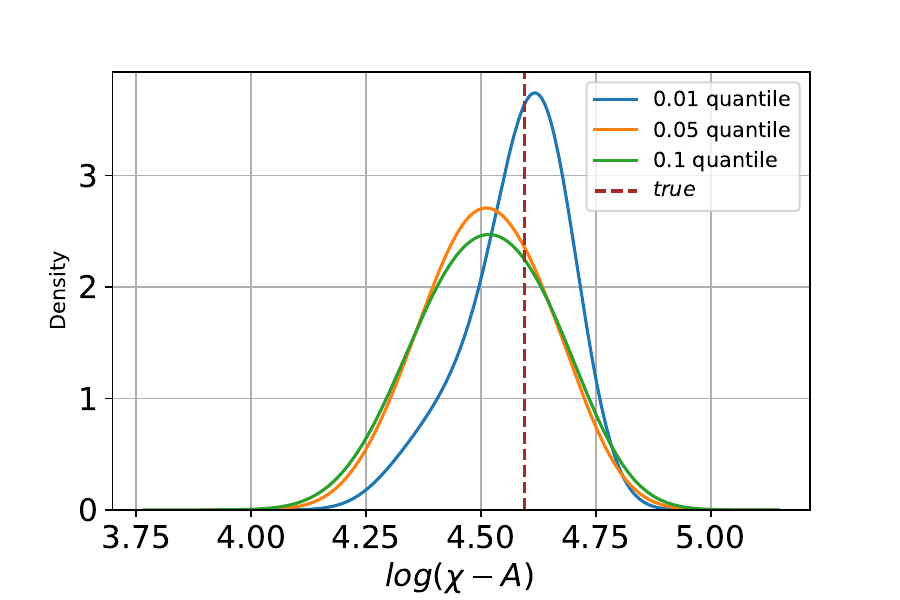}
         \label{fig: MLBA MOBOLFI vs tol threshold}
\end{subfigure}

\begin{subfigure}{0.4\linewidth}
         \includegraphics[width=1.0\linewidth,height=4cm]{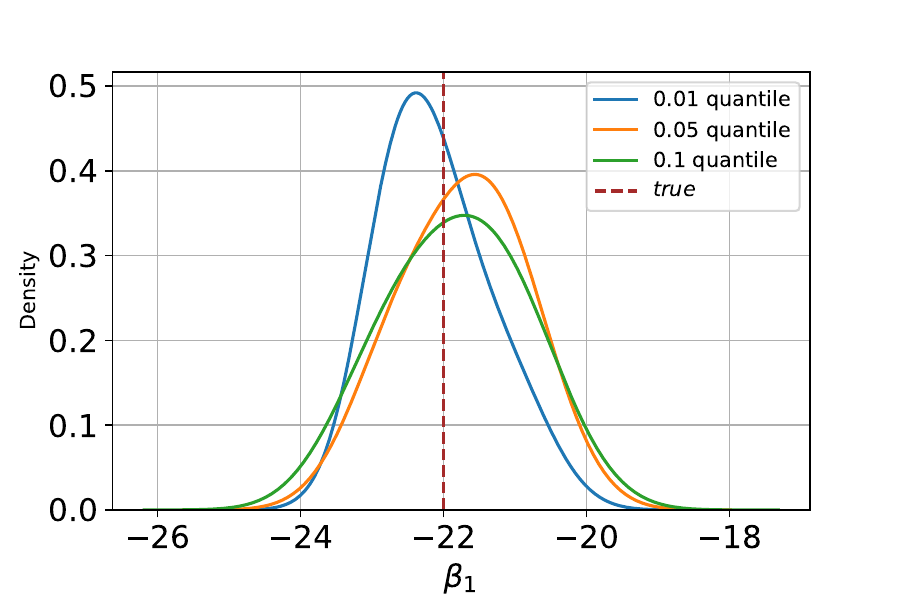}
         \label{fig: MLBA MOBOLFI vs tol beta1}
\end{subfigure}
\begin{subfigure}{0.4\linewidth}
         \includegraphics[width=1.0\linewidth,height=4cm]{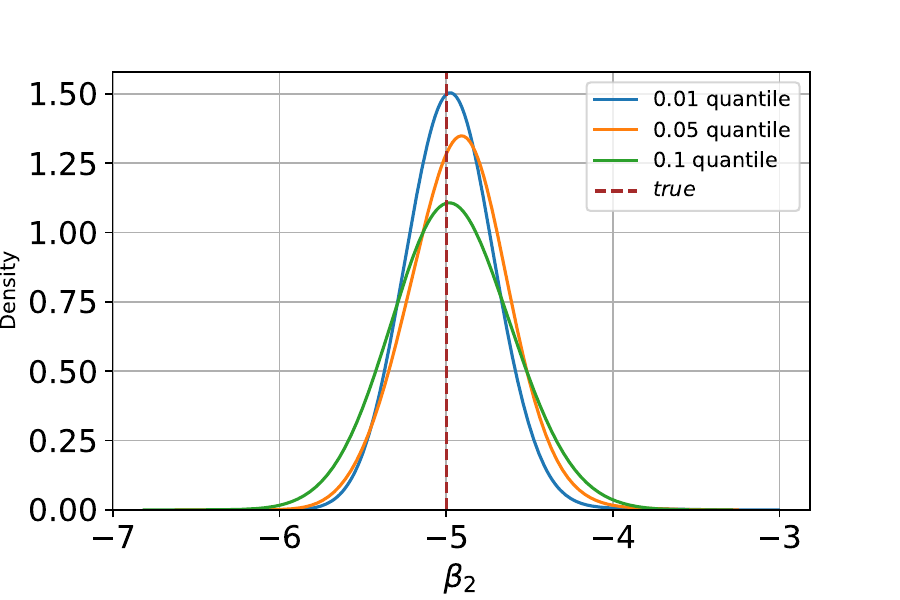}
         \label{fig: MLBA MOBOLFI vs tol beta2}
\end{subfigure}

\begin{subfigure}{0.4\linewidth}
         \includegraphics[width=1.0\linewidth,height=4cm]{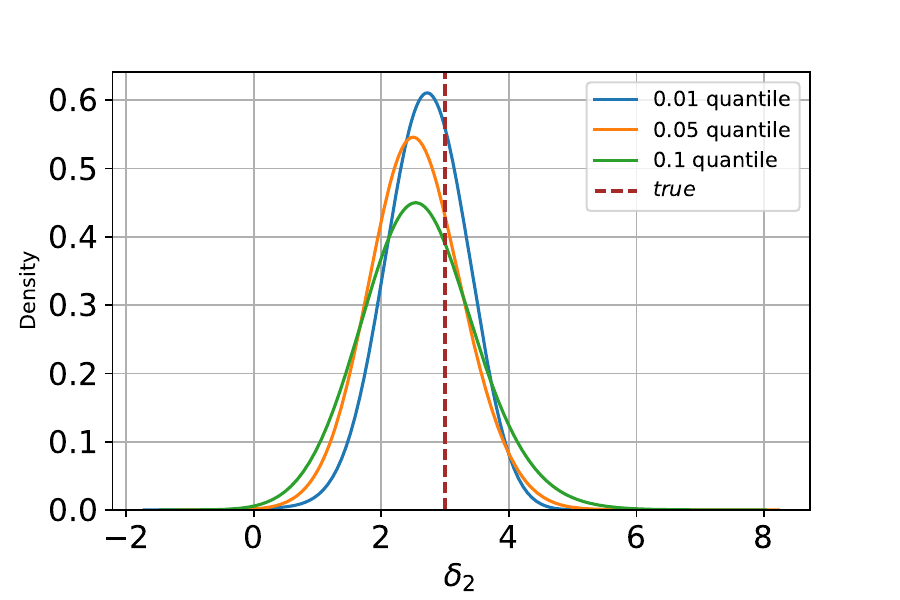}
         \label{fig: MLBA MOBOLFI vs tol delta2}
\end{subfigure}
\begin{subfigure}{0.4\linewidth}
         \includegraphics[width=1.0\linewidth,height=4cm]{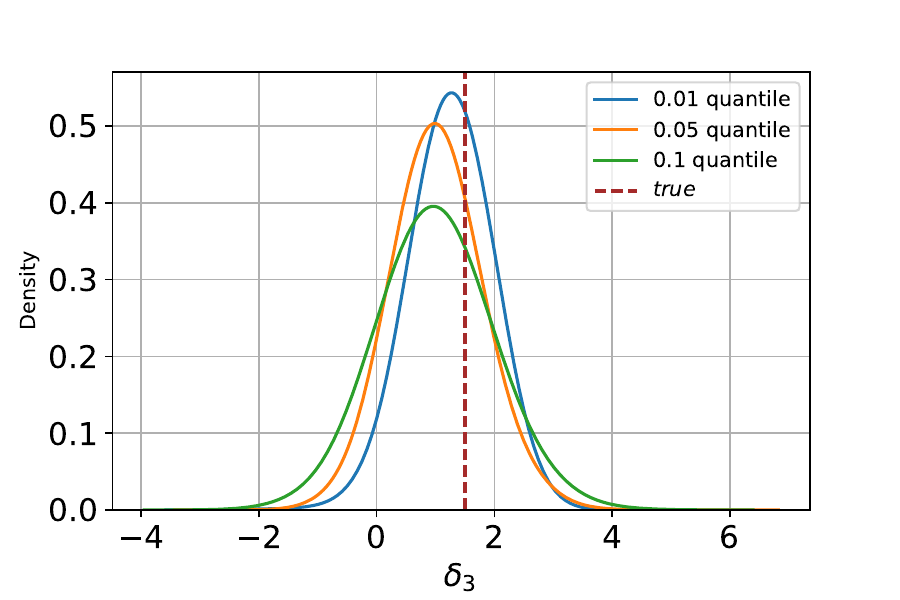}
         \label{fig: MLBA MOBOLFI vs tol delta3}
\end{subfigure}
\caption{Approximate posteriors of MLBA example given different levels of quantile of training dataset as the tolerance $t$. The blue/orange/green curve represents the MOBOLFI approximate posterior (with 1000 iterations training) using 10\%/5\%/1\% quantile of the training dataset as the tolerance $t$ respectively. The dashed red line shows the location of the true parameter.}
\label{MLBA_MOBOLFIvstol}
\end{figure}

~~~~We also investigate the affect of tolerance $t$ on MOBOLFI performance in MLBA. In Figure \ref{MLBA_MOBOLFIvstol}, we compare the MOBOLFI approximate posteriors for tolerances specified as 10\%/5\%/1\% quantiles of the training discrepancies. Given enough BO iterations, the 1\% quantile is the best choice. 
Although the performance of MOBOLFI is sensitive to the choice of $t$, setting the tolerance to a 1\% or 5\% quantile of the training data works well across many problems.

\begin{figure}[h]
\centering
\begin{subfigure}{0.4\linewidth}
         \includegraphics[width=1.0\linewidth,height=4cm]{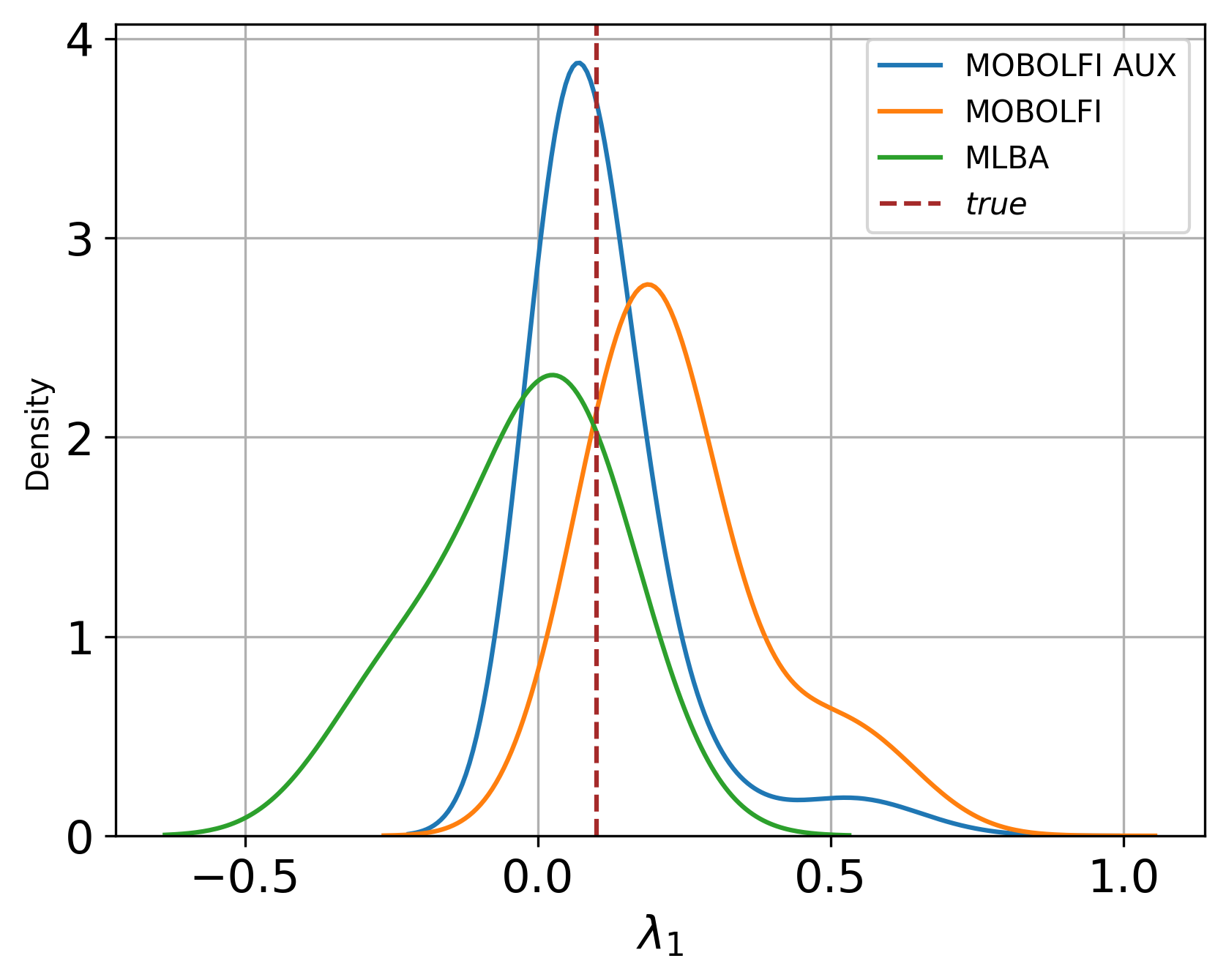}
         \label{fig: MLBA MOBOLFI MAD vs MLBA MOBOLFI AUX0}
\end{subfigure}
\begin{subfigure}{0.4\linewidth}
         \includegraphics[width=1.0\linewidth,height=4cm]{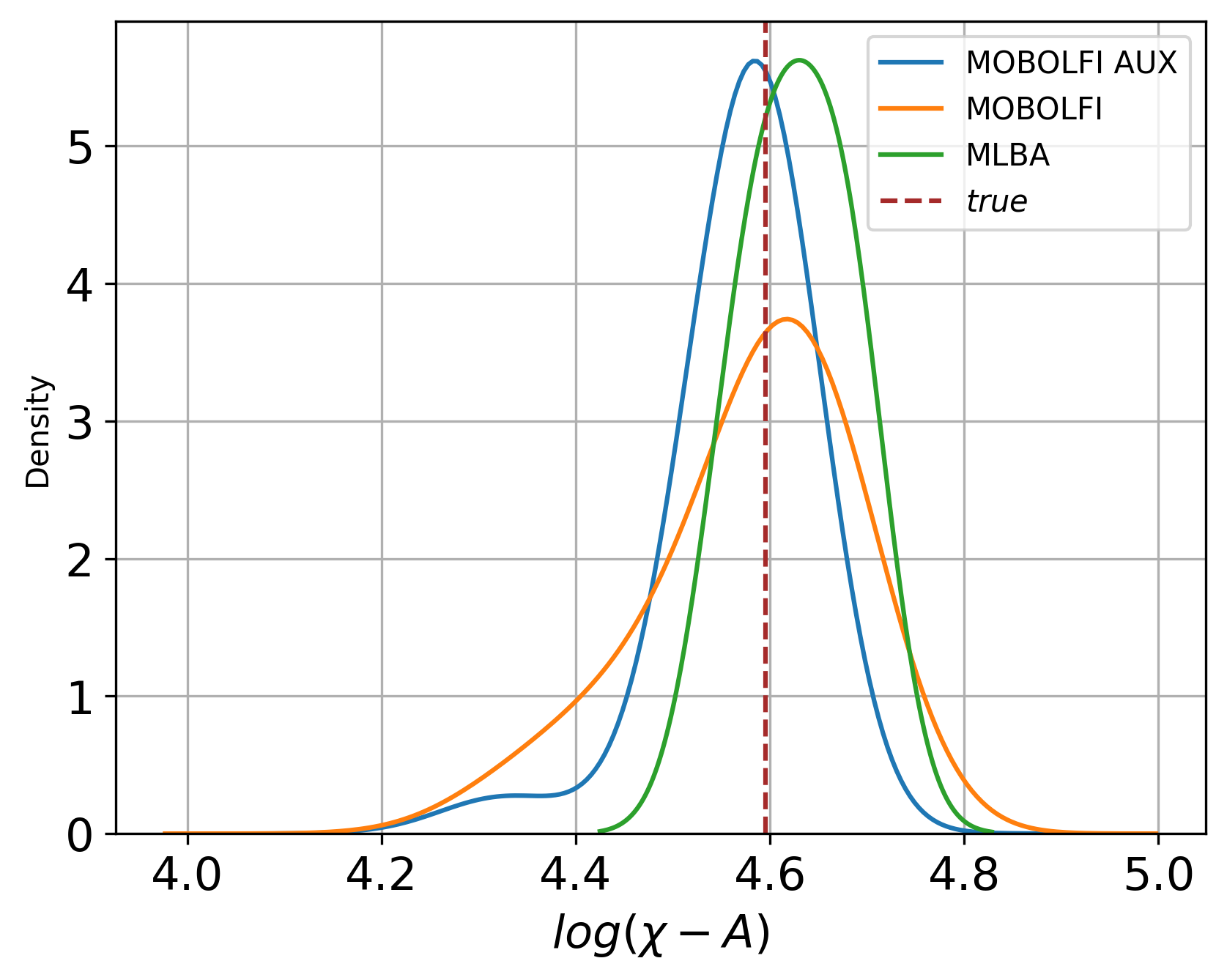}
         \label{fig: MLBA MOBOLFI MAD vs MLBA MOBOLFI AUX1}
\end{subfigure}

\begin{subfigure}{0.4\linewidth}
         \includegraphics[width=1.0\linewidth,height=4cm]{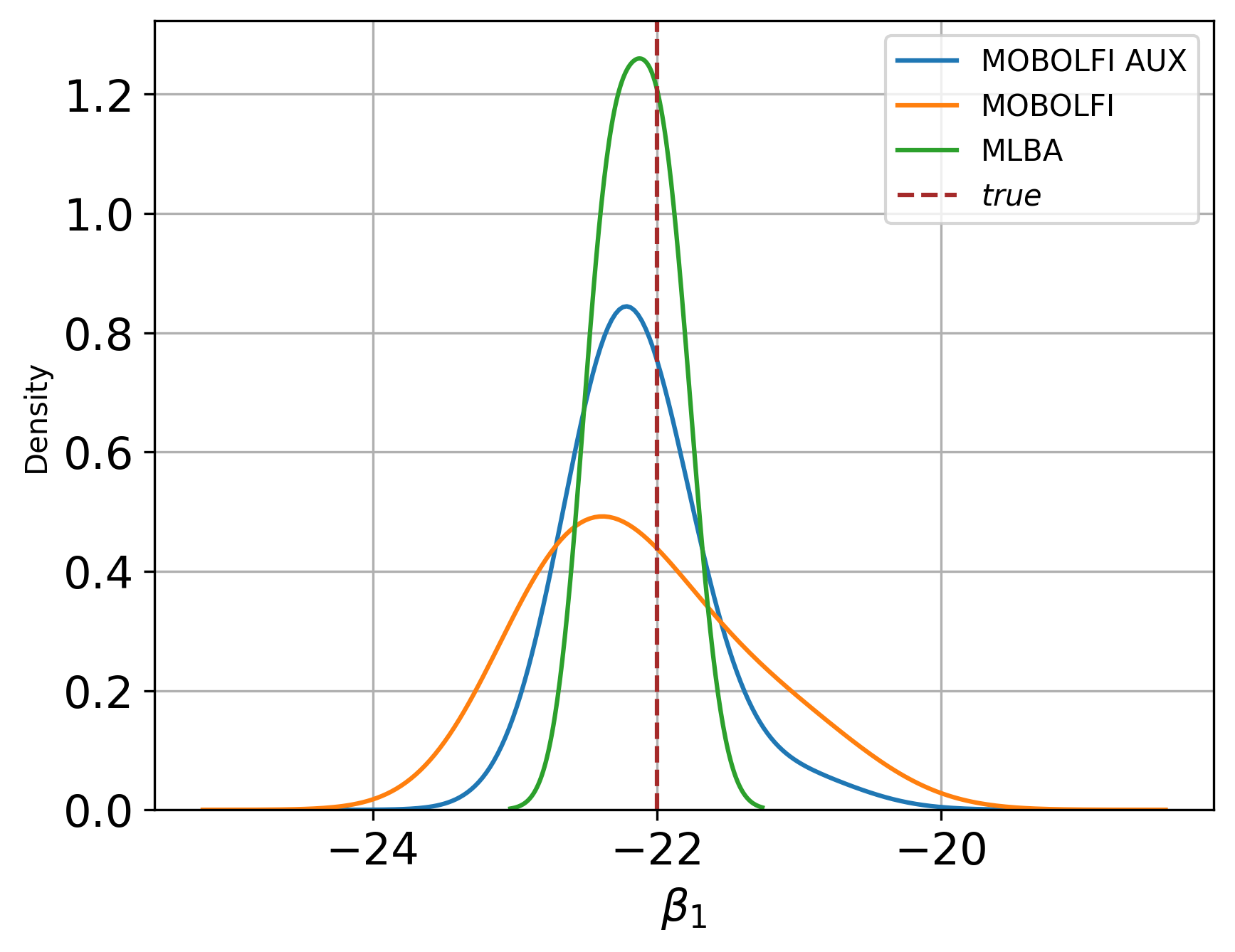}
         \label{fig: MLBA MOBOLFI MAD vs MLBA MOBOLFI AUX2}
\end{subfigure}
\begin{subfigure}{0.4\linewidth}
         \includegraphics[width=1.0\linewidth,height=4cm]{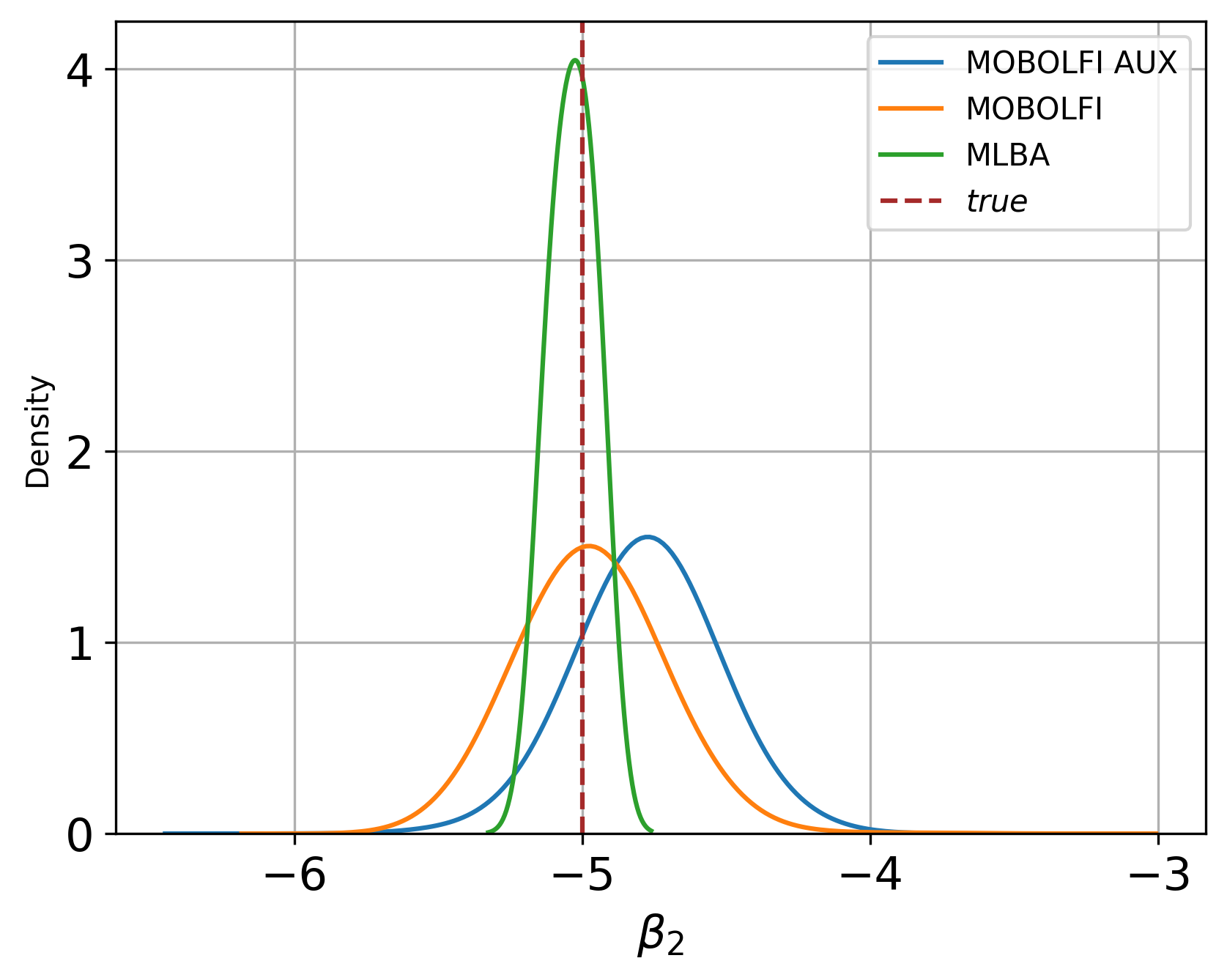}
         \label{fig: MLBA MOBOLFI MAD vs MLBA MOBOLFI AUX3}
\end{subfigure}

\begin{subfigure}{0.4\linewidth}
         \includegraphics[width=1.0\linewidth,height=4cm]{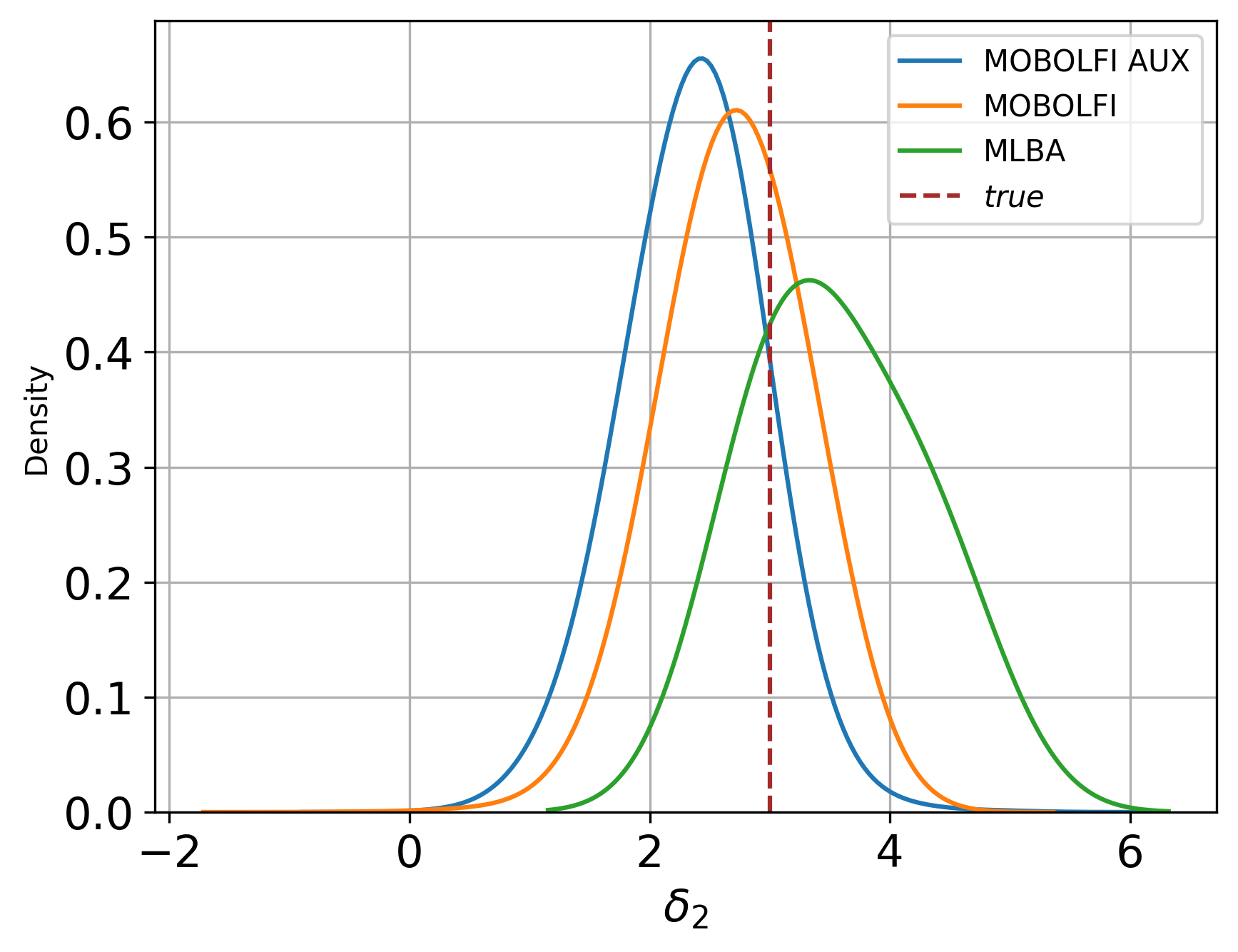}
         \label{fig: MLBA MOBOLFI MAD vs MLBA MOBOLFI AUX4}
\end{subfigure}
\begin{subfigure}{0.4\linewidth}
         \includegraphics[width=1.0\linewidth,height=4cm]{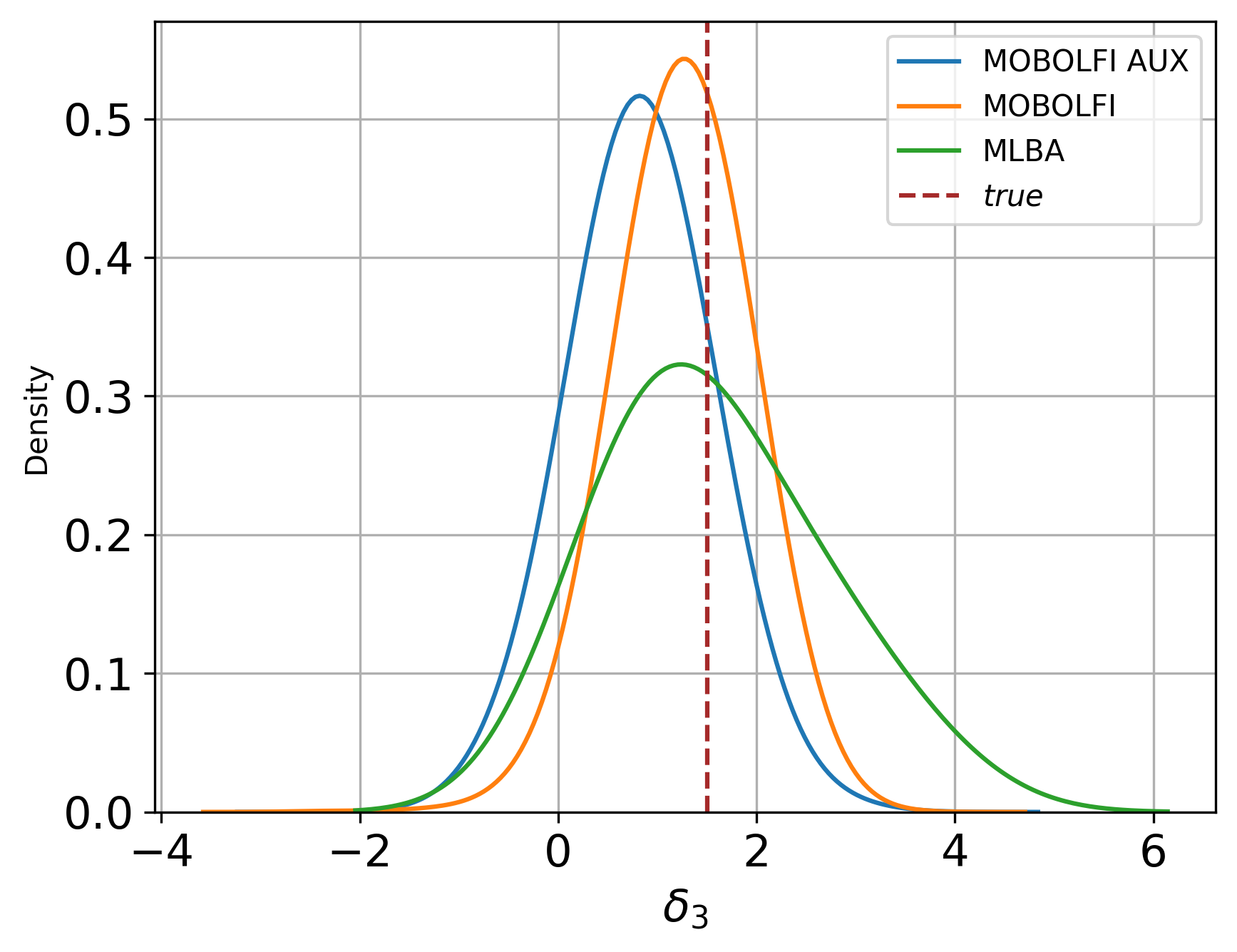}
         \label{fig: MLBA MOBOLFI MAD vs MLBA MOBOLFI AUX5}
\end{subfigure}
\caption{Approximate posteriors of MLBA example estimated by different data summary for response choice. The blue and orange curves represent the MOBOLFI approximate posterior (with 1000 iterations training) using the original distance (MOBOLFI) and score function of  an auxiliary model (MOBOLFI AUX) as the data summary, respectively. The red dash line is the value of true parameter $\theta^{\text{true}}$.}
\label{MLBA_MOBOLFIvsAUX}
\end{figure}

~~~~When a summary statistic is used in the definition of
the discrepancy, its choice is another factor that
affects the performance of MOBOLFI.  
For summary statistic based LFI methods, the choice
of summary is important, because good summaries can reduce the computational burden with little
loss of information. In Figure \ref{MLBA_MOBOLFIvsAUX}, we evaluate MOBOLFI approximate posteriors obtained using
two approaches.  The first approach uses the discrepancies
discussed in the main text.  The second approach 
changes the discrepancy used for the choice data, by
defining summary statistics using an auxiliary model
\citep[e.g., ][]{drovandi+pl15}.  
Specifically, we adopt the score vector evaluated at
the maximum likelihood estimate (MLE) for the observed data for a multinomial logit (MNL) model as the choice data discrepancy. 

~~~~To explain further, 
the MNL model is a discrete choice model based on random utility maximization theory (McFadden, 1974).  In the MNL, the choice probability for the alternative $a$ is  
\begin{equation}\label{eq:P_a}
P_a=\frac{\exp \left(V_a\right)}{\sum_b \exp \left(V_a\right)}
\end{equation}
where $V_a$ is a systematic utility of the alternative $a$, represented by a set of attributes ${X}=(X_1,...,X_K)$ and corresponding parameters $\xi=(\xi_1,...,\xi_K)$, where $K$ is the total number of parameters. Since it has a closed-form likelihood function, the parameters can be estimated by maximum likelihood estimation. Write $p_A(CH;\xi)$ for
the likelihood for the MNL auxiliary model for choice
data $CH$ and parameter $\xi$.  The score function
is $S_A(CH;\xi)=\nabla_\xi \log p_A(CH:\xi)$. 
Writing $CH^o$ for the observed choice data, and 
$\hat{\xi}$ for the MLE for the observed choice data, 
we have $S_A(CH^0;\hat{\xi})\approx0$, and for simulated
choice data $CH$ we use as summary statistics
$S_A(CH;\hat{\xi})$ (i.e. we use the ($K \times 1$) score vector
for the data $CH$ evaluated at the MLE $\hat{\xi}$ for
the observed choice data as the vector of summary
statistics). 

~~~~Each component of the score vector has a different scale from the others because the parameters $\xi$ correspond to different scales of attributes (e.g., driving range and purchase price). Therefore, the score vector $S_A(CH;\hat{\xi})$ should be converted into a scalar value to be used as choice data discrepancy $\Delta_2(CH,CH^o)$.  We scale parameter-specific components of the score vector to put all components on a similar scale as follows. Sample from the prior $\theta^{(i)} \sim \pi(\theta)$, $i=1,\ldots,n$. For each $i$, we evaluate the corresponding score vector $S^{(i)} = S_A(CH^{(i)};\hat{\xi})$, where $CH^{(i)}$ is the choice data simulation given $\theta^{(i)}$. Then, we apply Algorithm \ref{alg:scaling} to obtain the scaling weight $V_1^{-1}$ of the parameter-specific K-components of the score vectors with sample size $n=100$ and target function $S_A(\cdot;\hat{\xi})$. Finally, the choice data discrepancy is defined as 
$\Delta_2(CH,CH^o)=V_1^{-1}S_A(CH;\hat{\xi})$.

~~~~To implement the MOBOLFI approach, we need to calculate the training objective by calculating two data source-specific discrepancies. 
The response data discrepancy is the one used in the main text, and the joint discrepancy is $(\Delta_1(RT,RT^o),V_2 ^{-1}\Delta_2(CH,CH^o))$ where $V_2 ^{-1}$ is obtained by Algorithm \ref{alg:scaling}. 

~~~~In our experiment, we generated $n=100$ MLBA datasets for $K=5$ parameters of the MNL auxiliary model and obtained 
$V_1 ^{-1} =(128.4,  226.2, 1656.6 ,  519.9, 2269.1)$ and $V_2 ^{-1} =0.0007$, respectively.

~~~~In Figure \ref{MLBA_MOBOLFIvsAUX}, we compare MOBOLFI using the choice discrepancy from the main text with the auxiliary model choice discrepancy (MOBOLFI AUX). The closed-form likelihood (MLBA) is used as a benchmark model. 
Given enough training, the MOBOLFI and MOBOLFI AUX show different advantages in approximating the posterior with regard to point estimate and posterior variance. The lower posterior variance indicates more robust estimates, while the point estimate closer to the true parameter indicates accuracy for decision-making in practice (e.g., an estimate of electric vehicle adoption rate in this MLBA case). 
Since the MLBA parameters are behaviorally related to both reaction time and choice decision, the MOBOLFI AUX affects all parameters. For $\lambda_1$, $\beta_1$, and $\log(\chi- \mathcal{A})$, the MOBOLFI AUX outperforms the MOBOLFI by providing a slightly better point estimate while reducing posterior variance in parameter inference, indicating the validity of adopting summary statistics based on an auxiliary model. For  $\beta_2$, $\delta_2$, and $\delta_3$, the approximate posterior distributions from the MOBOLFI are superior to those from MOBOLFI AUX with similar posterior variance but much better point estimates. This result suggests that the performance of MOBOLFI and MOBOLFI AUX are comparable.


\subsubsection{Misspecified MLBA}

\begin{figure}[h]
\centering
\begin{subfigure}{0.4\linewidth}
         \includegraphics[width=1.0\linewidth,height=4cm]{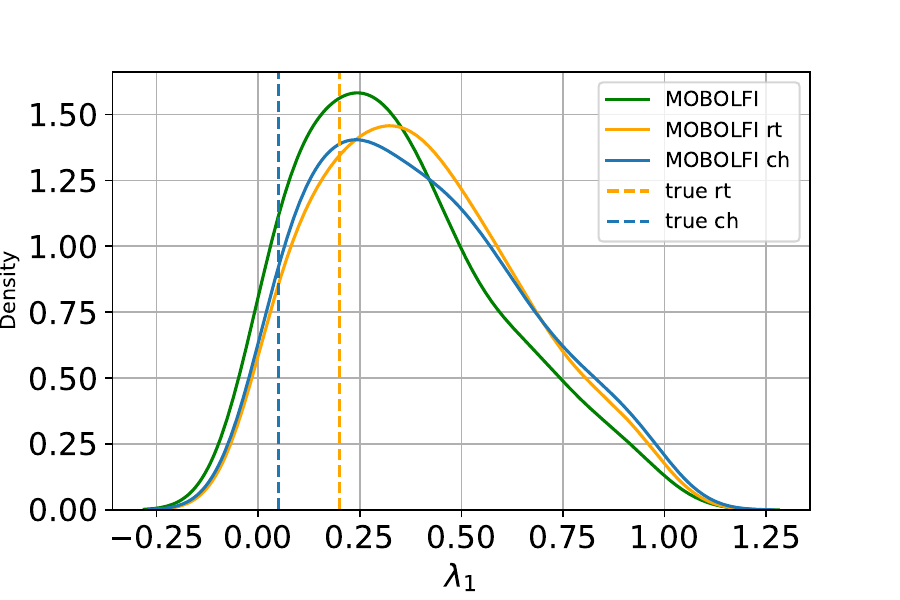}
         \label{fig: MLBA MOBOLFI mis lambda1}
\end{subfigure}
\begin{subfigure}{0.4\linewidth}
         \includegraphics[width=1.0\linewidth,height=4cm]{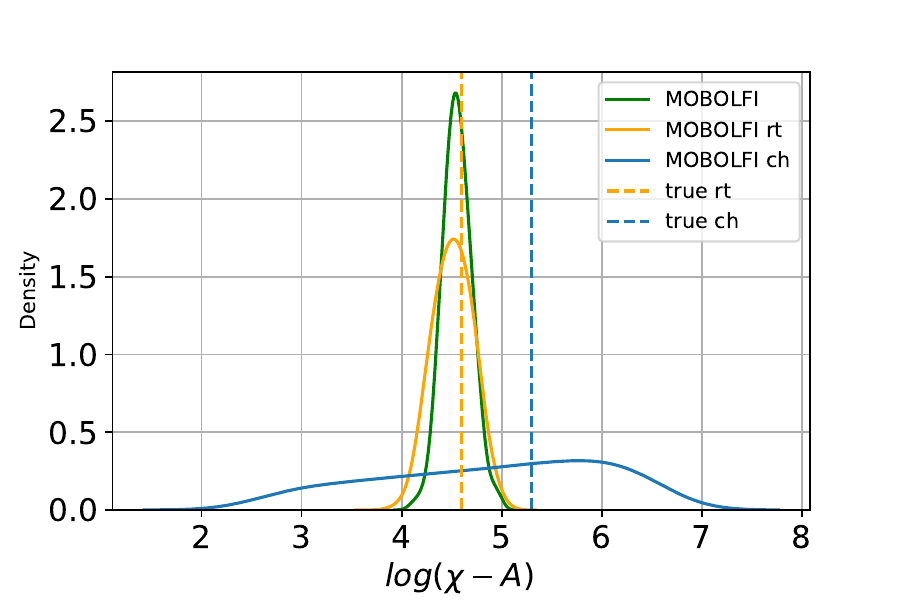}
         \label{fig: MLBA MOBOLFI mis threshold}
\end{subfigure}

\begin{subfigure}{0.4\linewidth}
         \includegraphics[width=1.0\linewidth,height=4cm]{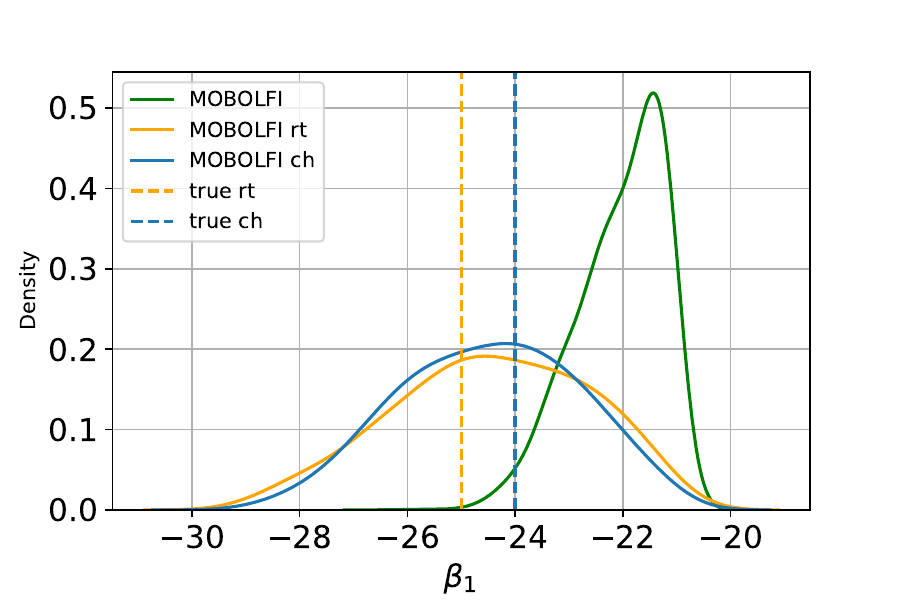}
         \label{fig: MLBA MOBOLFI mis beta1}
\end{subfigure}
\begin{subfigure}{0.4\linewidth}
         \includegraphics[width=1.0\linewidth,height=4cm]{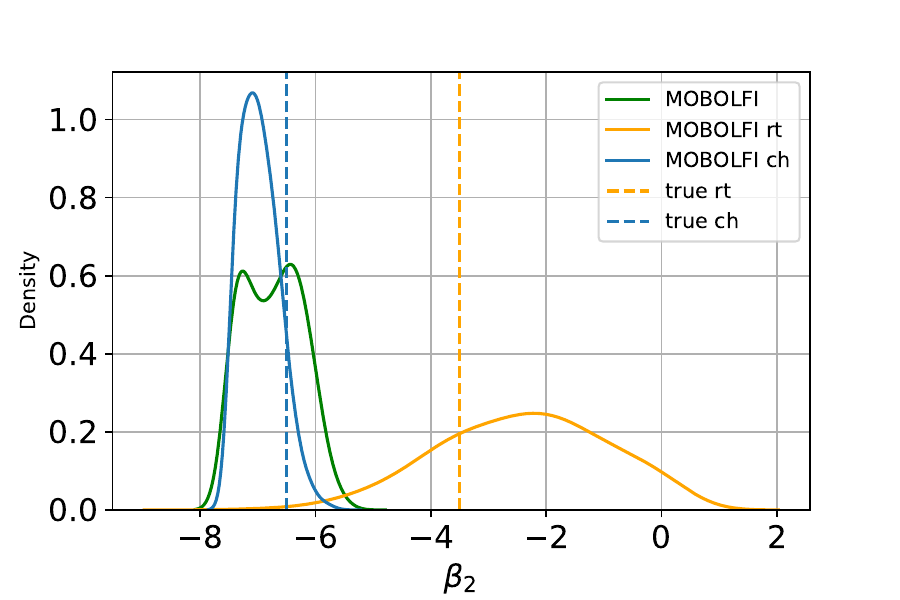}
         \label{fig: MLBA MOBOLFI mis beta2}
\end{subfigure}

\begin{subfigure}{0.4\linewidth}
         \includegraphics[width=1.0\linewidth,height=4cm]{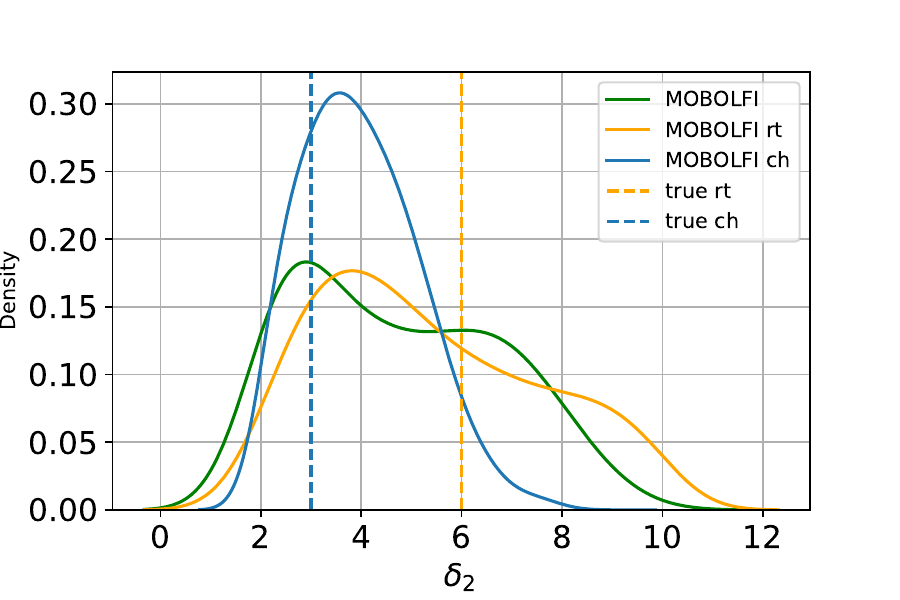}
         \label{fig: MLBA MOBOLFI mis delta2}
\end{subfigure}
\begin{subfigure}{0.4\linewidth}
         \includegraphics[width=1.0\linewidth,height=4cm]{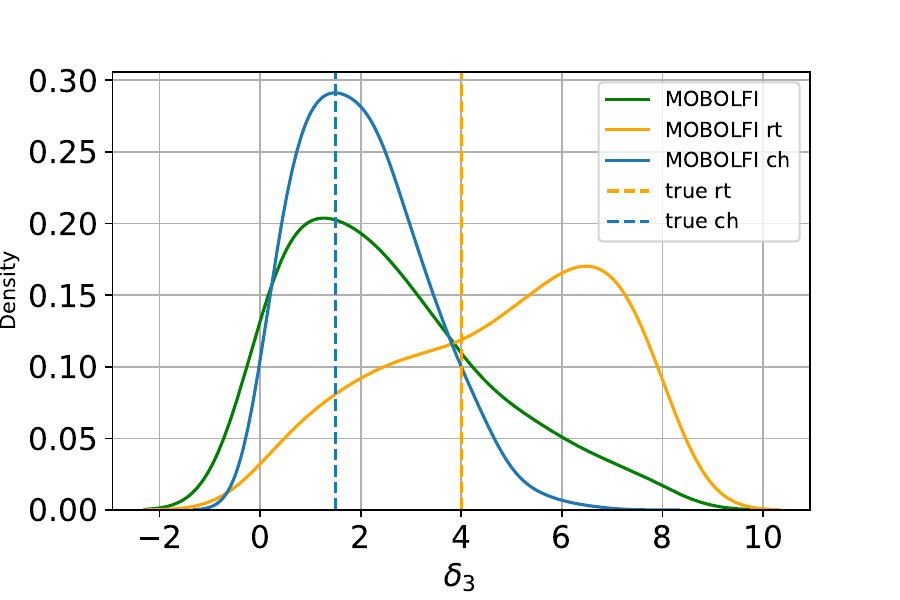}
         \label{fig: MLBA MOBOLFI mis delta3}
\end{subfigure}

\caption{Approximate marginal posteriors for MLBA example under misspecification. The green curve represents the MOBOLFI approximate posterior calculated using the joint likelihood of two data sources. The orange/blue curve are the MOBOLFI approximate posteriors calculated by the approximate likelihood of $RT$/$CH$ data only. The orange/blue dash line shows the location of the true parameters in
$\theta^{\text{true}}_{RT}$ and $\theta^{\text{true}}_{CH}$.}
\label{MLBA_MOBOLFImis}
\end{figure}

~~~~MOBOLFI approximate posteriors of the parameters 
for the misspecified MLBA scenario of Section 4.2.3 are shown in Figure \ref{MLBA_MOBOLFImis}. For $\lambda_1$, both data sources contribute to the inference, and the posterior calculated using single data sources are similar to the posterior calculated by the joint likelihood. For $\log(\chi-\mathcal{A})$, the definition of $\chi$ determines that inference of it mainly depends on the response time data $RT$. It is unsurprising to see that the posterior calculated by the joint likelihood has similar location to the posterior calculated conditional on $RT$ only. The posterior calculated conditional on $CH$ only provides much less information, having a large variance. For $\beta_1$, 
the true values for the different data sources lie in
the tail of the posterior calculated using the joint likelihood, while the posterior calculated conditioning on only one data source are similar and consistent with the true values from both data sources. As discussed above, there is a set of local optimum points that could obtain similar performance in evaluating the discrepancy to 
$\theta^{\text{true}}_{RT}$ and $\theta^{\text{true}}_{CH}$.
The marginal approximate posterior of $\beta_1$ calculated by the joint likelihood puts weight on points other than 
$\theta^{\text{true}}_{RT}$ and $\theta^{\text{true}}_{CH}$ consistent with
the data.   
For $\beta_2$, we found the posterior calculated using the joint likelihood is closer to the posterior calculated conditioning only on the $CH$ data but with variance larger.  This differs from the behaviour for $\beta_1$. We believe this is because given the fixed $\beta_3 = -6$, the values of $\beta_2$ in $\theta^{\text{true}}_{RT}$ and $\theta^{\text{true}}_{CH}$ are close to the value of $\beta_3$ but away from $\beta_1$. This suggests that in the synthetic data example, attribute 1 is more important in evaluating evidence of alternatives which directly affects the response time.  It is not surprising then to see that $\beta_2$ measures influence of the less important attribute 2 and appears to be more sensitive to the choice data $CH$ than $\beta_1$. For $\delta_2$ and $\delta_3$, the alternative specific constant is added into the drift rate mean calculation directly, unrelated to the attribute pairwise-comparison, which is the main source of variation of drift rate mean. Therefore, the posterior calculated using the joint likelihood appears similar to the posterior calculated conditional on $CH$ only. Inference of $\delta_2$ and $\delta_3$ should mainly depend on the $CH$ data. However, we do observe that the conflicting information from $RT$ does brings more uncertainty to the approximate posterior calculated by joint likelihood, which shows larger variance than the approximate posterior calculated conditional on $CH$ only.

\subsection{Extra findings - MLBA example (empirical case)}
\begin{figure}[h]
    \centering
    \begin{subfigure}{0.4\linewidth}
\includegraphics[width = \linewidth,height = 3.8cm]{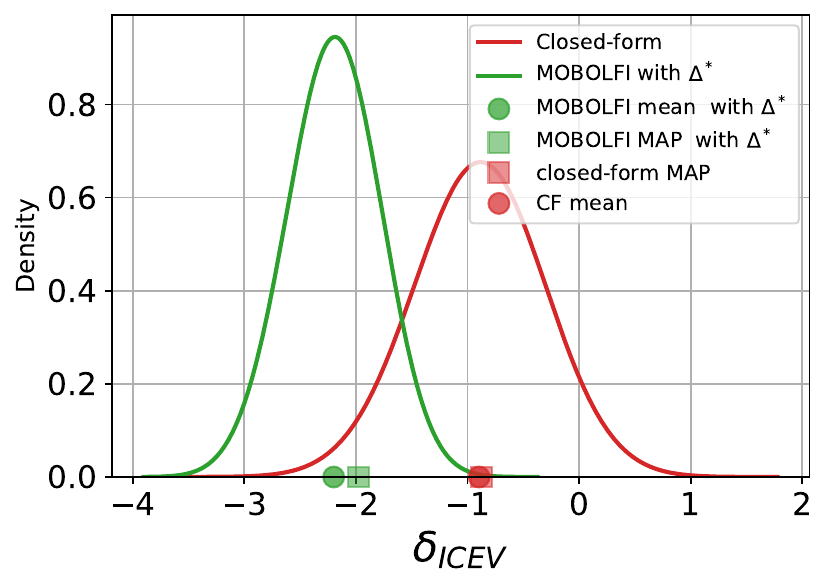}
         \label{fig: MLBA_em_delta2}
    \end{subfigure}~
    \begin{subfigure}{0.4\linewidth}
    \includegraphics[width = \linewidth,height = 3.8cm]{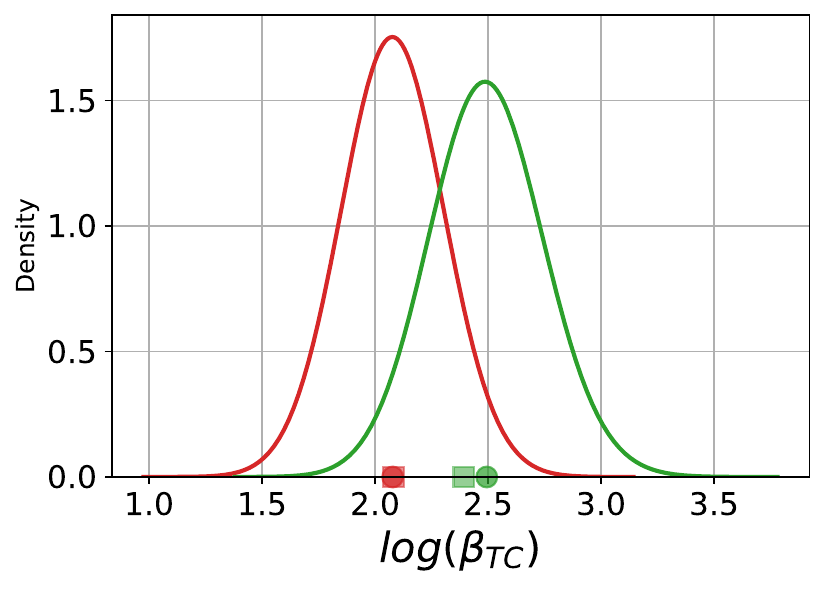}
         \label{fig: MLBA_em_logTC2}
    \end{subfigure}
    \caption{The posterior density and point estimates of MOBOLFI and closed-form based on CH and RT data sources.}
    \label{fig:MLBA_post_part}
    
\end{figure}

Figure \ref{fig:MLBA_post_part} shows the difference between the posterior mean and MAP estimates from the MOBOLFI approximate posterior and closed-form true posterior.  The 
largest differences are for $\delta_{ICEV}$ and $\log{\beta_{TC}}$.  Point estimates from the MOBOLFI approximate posterior have smaller value of $\hat{\delta}_{ICEV}$ and larger value on $\log{\hat{\beta}_{TC}}$, when compared to estimates from closed-form posterior. 
\FloatBarrier
\FloatBarrier

\section{Day care center example}

~~~~Next we consider data from \citet{numminen2013estimating} measuring infections of bacteria strains over individuals
in a number of day care centres.  
We refer the reader to \citet{numminen2013estimating} for a detailed description
of the model. The data are indicators for whether individuals are infected
with different strains at different times.  We use the simplified model in Example 3,
\citet{gutmann+c16}, and in their notation $R_s(t)$ is the rate of infection
with strain $s$ at time $t$.  $R_s(t)$ is informed as a weighted sum of a probability
for infection from within the day care centre (which may vary by strain and time)
and a probability of infection from outside (which varies by strain).  The weights
in the weighted sum are unknown parameters, denoted $\beta$ and $\Lambda$.  
There is a further unknown parameter, which we denote by $\vartheta$, which
controls the relative rate of infection with strain $s$ between situations
when an individual is already infected with another strain, compared to when
they are not.  The set of all unknowns is $\theta=(\beta,\Lambda,\vartheta)$.

\begin{figure}[h]
\centering
\begin{subfigure}{0.9\linewidth}
         \includegraphics[width=1.0\linewidth,height=3.8cm]{figures/dcc_MOBOLFIvsBOLFI.pdf}
         \caption{Joint Likelihood}
         \label{dcc:BOLFI VS MOBOLFI}
\end{subfigure}
\begin{subfigure}{0.9\linewidth}
         \includegraphics[width=1.0\linewidth,height=3.8cm]{figures/dcc_MOBOLFI_vsiter.pdf}
         \caption{Varying the number of BO iterations}
         \label{dcc:MOBOLFI_vsiter}
\end{subfigure}
\begin{subfigure}{0.9\linewidth}
         \includegraphics[width=1.0\linewidth,height=3.8cm]{figures/dcc_MOBOLFI_vstol.pdf}
         \caption{Varying tolerances}
         \label{dcc:MOBOLFI_vstol}
\end{subfigure}
\caption{Approximate posteriors for the day care center example:  (a) estimated posteriors using the joint likelihood, and the blue/orange curves show MOBOLFI/BOLFI approximate posterior; 
(b) estimated posteriors using varying number of BO iterations, and the blue/orange/green curves show MOBOLFI approximate posteriors with 250/150/50 BO iterations;  (c) estimated posteriors for different tolerances and the blue/orange/green curves compare MOBOLFI approximate posteriors using 1\%/5\%/20\% tolerance levels. The dashed blue lines show the location of the true parameter values.The approximate
posteriors are kernel density estimates based on importance sampling.}
\label{dcc_BOLFIvsMOBOLFIgroup}
\end{figure}

~~~~Following \citet{gutmann+c16} we consider synthetic data where the 
the number of day care centres, the number of attendee individuals sampled in each day care center and the number of strains are set to 29, 36, 33 respectively. We generate the observed data from the model with $\theta^{\text{true}} = (3.6, 0.6, 0.1)$, and implement MOBOLFI with 50/150/250 BO iterations and tolerances set as 20\%/5\%/1\% quantiles of the training data discrepancies. We consider the same four summary statistics $S=(S_1,S_2,S_3,S_4)^\top$ as
\citet{numminen2013estimating}, which can be computed for each daycare centre and 
averaged. The prior is the Uniform distribution illustrated in Example 8, \citet{gutmann+c16}. A discrepancy for BOLFI is then obtained by the Euclidean distance between
simulated and observed summary statistics.  
For the MOBOLFI method, we consider two discrepancies: the first
is the Euclidean distance between the first and fourth summary statistics for simulated and observed data, and
the second is the Euclidean distance between the second and third summaries.  
The reason for this partitioning is that the first and fourth summaries
represent diversity of prevalence of strains, and hence it could be
beneficial to consider a discrepancy for these summaries separately. In this example, discrepancies are not scaled, yet a MAD scaling (see algorithm \ref{alg:scaling}) is applied to the four data summaries in simulation. We consider approximate posterior densities obtained as kernel density estimates from an importance sampling with size 20000, and compare MOBOLFI approximate posterior densities with those obtained
from BOLFI. The choice of auxiliary density for importance sampling has some flexibility, for results here we use a multivariate student-t distribution, with location middle points of the uniform prior and degree of freedom 3. No rejection criteria are used.

~~~~Figure \ref{dcc_BOLFIvsMOBOLFIgroup} shows approximate BOLFI and MOBOLFI posteriors. In Figure \ref{dcc:BOLFI VS MOBOLFI}, the MOBOLFI approximate posterior exhibits similar variance compared to BOLFI.  Figure \ref{dcc:MOBOLFI_vsiter} compares approximate posteriors trained with different numbers of BO iterations. With more training iterations, the approximate marginal posterior densities stabilize. Figure \ref{dcc:MOBOLFI_vstol} shows approximate posteriors with different tolerance levels, using 250 BO iterations. Inference with a 1\% quantile tolerance appears superior.

\end{document}